\newif\ifnotblinded
\notblindedtrue  % Comment out to hide authors

\documentclass[a4paper, 11pt]{article}
\usepackage{amsmath}
\usepackage{mathtools}
\usepackage{amssymb}
\usepackage{mathrsfs}
\usepackage{algorithm}
\usepackage{algpseudocode}
\usepackage{setspace}
\usepackage{xcolor}
\usepackage{graphicx}
\usepackage{tikz}
\usepackage{natbib}
\usepackage{caption}
\usepackage{subcaption}
\usepackage{booktabs}
\usepackage{multirow}
\usepackage{minibox}
\usepackage{placeins}
\usepackage{authblk}
\usepackage{enumitem}
\usepackage{hyperref}
\usepackage[normalem]{ulem}
\usepackage[hmargin = 3cm, vmargin = 2.5cm]{geometry}

\newcommand{\E}{\mathop{\mathbb{E}}}
\newcommand{\Var}{\mathop{\mathrm{Var}}}

\newcommand{\ddd}{,\ldots,}
\newcommand{\vbar}{\:\vert\:}
\newcommand{\bvbar}{\:\big\vert\:}

\newcommand{\vech}{\mathop{\mathrm{vech}}}
\newcommand{\eig}{\mathop{\mathrm{eig}}}

\newcommand{\rd}[1]{\textcolor{red}{#1}}

\newcommand{\change}[1]{\textcolor{black}{#1}}

\DeclareMathOperator*{\argmin}{arg\,min}

\mathtoolsset{showonlyrefs}

\begin{document}
\title{Dynamic Quantile Function Models}
\ifnotblinded
\author[1]{Wilson Ye Chen}
\author[2]{Gareth W. Peters}
\author[3]{Richard H. Gerlach}
\author[4]{Scott A. Sisson}
\affil[1]{University of Sydney, Australia}
%\affil[1,4]{Australian Research Council Centre of Excellence for Mathematical and Statistical Frontiers} % [Scott] This is not an affiliation, but rather a grant you acknowledge
\affil[2]{Heriot-Watt University, UK}
\affil[3]{University of Sydney, Australia}
\affil[4]{University of New South Wales, Australia}
\else
\author{}
\fi
%\date{September 5, 2017}
\maketitle

\begin{abstract}
\change{Motivated by the need for effectively summarising, modelling, and forecasting the distributional characteristics of intra-daily returns, as well as the recent work on forecasting histogram-valued time-series in the area of symbolic data analysis, we develop a time-series model for forecasting quantile-function-valued (QF-valued) daily summaries for intra-daily returns. We call this model the dynamic quantile function (DQF) model. Instead of a histogram, we propose to use a $g$-and-$h$ quantile function to summarise the distribution of intra-daily returns. We work with a Bayesian formulation of the DQF model in order to make statistical inference while accounting for parameter uncertainty; an efficient MCMC algorithm is developed for sampling-based posterior inference. Using ten international market indices and approximately 2,000 days of out-of-sample data from each market, the performance of the DQF model compares favourably, in terms of forecasting VaR of intra-daily returns, against the interval-valued and histogram-valued time-series models. Additionally, we demonstrate that the QF-valued forecasts can be used to forecast VaR measures at the daily timescale via a simple quantile regression model on daily returns (QR-DQF). In certain markets, the resulting QR-DQF model is able to provide competitive VaR forecasts for daily returns.}

\vspace{0.5cm}
\end{abstract}

\begin{onehalfspace}
\section{Introduction}
\change{Modelling and forecasting the distributional characteristics of financial asset returns is an important task underlying many areas of financial econometrics. For example,  volatility, which corresponds to the second moment of the conditional distribution of returns, is a crucial input to pricing models for financial instruments and asset allocation strategies. Another example is Value-at-Risk (VaR), which is given by a quantile of the conditional return distribution. Since its introduction in 1994 as an integral part of J.P.~Morgan's RiskMetrics, VaR has become a standard risk measure for guiding investment decisions and regulatory capital allocation.}

\change{Over the past two decades, the rapidly increasing availability of high-frequency data has led to the development of a variety of approaches for modelling and forecasting a daily summary of intra-daily returns. This daily summary can then be incorporated into a model for a distributional characteristic of daily returns, often resulting in an improved forecasting performance compared to models using only daily returns. The most notable example of such summary is the daily realised volatility (RV), obtained as the sum of the squared intra-daily returns within the day. Compared to squared daily returns, RV is a much less noisy estimator of unobserved daily return volatility accompanied by extensive theoretical developments based on quadratic variation \citep{AndersenBollerslev1998, AndersenEtAl2001, BarndorffNielsenShephard2002, Meddahi2002}. While some authors \citep{MartensVanDijkDePooter2004, GhyselsSantaClaraValkanov2006, Corsi2009} focused on the problem of modelling the dynamic behaviour of the time-series of RV measurements, others \citep{AndersenEtAl2003, GiotLaurent2004, ClementsGalvaoKim2008, MaheuMcCurdy2011} considered incorporating a point forecast of RV, produced by a parametric time-series model, into a model for VaR or the conditional distribution of daily returns.}

\change{RV reduces intra-daily returns to a single-numbered summary characterising only the scale of the return distribution; any information provided by the sign or proportion of extreme observations of intra-daily  returns is lost. More recently, few papers considered summaries of intra-daily returns beyond the second-moment, and how such summaries can be modelled and forecasted. For example, in \citet{ArroyoMate2009}, \citet{ArroyoGonzalezRiveraMate2010}, \citet{ArroyoEtAl2011}, and \citet{GonzalezRiveraArroyo2012}, histograms and intervals are used as lower-frequency summaries of high-frequency returns. Specifically, \citet{GonzalezRiveraArroyo2012} modelled the time-series of daily histograms of intra-daily returns of the S\&P 500 equity index, where each histogram is partitioned into ten bins, each containing 10\% of the intra-daily returns. Additionally, the authors analysed each bin of the histogram separately as an interval-valued time-series. \citet{ArroyoEtAl2011} employed an exponential smoothing filter to forecast the daily histograms of intra-daily returns for the S\&P~500 and IBEX~35 indices, and compared models in terms of VaR forecasts retrieved from the histogram forecasts.}

\change{In an emerging area of statistics known as ``symbolic data analysis" (SDA) \citep{BillardDiday2003, Billard2011, BerangerLinSisson2020}, random intervals and histograms are commonly employed symbol types as group-level summaries \citep{DiasBrito2015, HronBritoFilzmoser2017}. SDA is concerned with performing exploratory analyses, forecasts, and statistical inference based on a collection of group-level distributional-valued summaries (i.e., symbols), where the statistical unit of interest is the symbol, and the analysis is required at the symbol level.  A key advantage of SDA is that a symbol represents a compressed version of a set of individual-level observations; large and complex datasets are replaced by much smaller sets of group-level symbol-valued observations \citep{BerangerLinSisson2020}. Admittedly, unless sufficient statistics are available, compression is at the cost of information loss. An optimally chosen symbol type should maximise the information content with respect to the population parameter of interest. Because SDA, at its core, is concerned with the modelling of distributional summaries, it provides a natural framework for analysing the distributional characteristics of intra-daily asset returns. In this paper, we explore the idea of employing the symbolic data approach for modelling and forecasting the daily distributions of intra-daily returns of ten major international equity indices.}

\change{While an interval is able to encode information on both the location and scale of the individual observations, a histogram provides a nonparametric summary for the entire distribution of intra-daily returns, and thus can capture information on higher moments. There are three potential weaknesses with histogram summaries. Firstly, the construction of histogram is not unique for a given dataset; it is not clear how to optimally set the number and the locations of the bin boundaries, which strongly influence the shape of the distributional summary. Some authors considered constructing unique histograms with the least number of (possibly unequal-width) bins for a chosen level of information content \citep{Herrholz2010, LiEtAl2020}. Secondly, it is difficult to model extreme quantiles since there is not sufficient data to fill the bins located too far into the tails. For this reason, \citet{ArroyoEtAl2011} chose to exclude both the 1\% and 99\% quantiles from their study. Thirdly, in general, a histogram does not characterise the distribution of a random variable.}

\change{Given the drawbacks of histograms, we propose to consider a flexible parametric quantile function as a new symbol type for summarising the intra-daily returns on a given day. Specifically, a quantile-function-valued (QF-valued) time-series model, termed the Dynamic Quantile Function (DQF) model, is developed, where each observation is a quantile function of the $g$-and-$h$ distribution. There are several advantages for employing the $g$-and-$h$ quantile function as the symbolic representation. Firstly, via the L-moment estimator \citep{PetersChenGerlach2016}, the intra-daily returns are uniquely and parsimoniously summarised by the four parameters of the $g$-and-$h$ quantile function. Secondly, the $g$-and-$h$ distribution (defined only through its quantile function) is able to accommodate a wide range of tail-thickness and skewness, and closely approximate a broad spectrum of distributions \citep{DuttaPerry2006}. For example, compared to other flexible parametric families such as the generalised Beta, exponential generalised Beta, skewed generalised t, and inverse hyperbolic sine, the $g$-and-$h$ distribution has the least restrictive bound on the skewness-kurtosis combination \citep{McDonaldMichelfelder2016}. Thirdly, the $g$-and-$h$ quantile function can accurately summarise the tail behaviour of the intra-daily returns by utilising all the realised returns in a day.}

\change{By adopting the approach of \citet{LeRademacherBillard2011} and \citet{BritoDuarteSilva2012} for constructing likelihood functions for symbolic data (see \citet{ZhangSisson2017} for an alternative approach), we are able to perform parameter estimation and inference in a computational Bayesian framework via a carefully designed adaptive Markov chain Monte Carlo sampling algorithm. Bayesian posterior sampling provides a natural mechanism for dealing with parameter uncertainty. For example, when computing a point forecast from the proposed DQF model, as opposed to plugging in the single best estimate for the parameter vector $\boldsymbol{\theta}$, a Bayesian forecast is given by averaging over all possible values of $\boldsymbol{\theta}$ weighted by the posterior distribution. Furthermore, the Bayesian formulation allows us to impose parameter constraints and shrinkage in a coherent and flexible manner via prior distributions. Finally, compared to maximum likelihood, which involves solving a numerically challenging high-dimensional constrained optimisation problem, the Bayesian approach allows us to take advantage of the numerical robustness associated with sampling-based methods for parameter estimation \citep{BriolEtAl2017, GerlachWang2020}.}

\change{The DQF model is applied to model and forecast one-step-ahead the daily quantile functions of intra-daily returns. Using ten international market indices, and approximately 2,000 days of out-of-sample data from each market, the performance of the proposed DQF model is compared, in terms of VaR forecasts, against the interval- and histogram-valued time-series models of \cite{ArroyoGonzalezRiveraMate2010}, \cite{ArroyoEtAl2011}, and \cite{GonzalezRiveraArroyo2012}. Forecasting of VaR is a natural application for the DQF model, since quantile forecasts at any threshold are directly available from the QF-valued forecasts by evaluating the predicted $g$-and-$h$ quantile functions. Focusing on the lower-tail of the return distribution, VaR forecasts are compared at 5\% and 1\% threshold levels. The DQF model significantly outperforms the previous models based on interval- and histogram-valued summaries, and even more so at the more extreme 1\% level. In addition to modelling and forecasting QF-valued summaries of intra-daily returns, we show that the QF-valued forecasts can be used as an input to a model of lower-frequency returns. To demonstrate, we forecast the lower-tail VaR measures of daily returns using a standard simple quantile regression as the lower-frequency model.}

\change{The sections are organised as follows. In Section~\ref{sec:dqf_models}, we set the notation and introduce the DQF model class, focusing on the $gh$-DQF model. The associated likelihood, Bayesian formulation, and the adaptive MCMC sampling algorithm used for parameter estimation are discussed in Section~\ref{sec:bayesian}. A simulation study of the proposed MCMC estimator is presented in Section~\ref{sec:sim_study}. In Section~\ref{sec:emp_study}, the $gh$-DQF model is applied to analyse the equity indices of ten international markets, and the out-of-sample performance is assessed in terms of VaR forecasts. Section~\ref{sec:conclusion} summarises the findings and concludes the article.}

\section{Dynamic Quantile Function Models}
\label{sec:dqf_models}
\subsection{Preliminaries}
\label{sec:preliminaries}
\change{In this section, we set up the notation and probabilistic framework for QF-valued time-series. Early papers on symbol-valued time-series models considered intervals and histograms as the daily summaries of intra-daily returns \citep{ArroyoMate2009, ArroyoGonzalezRiveraMate2010, ArroyoEtAl2011, GonzalezRiveraArroyo2012}, where forecasting was performed without specifying a probabilistic model for the symbols that approximates the underlying data generating process. \citet{LeRademacherBillard2011} and \citet{BritoDuarteSilva2012} proposed a method for indirectly defining a probabilistic generative model for symbolic data $\{X_t \in \mathbb{S}\}$ as the push-forward measure of a distribution on $\mathbb{R}^p$ under the mapping $\mathcal{M}^{-1}:\mathbb{R}^p \to \mathbb{S}$. Here we provide a summary of the likelihood construction method of \cite{LeRademacherBillard2011} and \citet{BritoDuarteSilva2012}, which we employed for designing a Bayesian model for QF-valued time-series.} 

We consider a \emph{quantile function} as an observation. Let $(\Omega,\, \mathcal{A},\, P)$ be a probability space, where $\Omega$ is the reference space with $\omega \in \Omega$ being an element, $\mathcal{A}$ is a $\sigma$-algebra of subsets of $\Omega$, and $P$ is a probability measure over $\mathcal{A}$. Let $X$ be a real-valued function of two variables, $u \in [0,\, 1]$ and $\omega \in \Omega$. If $u$ is fixed, $X(u,\, \cdot)$ is a real-valued random variable defined on $(\Omega,\, \mathcal{A},\, P)$. If $\omega$ is fixed, $X(\cdot,\, \omega)$ is a real-valued function on $[0, 1]$ belonging to some function space $\mathbb{S}$. If $\mathbb{S}$ is restricted to be the subset of all quantile functions, i.e.
\begin{equation}
\label{eq:symbol_set}
\mathbb{S} \subseteq \{(Q \colon\, [0,1] \to \mathbb{R})\colon\, Q(a) < Q(b),\, \forall a < b\},
\end{equation}
and $\omega$ is allowed to vary, then $X$ is called a \emph{quantile-function-valued (QF-valued) random variable}. A \emph{QF-valued discrete-time stochastic process} is a set of QF-valued random variables indexed by integers. We use the notation $\{X_{t}\}$ to refer to such processes. I.e. $\{X_{t}\}$ is the abbreviated notation for $\{X_{t}\colon\, t \in \mathbb{Z}\}$.

We can view a dynamic model for \emph{QF-valued time-series} in general as
\begin{equation}
\label{eq:Xt_tilde}
\tilde{X}_{t} = \rho(\{X_{s}\colon\, s \le t-1\}),\;\: \tilde{X}_{t} \in \mathbb{S},
\end{equation}
where $\tilde{X}_{t}$ is an one-step-ahead forecast generated by a function $\rho$ of the set of all QF-valued observations up to $t-1$. Given a sample of QF-valued realisations $\{X_{1} \ddd X_{T}\}$, a forecasting tool can be found by first defining a loss functional for quantile functions $L\colon \mathbb{S}\times\mathbb{S} \to \mathbb{R}$, then minimising the overall loss $\sum_{t=1}^{T} L(\tilde{X}_{t},\, X_{t})$ with respect to $\rho$. A generative model is constructed by defining a conditional distribution $F_{\mathrm{X},\, t}$ such that
\begin{equation}
\label{eq:gen_Xt}
X_{t} \vbar \mathcal{G}_{t-1} \sim F_{\mathrm{X},\, t},
\end{equation}
where $\mathcal{G}_{t-1} = \sigma(\{X_{s}\colon\, s \le t-1\})$ denotes the smallest $\sigma$-algebra containing the past observations of the process, and represents the available information at $t - 1$. The filtration $\mathcal{G}_{t-1}$ is referred to as the \emph{natural filtration}. \change{Analogous to scalar-valued time-series, where a point forecast may be taken as the mean-, median-, or a quantile-functional of the predictive distribution \citep{Gneiting2011}, a QF-valued point forecast} $\tilde{X}_{t}$ can then be taken as a functional of the predictive distribution $F_{\mathrm{X},\, t}$. Looking for a generative model is a more challenging task than building a forecasting tool, as the notion of a distribution function defined on a function space is in general not straightforward. A detailed explanation can be found in \citet{DelaigleHall2010} and \citet{Cuevas2014}.

The approach adopted in this article is to develop a generative model for a QF-valued time-series indirectly, by first finding a suitable low-dimensional parameterisation for the observed quantile functions, and then specifying a generative model for the time-series of mapped low-dimensional vectors. 
% The strategy of indirectly modelling the symbolic observations through the modelling of their mapped vectors was proposed by \citet{LeRademacherBillard2011} and \citet{BritoDuarteSilva2012}, where the authors applied it to interval-valued and histogram-valued data.
Suppose that we define a parameterisation of $X_{t}$ that maps a symbolic observation to a $p$-dimensional vector,
\begin{equation}
\label{eq:param_Xt}
\mathcal{M}: \mathbb{S} \rightarrow \mathbb{R}^{p},
\end{equation}
so that we are able to obtain a vector
\begin{equation}
\label{eq:xit}
\boldsymbol{\xi}_{t} = \mathcal{M}(X_{t}),
\end{equation}
and define a conditional distribution $F_{t}$ on $\mathbb{R}^{p}$ such that
\begin{equation}
\label{eq:gen_xit}
\boldsymbol{\xi}_{t} \vbar \mathcal{F}_{t-1} \sim F_{t},
\end{equation}
where $\mathcal{F}_{t-1} = \sigma(\{\boldsymbol{\xi}_{s}\colon\, s \le t-1\})$. \change{If $\mathcal{M}$ is a measurable mapping}, the distribution $F_{t}$ corresponds to a generative model for $X_{t}$, namely, the push-forward of $F_{t}$ under $\mathcal{M}^{-1}$. \change{In other words, for any fixed $u \in [0,1]$, the distribution of the random variable $X_t(u)$ corresponds to the distribution of the transformed random vector $\mathcal{M}^{-1}(\boldsymbol{\xi}_t)$. A possible choice for the QF-valued point forecast is the conditional point-wise mean function given by $\mu_{\mathrm{X},t}(u) := \E[X_t(u) \vbar \mathcal{G}_{t-1}],\, \forall u \in [0,1]$, where the expectation is with respect to $F_{\mathrm{X},t}$. Although $\mu_{\mathrm{X},t}$ is a valid quantile function, it is not necessarily still an element of $\mathbb{S}$. Instead of $\mu_{\mathrm{X},t}$, we choose the QF-valued one-step-ahead point forecast to be the quantile function whose mapped vector is the conditional expectation of $\boldsymbol{\xi}_{t}$ with respect to $F_t$,}
\begin{equation}
\label{eq:xt_tilde_via_xi}
\tilde{X}_{t} := \mathcal{M}^{-1}(\E[\boldsymbol{\xi}_{t} \vbar \mathcal{F}_{t-1}]).
\end{equation}
\change{It can be shown via simulation that, under the proposed $gh$-DQF model (introduced later in Section~\ref{sec:gh-dqf}), $\mu_{\mathrm{X},t}$ is closely approximated by $\tilde{X}_t$.} The \emph{modelling task} is then to define the collection $\{\mathbb{S},\, \mathcal{M},\, F_{t}\}$. Notice that the mapping $\mathcal{M}$ is non-unique and the properties of the model greatly depends on its choice. In the subsequent sections, we will introduce one particular choice of $\mathcal{M}$ based on the family of $g$-and-$h$ distributions.

\subsection{The $g$-and-$h$ Distributions}
\label{sec:gh}
We adopt a parametric approach and assume that $X_{t}$ has the functional form of the quantile function of a $g$-and-$h$ distribution, that is, we define the symbol set as $\mathbb{S} := \{\text{$g$-and-$h$ quantile functions}\}$. The $g$-and-$h$ family of distributions was first introduced by \citet{Tukey1977} and further developed by \citet{MartinezIglewicz1984} and \citet{Hoaglin1985}. It is generated by a transformation of a standard normal random variable which allows for asymmetry and heavy tails. Specifically, let $z$ be a standard normal random variable, and let $a \in \mathbb{R}$, $b \in (0, \infty)$, $g \in \mathbb{R}$, and $h \in [0, \infty)$ be constants. The random variable $y$ is said to follow a $g$-and-$h$ distribution if it is given by the transformation,
\begin{equation}
\label{eq:gh_trans}
y := a + bG(z)H(z)z,
\end{equation}
where
\begin{equation}
\label{eq:g_func}
G(z) := \frac{\exp(gz) - 1}{gz}
\end{equation}
and
\begin{equation}
\label{eq:h_func}
H(z) := \exp\left(\frac{hz^{2}}{2}\right).
\end{equation}
Note that the same transformation can be applied to any ``base" random variable. It can be seen from \eqref{eq:gh_trans} that $a$ and $b$ account for location and scale, respectively. It can be checked from \eqref{eq:g_func} that the reshaping function $G$ is bounded from below by zero, that it is either monotonically increasing or monotonically decreasing for $g$ being, respectively, positive or negative, and that by rewriting it as its series expansion,
\begin{equation}
\label{eq:g_ex}
G(z) = 1 + \frac{gz}{2!} + \frac{(gz)^{2}}{3!} + \frac{(gz)^{3}}{4!} + \cdots,
\end{equation}
$G$ is equal to one at zero for all $g$. Thus $G$ generates asymmetry by scaling $z$ differently for different sides of zero via the parameter $g$. Furthermore, as $G(z; g) = G(-z; -g)$, the sign of $g$ affects only the direction of skewness. For $g = 0$, by equation \eqref{eq:g_ex}, the constant function $G(z) = 1$ is obtained, and thus the symmetry remains unmodified. For $h > 0$, $H$ is a strictly convex even function with $H(0) = 1$, and thus it generates heavy tails by scaling upward the tails of $z$ while preserving the symmetry. When $h = 0$, the transformation given by \eqref{eq:gh_trans} generates the subfamily of $g$-distributions, which coincides with the family of shifted log-normal distributions for $g > 0$. When $g = 0$, the transformation generates the subfamily of $h$-distributions, which is symmetric and has heavier tails than normal distributions.

% \textcolor{red}{Add a paragraph on the theoretical work on characterising the tail behavior of heavy-tailed distributions, i.e., the relationship between $x$ and $1 - F(x)$ as $x$ becomes large. E.g., $\alpha = 1/h$, and Cauchy distribution is well approximated by an h-distribution with $h$ close to 1.}

\subsection{The $gh$-DQF Model}
\label{sec:gh-dqf}
As the transformation given by \eqref{eq:gh_trans} is monotonically increasing as long as $h > 0$, the quantile function of the $g$-and-$h$ distributions is explicitly available. As discussed in Section~\ref{sec:gh}, we assume that $X_{t}$ is the quantile function of a $g$-and-$h$ distribution, so that
\begin{equation}
\label{eq:gh}
X_{t}(u) :=
\begin{dcases}
a_{t} + b_{t} \frac{\exp(g_{t} Z(u)) - 1}{g_{t}} \exp\left(\frac{h_{t} Z(u)^{2}}{2}\right) & \text{if $g_{t} \neq 0$}, \\
a_{t} + b_{t} Z(u) \exp\left(\frac{h_{t} Z(u)^{2}}{2}\right) & \text{if $g_{t} = 0$},
\end{dcases}
\end{equation}
where $Z$ is the quantile function of the standard normal distribution, $a_{t} \in \mathbb{R}$, $b_{t} \in (0, \infty)$, $g_{t} \in \mathbb{R}$, and $h_{t} \in [0, \infty)$ are parameters responsible for location, scale, asymmetry, and heavy-tailedness, respectively. We then choose the parameterisation $\mathcal{M}$ to be
\begin{equation}
\label{eq:xit_param}
\boldsymbol{\xi}_t := (\xi_{1,\, t},\, \xi_{2,\, t},\, \xi_{3,\, t},\, \xi_{4,\, t}) := (a_{t},\, b^{*}_{t},\, g_{t},\, h_{t}),
\end{equation}
where $b^{*}_{t} := \log(b_{t})$. As $b_{t}$ is a positive scale parameter, its natural logarithm is used for subsequent modelling convenience.

\subsection{Estimating the $g$-and-$h$ Parameters}
\label{sec:fitting_gh}
\begin{figure}[h]
\centerline{\includegraphics[width = 0.7 \textwidth]{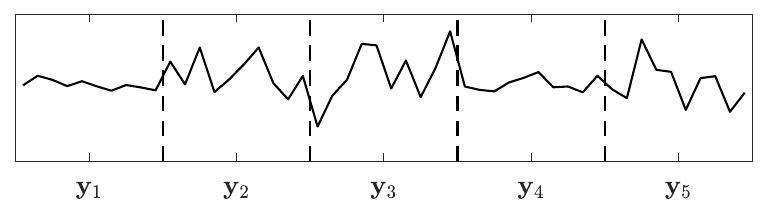}}
\caption{\small Illustration of constructing a sequence of vectors $\{\mathbf{y}_{t}\}$ by cutting a long time-series into pieces.}
\label{fig:y_vec}
\end{figure}

Up until this point, we have been treating $\{X_{1} \ddd X_{T}\}$ as data that are directly observable. However, infinite dimensional quantile functions can never be observed in reality; only the realised order statistics are observed. QF-valued observations must therefore be constructed using scalar-valued observations. Let $\{\mathbf{y}_{1} \ddd \mathbf{y}_{T}\}$ denote a sequence of vectors, where, for each $t \in \{1 \ddd T\}$, the vector $\mathbf{y}_{t} \in \mathbb{R}^{n_{t}}$ denotes a sample of $n_{t}$ scalar-valued observations. One way of constructing the sequence $\{\mathbf{y}_{t}\}$ is by partitioning a long time-series $\{y_{1} \ddd y_{N_{T}}\}$ into $T$ consecutive pieces, where $N_{t} = \sum_{i=1}^{t} n_{i}$, so that $\mathbf{y}_{t} = (y_{N_{t-1}+1} \ddd y_{N_{t}})$ contains the $n_{t}$ observations belong to the $t$-th piece, as illustrated in Figure~\ref{fig:y_vec}.
\FloatBarrier

Let $S: \mathbb{R}^{n} \to \mathbb{S}$, for $n \in \mathbb{N}$, denote a summarisation function. The sequence $\{X_{1} \ddd X_{T}\}$ is obtained via
\begin{equation}
\label{eq:Xt_const}
X_{t} = S(\mathbf{y}_{t}),
\end{equation}
for each $t \in \{1 \ddd T\}$. In the case where $\mathbb{S}$ is the set of $g$-and-$h$ quantile functions, the summarisation function $S$ corresponds to an estimator for the parameters of the $g$-and-$h$ quantile function. Several studies in the statistics literature have been performed on the estimation of parametric quantile functions, such as those based on numerical likelihood \citep{RaynerMacGillivray2002, HossainHossain2009}, matching quantiles \citep{XuIglewiczChervoneva2014}, matching moments \citep{HeadrickKowalchukSheng2008}, and Bayesian methods \citep{HaynesMengersen2005, PetersSisson2006, AllinghamKingMengersen2009}. We employ a method developed by \citet{PetersChenGerlach2016} based on L-moments which shows favourable statistical properties while being computationally simple compared to previously proposed methods. It is shown via simulations that the parameter estimates from the L-moment method have the smallest mean-squared-error compared to those from methods based on numerical likelihood, conventional moments, and quantiles. The details on the L-moment method is given in Appendix~\ref{app:lmoment}.

\subsection{Modelling the Conditional Joint Distribution of $\boldsymbol{\xi}_t$}
\label{sec:xit_dist}
We assume a flexible model in which the conditional joint distribution of $\boldsymbol{\xi}_{t}$ is defined by a copula and univariate conditional marginal distributions, denoted by
\begin{equation}
\label{eq:xi_cdf}
\boldsymbol{\xi}_{t} \vbar \mathcal{F}_{t-1} \sim F_{t} := C(F_{1,t} \ddd F_{4,t}),
\end{equation}
where $C:[0,1]^{4} \to [0,1]$ is the copula function that maps the conditional marginal distributions $\{F_{i,t}\}$ to the conditional joint distribution $F_{t}$. \change{To account for central and tail dependence while being parsinonious, we choose $C$ as the Student-$t$ copula \citep{DemartaMcNeil2005}.} Let $u_{i,t} := F_{i,t}(\xi_{i,t})$ and $\mathbf{u}_{t} := (u_{1,t} \ddd u_{4,t})$. The conditional joint density of $\boldsymbol{\xi}_t$ implied by the distribution function in \eqref{eq:xi_cdf} is
\begin{equation}
\label{eq:xi_pdf}
f_{t}(\boldsymbol{\xi}_{t}) = \varpi(\mathbf{u}_{t}) \prod_{i=1}^{4}f_{i,t}(\xi_{i,t}),
\end{equation}
where $\varpi$ is the Student-$t$ copula density and $f_{i,t}$ is the conditional marginal density. The $t$ copula density is given by
\begin{equation}
\label{eq:c_pdf}
\varpi(\mathbf{u}_{t}) := \frac{f_{\mathrm{MSt}}\left(F_{\mathrm{St}}^{-1}(u_{1,t}; \nu) \ddd F_{\mathrm{St}}^{-1}(u_{4,t}; \nu);\mathbf{R}, \nu\right)}{\prod_{i=1}^{4}f_{\mathrm{St}}\left(F_{\mathrm{St}}^{-1}(u_{i,t}; \nu); \nu\right)},
\end{equation}
where $f_{\mathrm{MSt}}$ is the multivariate $t$ density parameterised a correlation matrix $\mathbf{R}$ and degree-of-freedom $\nu$, $f_{\mathrm{St}}$ is the univariate $t$ density with $\nu$ degrees-of-freedom, and $F_{\mathrm{St}}^{-1}$ is the univariate $t$ distribution function with $\nu$ degrees-of-freedom.

\subsubsection{Conditional Marginal Distribution of $a_{t}$, $b^{*}_{t}$, and $g_{t}$}
\label{sec:abg_dist}
We model the conditional marginal distributions $F_{i,t}$ for $i \in \{1, 2, 3\}$, which correspond to the parameters $a_{t}$, $b^{*}_{t}$, and $g_{t}$, as follows.
\begin{equation}
\label{eq:sktmar}
\begin{aligned}
\xi_{i,t} &= \mu_{i,t} + \epsilon_{i,t}, \\
\epsilon_{i,t} &= \sigma_{i,t} v_{i,t}, \\
v_{i,t} &\sim F_{\mathrm{skt}}(\cdot; \eta_{i}, \lambda_{i}), \\
\mu_{i,t} &= \delta_{i} + \psi_{i} \xi_{i,t-1} + \phi_{i} \mu_{i,t-1}, \\
\sigma_{i,t}^{2} &= \omega_{i} + \alpha_{i} \epsilon_{i,t-1}^{2} + \beta_{i} \sigma_{i,t-1}^{2}. \\
\end{aligned}
\end{equation}
The innovation $v_{i,t}$ is generated from the skewed Student $t$ distribution of \citet{Hansen1994}, denoted by $F_{\mathrm{skt}}(\cdot; \eta_{i}, \lambda_{i})$, where $\eta_{i} \in (2,\infty)$ is the degrees-of-freedom parameter and $\lambda_{i} \in (-1, 1)$ the asymmetry parameter. The special cases $F_{\mathrm{skt}}(\cdot; \eta_{i}, 0)$ and $F_{\mathrm{skt}}(\cdot; \infty, 0)$ are the Student $t$ and the standard normal distributions, respectively. Furthermore, the skewed $t$ distribution is standardised so that $\E(v_{i,t}) = 0$ and $\Var(v_{i,t}) = 1$. More properties of Hansen's skewed $t$ distribution can be found in \citet{JondeauRockinger2003}. The model in \eqref{eq:sktmar} implies that the mean and variance of the conditional marginal distribution $F_{i,t}$ are given by $\E(\xi_{i,t} \vbar \mathcal{F}_{t-1}) = \mu_{i,t}$ and $\Var(\xi_{i,t} \vbar \mathcal{F}_{t-1}) = \sigma_{i,t}^{2}$, respectively. The dynamic properties of both $\{\mu_{i,t}\}$ and $\{\sigma_{i,t}^{2}\}$ are characterised by an extended form of exponential smoothing \citep{Bosq2015}. For the conditional variance to be positive, the conditions $\omega_{i} > 0$, $\alpha_{i} \ge 0$, and $\beta_{i} \ge 0$ are sufficient. Given that the positivity conditions are satisfied, the process $\xi_{i,t}$ is covariance stationary if $-1 < \psi_{i} + \phi_{i} < 1$ and $\alpha_{i} + \beta_{i} < 1$.

\subsubsection{The Family of Apatosaurus Distributions}
\label{sec:apat}
The tail shape parameter $h_{t}$ must be non-negative for the transformation in \eqref{eq:gh_trans} to be monotonically increasing in $z_{u}$ and thus one-to-one. It can be challenging to model the conditional distribution of $h_{t}$ or its logrithmic transformation, because $h_{t}$ can become empirically very close to zero for many days. Furthermore $h_{t}$ can also become very large (close to 0.5) occasionally, for example, on days when so-called ``flash-crashes" occur. For the above reasons, we develop a novel family of distributions, termed the Apatosaurus family, for the modelling of $h_{t}$, which shows a good fit to the data empirically.

The Apatosaurus is a family of non-negative distributions constructed using a mixture of a truncated-skewed-$t$ and an Exponential distribution. Depending on its parameters, the distribution can take on a variety of general shapes including having one mode at zero, one mode away from zero, and two modes. The random variable $h \sim F_{\mathrm{Apat}}(h; \mu, \sigma, \eta, \lambda, \iota, w)$ has a density function given by
\begin{equation}
\label{eq:apat_pdf_main_text}
f_{\mathrm{Apat}}(h; \mu, \sigma, \eta, \lambda, \iota, w) = wf_{\mathrm{TrSkt}}(h; \mu,\sigma,\eta,\lambda) + (1 - w)f_{\mathrm{Exp}}(h; \iota)
\end{equation}
for $h \in [0, \infty)$, where $f_{\mathrm{TrSkt}}$ and $f_{\mathrm{Exp}}$ are the density functions of a truncated-skewed-$t$ distribution and an Exponential distribution, and $w \in [0,1]$ is the mixing weight. The parameters $\mu$, $\sigma$, $\eta$, and $\lambda$ correspond to the mode, scale, degrees-of-freedom, and asymmetry of the truncated-skewed-$t$ component. The parameter $\iota$ is the mean of the Exponential component. Examples of the Apatosaurus density and distribution functions are plotted in Figure~\ref{fig:apat_pdf_cdf}. Additional details on the Apatosaurus family are presented in Appendix~\ref{app:apatosaurus}. The Halphen distribution system \citep{PerreaultBobeeRasmussen1999, PerreaultBobeeRasmussen1999b} is able to take on a qualitatively similar set of shapes, however for our purpose of modelling $h_{t}$, the Apatosaurus distributions show a better fit to the data compared to the Halphen system.

\begin{figure}[h]
    \centering
    \begin{subfigure}[t]{0.35\textwidth}
        \centering
        \includegraphics[height=5cm]{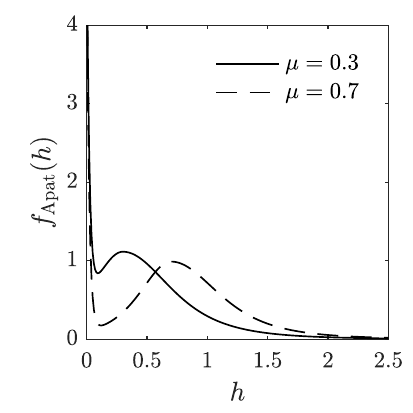}
        \caption{Density function}
    \end{subfigure}%
    ~
    \begin{subfigure}[t]{0.35\textwidth}
        \centering
        \includegraphics[height=5cm]{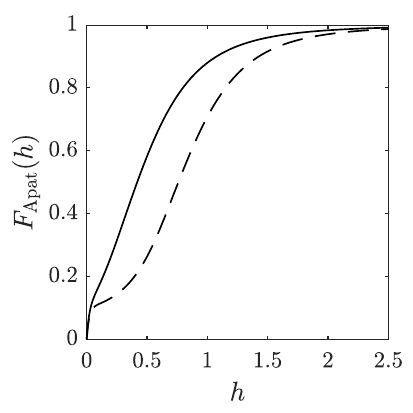}
        \caption{Distribution function}
    \end{subfigure}
    \caption{\small Plots of the density function $f_{\mathrm{Apat}}(h; \mu, 0.6, 3, 0.2, 0.02, 0.9)$ and the distribution function $F_{\mathrm{Apat}}(h; \mu, 0.6, 3, 0.2, 0.02, 0.9)$ of the Apatosaurus distribution for $\mu = 0.3$ (solid lines) and $\mu = 0.7$ (dashed lines). The distribution function $F_{\mathrm{Apat}}$ is defined in Appendix~\ref{app:apatosaurus}.}
\label{fig:apat_pdf_cdf}
\end{figure}
\FloatBarrier

\subsubsection{Conditional Distribution of $h_{t}$}
\label{subsubsec:h_dist}
Employing the Apatosaurus distribution family developed in section~\ref{sec:apat}, we model the conditional marginal distribution $F_{4,t}$, which correspond to the tail shape parameter $h_{t}$, as follows:
\begin{equation}
\label{eq:apatmar}
\begin{aligned}
\xi_{4,t} &\sim F_{\mathrm{Apat}}(\cdot; \mu_{t}, \sigma, \eta_{4}, \lambda_{4}, \iota, w_{t}), \\
\mu_{t} &= \delta_{4} + \psi_{4} \xi_{4,t-1} + \phi_{4} \mu_{t-1}, \\
w_{t} &= 0.5 + 0.5 / \left\{1 + \exp[-\gamma (\mu_{t} - c)]\right\}, \\
\gamma &= \exp(\gamma^{*}).
\end{aligned}
\end{equation}
We assume that $h_{t}$ follows an Apatosaurus distribution with time varying location and mixing weight parameters. Here the location parameter, $\mu_{t}$, is the \emph{mode} of the distribution, whose dynamics is given by the extended form of exponential smoothing. The conditional weight $w_{t}$ is linked to $\mu_{t}$ via a logistic function, parametrised by $\gamma > 0$ and $c \in [0,\, 1]$, where $\gamma$ controls how sensitive $w_{t}$ is to the changes in $\mu_{t}$, and $c$ determines the location at which $w_{t}$ is most sensitive to $\mu_{t}$. This logistic link function allows the conditional distribution of $h_{t}$ to shift more density to $h_{t} = 0$ for periods $\mu_{t}$ is closer to zero. It also restricts $w_{t}$ to lie inside $[0.5,\, 1]$, so that the truncated-skewed-t component stays dominant. The mean of the conditional marginal distribution $\E(\xi_{4,t} \vbar \mathcal{F}_{t-1})$ can be calculated using equation~\eqref{eq:apat_mean}. The conditions $\delta_{4} > 0$, $\psi_{4} \ge 0$, $\phi_{4} \ge 0$, and $\psi_{4} + \phi_{4} < 1$ are sufficient to ensure that $\mu_{t}$ is positive and non-divergent.

\section{Bayesian Inference of $gh$-DQF Model}
\label{sec:bayesian}
\subsection{Likelihood and Prior}
\label{subsec:likelihood_prior}
As we now have a generative model for a QF-valued time-series, following from \eqref{eq:xi_pdf}, the likelihood function of the $gh$-DQF model, denoted by $f$, can be written as
\begin{equation}
\label{eq:likelihood}
f(\boldsymbol{\xi}_{1} \ddd \boldsymbol{\xi}_T; \boldsymbol{\theta}) = \prod_{t=1}^{T} f_{t}(\boldsymbol{\xi}_{t}; \boldsymbol{\theta}),
\end{equation}
where $\boldsymbol{\theta}$ is the vector of model parameters. Let $\boldsymbol{\theta}_{i \in \{1, 2, 3\}} = (\delta_{i}, \psi_{i}, \phi_{i}, \omega_{i}, \alpha_{i}, \beta_{i}, \eta_{i}, \lambda_{i})$, $\boldsymbol{\theta}_{4} = (\delta_{4}, \psi_{4}, \phi_{4}, \gamma^{*}, c, \sigma, \eta_{4}, \lambda_{4}, \iota)$, and $\boldsymbol{\theta}_{\mathrm{c}} = (\vech(\mathbf{R})^{\top}, \nu)$, then $\boldsymbol{\theta} = (\boldsymbol{\theta}_{1} \ddd \boldsymbol{\theta}_{4}, \boldsymbol{\theta}_{\mathrm{c}})$. An improper prior is used for $\boldsymbol{\theta}$ over the allowable parameter region. Let the indicator function $I_{\mathbb{A}}$ take the value one if $\boldsymbol{\theta}$ is in the allowable region and zero otherwise. Specifically,
\begin{equation}
\label{eq:theta_ind}
I_{\mathbb{A}}(\boldsymbol{\theta}) =
\begin{dcases}
1, & \text{if } \boldsymbol{\theta} \in \mathbb{A} = \bigcap_{i=1}^{4} \mathcal{A}_{i} \\
0, & \text{otherwise},
\end{dcases}
\end{equation}
where
\begin{equation}
\label{eq:set_a}
\mathcal{A}_{i} = \left\{\boldsymbol{\theta} \:\left|\:
\minibox{
$-1 < \psi_{i} + \phi_{i} < 1$, \\
$\omega_{i} > 0,\, \alpha_{i} \ge 0,\, \beta_{i} \ge 0,\, \alpha_{i} + \beta_{i} < 1$, \\
$2 < \eta_{i} \le 40,\, -1 < \lambda_{i} < 1$, \\
$\delta_{4} \ge 0,\, \psi_{4} \ge 0,\, \phi_{4} \ge 0$, \\
$-6 \le \gamma^{*} \le 6,\, 0 \le c \le 1$, \\
$\vech(\mathbf{R}) \in [0,1]^{6},\, \min\{\eig(\mathbf{R})\} > 0$, \\
$2 < \nu \le 40$
}
\right.\right\}.
\end{equation}
The constraints on $\boldsymbol{\theta}_{\mathrm{c}}$ and $\mathbf{R}$ ensure that $\mathbf{R}$ is a valid correlation matrix. The prior density, denoted by $p$, can be written as
\begin{equation}
\label{eq:prior}
\boldsymbol{\theta} \sim p(\boldsymbol{\theta}) \propto I_{\mathbb{A}}(\boldsymbol{\theta}) \prod_{i=1}^{3} \omega_{i}^{-1} \prod_{i=1}^{4} \eta_{i}^{-2} \left[1 + \left(\frac{\iota}{10^{-5}}\right)^{2}\right]^{-1} \nu^{-2}.
\end{equation}
This prior is flat on most elements of $\boldsymbol{\theta}$ in $\mathbb{A}$ with the exceptions of $\omega_{1} \ddd \omega_{3}$, $\eta_{1} \ddd \eta_{4}$, $\iota$, and $\nu$. The marginal prior for $\omega_{i}$ reduces the upward bias typically observed for this intercept parameter in the conditional variance equation. The marginal prior for $\eta_{i}$ behaves similar to a half-Cauchy prior, and is obtained by being flat on $\eta_{i}^{-1}$. Using simulation, \citet{BauwensLubrano1998} show that the half-Cauchy prior results in a posterior mean closer to the true value than that obtained from a uniform prior. The marginal prior for $\iota$ is a half-Cauchy with scale  $10^{-5}$. For identifiability purposes, it is important to keep the mean parameter of the Exponential component of the Apatosaurus distribution close to zero. Finally, the marginal prior for $\nu$ is the same as that for $\eta_{i}$. With the likelihood and prior defined, the kernel of the posterior density, denoted by $\pi$, can be computed as
\begin{equation}
\label{eq:posterior}
\boldsymbol{\theta} \vbar \boldsymbol{\xi}_{1} \ddd \boldsymbol{\xi}_{T} \sim \pi(\boldsymbol{\theta}) \propto f(\boldsymbol{\xi}_{1} \ddd \boldsymbol{\xi}_{T}; \boldsymbol{\theta}) p(\boldsymbol{\theta}).
\end{equation}

% \subsection{Bayesian forecast}
% \label{subsec:fore}

\subsection{Adaptive MCMC Algorithm}
\label{sec:mcmc}
In order to evaluate the various integrals of interest involving the posterior density given by \eqref{eq:posterior}, we generate a sample of points from the posterior distribution using an adaptive Markov Chain Monte Carlo (MCMC) algorithm. We first describe the sampling scheme in general, and then discuss the specific steps aimed at improving the mixing of the Markov chain.

One approach is to use a symmetric random-walk Metropolis (RWM) algorithm where we generate the entire parameter vector $\boldsymbol{\theta}$ simultaneously from a symmetric proposal distribution whose number of dimensions is equal to that of $\boldsymbol{\theta}$, and accept the move with the usual Metropolis acceptance probability. However, as $\boldsymbol{\theta}$ has 40 dimensions in our case, it may be difficult to tune the proposal distribution to achieve a satisfactory level of mixing. To mitigate this problem, we employ the well known strategy of updating the parameter vector in blocks, where a 40-dimensional move is broken into lower dimensional sub-moves. The blocking strategy is known to work well if the dependencies between the parameters in different blocks are low. Our model specification given by \eqref{eq:xi_cdf}, \eqref{eq:sktmar}, and \eqref{eq:apatmar} offers a somewhat natural partition of parameters. Let $\boldsymbol{\theta}_{[i]}$ denote the vector of parameters allocated to the $i$-th block. The entire parameter vector is then partitioned into ten blocks $\boldsymbol{\theta} = (\boldsymbol{\theta}_{[1]} \ddd \boldsymbol{\theta}_{[10]})$. The specific blocking scheme is in Appendix~\ref{app:blocking}. Let $\boldsymbol{\cdot}\,^{(j)}$ denote any vector or scalar associated with the state of the Markov chain in period $j$. We then move from $\boldsymbol{\theta}^{(j)}$ to $\boldsymbol{\theta}^{(j+1)}$ according to the following scheme.
\begin{algorithmic}[1]
\For{$i \gets 1:10$}
\State Update $\boldsymbol{\theta}_{[i]}^{(j+1)} \bvbar \boldsymbol{\theta}_{[1]}^{(j+1)} \ddd \boldsymbol{\theta}_{[i-1]}^{(j+1)}, \boldsymbol{\theta}_{[i+1]}^{(j)} \ddd \boldsymbol{\theta}_{[10]}^{(j)}$.
\EndFor
\end{algorithmic}
Thus, a single sweep of the entire parameter vector consists of ten sub-moves, where each sub-move or block is updated by a Metropolis step.

We generate a block-wise proposal for each $\boldsymbol{\theta}_{[i]}$, denoted by $\boldsymbol{\theta}_{[i]}^{*}$ from a symmetric proposal distribution with density $q_{[i]}$, and accept the proposal with the usual Metropolis acceptance probability given by
\begin{equation}
\label{eq:accpr}
\text{min}\left\{\frac{\pi\left(\left(\boldsymbol{\theta}_{[1]}^{(j+1)} \ddd \boldsymbol{\theta}_{[i-1]}^{(j+1)}, \boldsymbol{\theta}_{[i]}^{*}, \boldsymbol{\theta}_{[i+1]}^{(j)} \ddd \boldsymbol{\theta}_{[10]}^{(j)}\right)\right)}{\pi\left(\left(\boldsymbol{\theta}_{[1]}^{(j+1)} \ddd \boldsymbol{\theta}_{[i-1]}^{(j+1)}, \boldsymbol{\theta}_{[i]}^{(j)}, \boldsymbol{\theta}_{[i+1]}^{(j)} \ddd \boldsymbol{\theta}_{[10]}^{(j)}\right)\right)}, 1\right\}.
\end{equation}
We choose the proposal density $q_{[i]}$ to be a mixture of multivariate normals with a different scale for each component,
\begin{equation}
\label{eq:q}
q_{[i]} = \sum_{j=1}^{n_{\mathrm{mix}}} w_{j}f_{\mathrm{mvn}}\left(\cdot; \boldsymbol{\theta}_{[i]}^{(j)}, \Delta_{[i]}^{2}s_{j}\boldsymbol{\Sigma}_{[i]}\right),
\end{equation}
where $s_{j}$ is the scale chosen for component $j$, and $\Delta_{[i]}$ is a tuning scale common to all components. Thus, all components are centred at $\boldsymbol{\theta}_{[i]}^{(j)}$, with covariance structures differing only by scale. We heuristically choose the vector of mixing weights $\boldsymbol{w} = (w_{1}, w_{2}, w_{3})$ to be $(0.7, 0.15, 0.15)$ and the corresponding vector of component scales $\boldsymbol{s} = (s_{1}, s_{2}, s_{3})$ to be $(1, 100, 0.01)$. The intuition is that mixing relatively large jumps with relatively small ones would lower the chance of the chain getting ``stuck", either all together or in some dimensions of the parameter space.

It is important to realise that the acceptance probability in \eqref{eq:accpr} is zero when a proposed jump lands outside of $\mathbb{A}$ (due to the indicator in the prior), so that such a move is guaranteed to be rejected. If the posterior density $\pi$ is small on the boundary of $\mathbb{A}$ (which is the case for all the empirical data considered in this paper), the proposed adaptive sampler is able to handle constrained parameter space without any noticeable loss of efficiency.

With $\boldsymbol{w}$ and $\boldsymbol{s}$ chosen a priori, and a fixed covariance matrix $\boldsymbol{\Sigma}_{[i]}$ (see below for how $\boldsymbol{\Sigma}_{[i]}$ is specified), the scale $\Delta_{[i]}$ is tuned automatically for each block during a tuning \emph{epoch}. For a given $\boldsymbol{\Sigma}_{[i]}$, we update the value of $\Delta_{[i]}$ every $n_{\Delta}$ iterations of the chain to target a specific acceptance rate, denoted by $r_{[i]}^{(\mathrm{tar})}$. Let $n_{\mathrm{epo}}$ denote the number of iterations spent in a tuning epoch, and let $k \in \{n_{\Delta}(1 \ddd \lfloor\frac{n_{\mathrm{epo}}-1}{n_{\Delta}}\rfloor)\}$ denote the iteration where a scale-update occurs. The new value of $\Delta_{[i]}$ is then given by
\begin{equation}
\label{eq:delta}
\Delta_{[i]}^{(k+1 \ddd k+n_{\Delta})} = \Upsilon(r_{[i]}^{(\mathrm{obs})};r_{[i]}^{(\mathrm{tar})})\Delta_{[i]}^{(k-n_{\Delta}+1 \ddd k)},
\end{equation}
where $\Upsilon$ is a sensibly chosen tuning function that takes as an argument the realised acceptance rate since the last update, denoted by $r_{[i]}^{(\mathrm{obs})}$. We choose $\Upsilon$ to be
\begin{equation}
\label{eq:upsilon}
\Upsilon(r_{[i]}^{(\mathrm{obs})};r_{[i]}^{(\mathrm{tar})}) = \frac{\Phi^{-1}(r_{[i]}^{(\mathrm{tar})}/2)}{\Phi^{-1}(r_{[i]}^{(\mathrm{obs})}/2)}.
\end{equation}
This particular tuning function exploits the relationship between the scale of the proposal distribution $\Delta$ and the acceptance rate of the RWM algorithm $r$ when both the proposal and target distributions are $d$-dimensional normal \citep{RobertsRosenthal2001}; $r = 2\Phi(-\Delta\sqrt{d}/2)$. Even if this relationship does not hold in practice, as $\Upsilon$ is both positive and monotonically increasing on the interval $(0, 1)$ while being equal to one for $r_{[i]}^{(\mathrm{obs})} = r_{[i]}^{(\mathrm{tar})}$, the tuning function in \eqref{eq:upsilon} will still behave sensibly. We set the target acceptance rate according to the size of the block $d_{[i]}$ by following the empirically successful heuristics based on the results of \citet{GelmanRobertsGilks1996}. Specifically, we choose $r_{[i]}^{(\mathrm{tar})} = 0.44$ for $d_{[i]} = 1$, $r_{[i]}^{(\mathrm{tar})} = 0.35$ for $2 \le d_{[i]} \le 4$, and $r_{[i]}^{(\mathrm{tar})} = 0.234$ for $d_{[i]} > 4$. The initial scale of each block $\Delta_{[i]}^{(1 \ddd n_{\Delta})}$ is set to $2.38/\sqrt{d_{[i]}}$.

We gradually improve the estimate of the covariance matrix of the proposal distribution for each block $\boldsymbol{\Sigma}_{[i]}$ by running multiple tuning epochs. With the exception of the first epoch, we initialise the chain of each epoch with the last generated parameter vector of the previous epoch, and set the covariance matrix of each block to the sample covariance matrix of the corresponding block of the previous epoch. A few initial iterations of each epoch is discarded when computing the sample covariance matrix; we denote this number by $n_{\mathrm{disc}}^{(\mathrm{epo})}$. The number of tuning epochs required is judged by a simple stopping criterion based on the mean absolute percentage change (MAPC) given by
\begin{equation}
\label{eq:mapc}
\mathrm{MAPC}_{j} = \frac{1}{40}\sum_{i=1}^{40}\left|\frac{\hat{\sigma}_{\theta,i}^{\langle j \rangle} - \hat{\sigma}_{\theta,i}^{\langle j-1 \rangle}}{\hat{\sigma}_{\theta,i}^{\langle j-1 \rangle}}\right|,
\end{equation}
where $\hat{\sigma}_{\theta,i}^{\langle j \rangle}$ denotes the sample standard deviation of the $i$-th dimension of the chain from the $j$-th epoch. Adaptation is stopped after $j$ epochs if $j_{\mathrm{min}} \le j \le j_{\mathrm{max}}$ and $\mathrm{MAPC}_{j} \le \varepsilon_{\mathrm{mapc}}$, where $j_{\mathrm{min}}$ and $j_{\mathrm{max}}$ are the least and most number of tuning epochs, and $\varepsilon_{\mathrm{mapc}}$ is a tolerance level. In all the applications, we set $j_{\mathrm{min}} = 2$, $j_{\mathrm{max}} = 30$, $\varepsilon_{\mathrm{mapc}} = 0.1$, $n_{\mathrm{epo}} = 12000$, and $n_{\mathrm{disc}}^{(\mathrm{epo})} = 2000$. Note that any convergence criterion can be used to terminate the adaptation, however we find the MAPC to be effective in practice while having the advantage of being computationally simple.

Once the adaptive phase ends, the RWM sampler transitions into the sampling phase where all adaptations are tuned-off. That is, we generate a Markov chain according to our sampling scheme with $\Delta_{[i]}$ and $\boldsymbol{\Sigma}_{[i]}$ fixed. The scale $\Delta_{[i]}$ and covariance matrix $\boldsymbol{\Sigma}_{[i]}$ are fixed at, respectively, the mean of the tuned scales and the sample covariance matrix of the generated parameters, over the iterations of the final tuning epoch after discarding the first $n_{\mathrm{disc}}^{(\mathrm{epo})}$ ones. The sample mean of the generated parameters over these iterations is used as the initial state of the sampling phase chain.

Notice that, in our sampling scheme, the posterior kernel in \eqref{eq:posterior} must be evaluated at least once when computing the acceptance probability in \eqref{eq:accpr} for each block update. Naively evaluating the entire posterior kernel for each sub-move can be computationally costly. However, because of the way in which the blocks are chosen and the special structure of the conditional joint density of $\boldsymbol{\xi}_{t}$ in \eqref{eq:xi_pdf}, computational cost can be reduced substantially by only updating the part of the likelihood related to each sub-move. For example, the vectors $(f_{1,1}(\xi_{1,1}) \ddd f_{1,T}(\xi_{1,T}))$, $(u_{1,1} \ddd u_{1,T}) = (F_{1,1}(\xi_{1,1}) \ddd F_{1,T}(\xi_{1,T}))$, and $(F_{\mathrm{St}}^{-1}(u_{1,1}) \ddd F_{\mathrm{St}}^{-1}(u_{1,T}))$ only need to be recomputed when updating the blocks $\boldsymbol{\theta}_{[1]}^{(j+1)}$ and $\boldsymbol{\theta}_{[2]}^{(j+1)}$.

\section{Simulation Study}
\label{sec:sim_study}
A simulation study is conducted to investigate the effectiveness of the adaptive MCMC sampling algorithm proposed in Section~\ref{sec:mcmc}. We generate 1000 independent datasets from the true data generating process (DGP), with each dataset containing 3000 observations. The true DGP is the model for the conditional joint distribution of $\boldsymbol{\xi}_{t}$ specified in Section~\ref{sec:xit_dist} whose parameters values are chosen to be similar to the parameter estimates from the real data. The sampling phase of the MCMC algorithm is set to run for 105,000 iterations. The posterior mean estimates of the parameters are computed using the last 100,000 iterations from the sampling phase.

In Table~\ref{tab:sim_post_mean}, for each parameter, we report the true parameter value (True), Monte Carlo (MC) mean of the 1000 posterior mean estimates (Mean), and 95\% MC interval (Lower, Upper). The posterior mean estimates are all reasonably close to the true values, with all the MC intervals covering the true values. 

\begin{table}[h]
\small
\center
\begin{tabular}{cccccccccc}
\toprule
$\boldsymbol{\theta}_{1}$ & $\delta_{1}$ & $\psi_{1}$ & $\phi_{1}$ & $\omega_{1}$ & $\alpha_{1}$ & $\beta_{1}$ & $\eta_{1}$ & $\lambda_{1}$ &  \\ 
\midrule
True & 0.000 & 0.060 & 0.910 & 6.000E-08 & 0.150 & 0.840 & 8.000 & -0.160 &  \\ 
Mean & 1.914E-07 & 0.062 & 0.902 & 6.679E-08 & 0.151 & 0.837 & 8.244 & -0.161 &  \\ 
Lower & -4.714E-06 & 0.048 & 0.869 & 4.440E-08 & 0.125 & 0.807 & 6.512 & -0.206 &  \\ 
Upper & 5.330E-06 & 0.078 & 0.925 & 9.616E-08 & 0.180 & 0.863 & 10.991 & -0.114 &  \\ 
\midrule
$\boldsymbol{\theta}_{2}$ & $\delta_{2}$ & $\psi_{2}$ & $\phi_{2}$ & $\omega_{2}$ & $\alpha_{2}$ & $\beta_{2}$ & $\eta_{2}$ & $\lambda_{2}$ &  \\ 
\midrule
True & -0.130 & 0.430 & 0.530 & 5.000E-03 & 0.060 & 0.880 & 15.000 & 0.000 &  \\ 
Mean & -0.135 & 0.432 & 0.527 & 5.981E-03 & 0.064 & 0.865 & 15.410 & 0.011 &  \\ 
Lower & -0.168 & 0.405 & 0.500 & 3.230E-03 & 0.043 & 0.799 & 10.300 & -0.038 &  \\ 
Upper & -0.106 & 0.456 & 0.556 & 1.039E-02 & 0.087 & 0.912 & 24.104 & 0.057 &  \\ 
\midrule
$\boldsymbol{\theta}_{3}$ & $\delta_{3}$ & $\psi_{3}$ & $\phi_{3}$ & $\omega_{3}$ & $\alpha_{3}$ & $\beta_{3}$ & $\eta_{3}$ & $\lambda_{3}$ &  \\ 
\midrule
True & 0.000 & 0.050 & 0.930 & 7.000E-05 & 0.070 & 0.920 & 18.000 & 0.140 &  \\ 
Mean & -9.664E-06 & 0.053 & 0.921 & 8.360E-05 & 0.073 & 0.915 & 19.011 & 0.138 &  \\ 
Lower & -2.251E-04 & 0.040 & 0.894 & 4.428E-05 & 0.055 & 0.893 & 11.922 & 0.089 &  \\ 
Upper & 2.197E-04 & 0.069 & 0.942 & 1.466E-04 & 0.092 & 0.937 & 28.197 & 0.187 &  \\ 
\midrule
$\boldsymbol{\theta}_{4}$ & $\delta_{4}$ & $\psi_{4}$ & $\phi_{4}$ & $\gamma^{*}$ & $c$ & $\sigma$ & $\eta_{4}$ & $\lambda_{4}$ & $\iota$ \\ 
\midrule
True & 3.000E-03 & 0.220 & 0.740 & 3.700 & 0.030 & 0.060 & 6.000 & 0.150 & 1.000E-04 \\ 
Mean & 3.804E-03 & 0.218 & 0.737 & 3.689 & 0.031 & 0.060 & 6.100 & 0.151 & 8.164E-05 \\ 
Lower & 2.063E-03 & 0.199 & 0.712 & 3.458 & 0.012 & 0.057 & 4.905 & 0.098 & 5.528E-05 \\ 
Upper & 5.906E-03 & 0.239 & 0.761 & 3.960 & 0.049 & 0.063 & 7.804 & 0.204 & 1.095E-04 \\ 
\midrule
$\boldsymbol{\theta}_{\mathrm{c}}$ & $\mathbf{R}_{2,1}$ & $\mathbf{R}_{3,1}$ & $\mathbf{R}_{4,1}$ & $\mathbf{R}_{3,2}$ & $\mathbf{R}_{4,2}$ & $\mathbf{R}_{4,3}$ & $\nu$ &  &  \\ 
\midrule
True & -0.300 & -0.100 & 0.200 & -0.220 & -0.600 & 0.120 & 15.000 &  &  \\ 
Mean & -0.299 & -0.099 & 0.193 & -0.220 & -0.580 & 0.115 & 14.583 &  &  \\ 
Lower & -0.331 & -0.136 & 0.158 & -0.257 & -0.606 & 0.076 & 11.344 &  &  \\ 
Upper & -0.267 & -0.062 & 0.227 & -0.184 & -0.553 & 0.152 & 19.737 &  &  \\ 
\bottomrule
\end{tabular}
\caption{\small True values and summary statistics for the posterior mean estimates of the DQF model parameters.}
\label{tab:sim_post_mean}
\end{table}
\FloatBarrier

The MC mean of the acceptance rates for the last $10^{4}$ iterations of the sampling phase is reported in Table~\ref{tab:acc_rate} for each parameter block. \change{The realised acceptance rates closely match the targets recommended in \citet{GelmanRobertsGilks1996}, indicating that the scale of the proposal distribution is effectively tuned by the updating mechanism of Equation \ref{eq:upsilon}, and that the Metropolis algorithm operates efficiently during the sampling phase.}

\begin{table}[h]
\small
\center
\begin{tabular}{ccccccccccc}
\toprule
Block (Size) & 1 (3) & 2 (5) & 3 (3) & 4 (5) & 5 (3) & 6 (5) & 7 (5) & 8 (4) & 9 (6) & 10 (1) \\ 
\midrule
Target & 0.350 & 0.234 & 0.350 & 0.234 & 0.350 & 0.234 & 0.234 & 0.350 & 0.234 & 0.440 \\ 
Mean & 0.345 & 0.223 & 0.344 & 0.230 & 0.345 & 0.230 & 0.231 & 0.348 & 0.230 & 0.438 \\ 
\bottomrule
\end{tabular}
\caption{\small Mean acceptance rate of each block during the sampling phase.}
\label{tab:acc_rate}
\end{table}
\FloatBarrier

\section{Empirical Studies}
\label{sec:emp_study}

\subsection{Cleaning of High Frequency Data}
\label{sec:data}
All of the following empirical studies rely on high-frequency price data of major international stock indices. As high-frequency intra-daily data are collected via real-time streaming of asynchronous messages, recording errors are present in the raw data. Furthermore, the raw data also contain artefacts due to events such as trading halts, lunch breaks, and special orders outside the normal trading hours. Therefore, it is important to pre-process the raw data to remove as many incorrectly recorded prices as possible \citep{BrownleesGallo2006}.

For our empirical analysis, we use transaction prices sampled at one-minute intervals. The raw data is provided by the Thomson Reuters Tick History. Let $\boldsymbol{\zeta}_{t} = (\zeta_{t,1} \ddd \zeta_{t,n_{t}})$ denote the vector of intra-daily prices for day $t$. To clean the data, we apply the following set of rules for each $t$:
\begin{enumerate}[label=\Roman*., align=left, leftmargin=*]
\item For $i \in \{1 \ddd n_{t}\}$, remove $\zeta_{t,i}$ if its timestamp is outside the normal trading hours.
\item For $i \in \{1 \ddd n_{t}\}$, remove $\zeta_{t,i}$ if $\zeta_{t,i} \le 0$.
\item For $j \in \{2 \ddd n_{t}\}$, remove $\zeta_{t,1} \ddd \zeta_{t,j-1}$ if $\zeta_{t,1} = \cdots = \zeta_{t,j}$.
\item For $j \in \{2 \ddd n_{t}\}$, remove $\zeta_{t,n_{t}-j+1} \ddd \zeta_{t,n_{t}-1}$ if $\zeta_{t,n_{t}-j+1} = \cdots = \zeta_{t,n_{t}}$.
\item For $i \in \{1 \ddd n_{t}-j+1\}$ and $j \in \{31 \ddd n_{t}\}$, remove $\zeta_{t,i} \ddd \zeta_{t,i+j-2}$ if $\zeta_{t,i} = \cdots = \zeta_{t,i+j-1}$.
\item For $i \in \{1 \ddd n_{t}\}$, remove $\zeta_{t,i}$ if its outlier score is greater than 20.
\item Remove $\zeta_{t,1} \ddd \zeta_{t,n_{t}}$ if $n_{t} < 60$.
\end{enumerate}
Note that the rules are applied in sequence. I.e., $\boldsymbol{\zeta}_{t}$ and $n_{t}$ are updated after each rule is applied. The difficulty in implementing rule I is that changes have been made over time to the normal trading hours for many major stock exchanges. Therefore, we must keep track of all the changes for each stock exchange over our sample period. A complete history of session times is documented in Appendix~\ref{app:session_times}. Rule II removes any obvious mistakes. Rules III and IV are responsible for removing static prices at the beginning and the end of a day, which usually indicate events such as late starts and trading halts. Similarly, rule V removes any static gaps that are longer than 30 minutes. For rule VI, an outlier score is computed for each $\zeta_{t,i}$, which is a scale-invariant distance measure between $\zeta_{t,i}$ and its neighbouring observations. See Appendix~\ref{app:outscr} for details on computing the outlier score. Finally, rule VII removes the entire trading day if there are less than 60 observations left after applying the first six rules.

Our data includes one-minute price series of ten major stock indices: S\&P~500 (SPX), Dow Jones Industrial Average (DJIA), NASDAQ Composite (Nasdaq), FTSE 100 (FTSE), DAX, CAC 40 (CAC), Nikkei Stock Average 225 (Nikkei), Hang Seng (HSI), Shanghai Composite (SSEC), and All Ordinaries (AORD). The sample period starts on January 3, 1996 and ends on May 24, 2016. Table~\ref{tab:data} shows the percentage proportion of removed observations after applying all the rules for each stock index. It also documents the number of trading days and the number of observations for each index before cleaning.

\begin{table}[h]
\small
\centering
\begin{tabular}{lccc}
\toprule
 & Days & Obs. & Del.~(\%) \\ 
\midrule
SPX & 5103 & 1,979,767 & 0.12 \\ 
DJIA & 5105 & 1,977,270 & 0.02 \\ 
Nasdaq & 5112 & 1,975,540 & 0.17 \\ 
FTSE & 5380 & 2,562,661 & 0.23 \\ 
DAX & 5036 & 2,470,843 & 0.09 \\ 
CAC & 5167 & 2,533,410 & 0.17 \\ 
Nikkei & 4977 & 1,345,816 & 0.04 \\ 
HSI & 4998 & 1,299,330 & 0.01 \\ 
SSEC & 4906 & 1,175,712 & 0.05 \\ 
AORD & 5138 & 1,779,671 & 0.09 \\ 
\bottomrule
\end{tabular}
\caption{\small Data cleaning summary}
\label{tab:data}
\end{table}
\FloatBarrier

\subsection{A Case Study of S\&P~500 One-Minute Returns}
\label{sec:sp500}
In this part of the empirical study, the goal is demonstrate the various aspects of the DQF model by focusing on arguably one of the most widely followed market indices -- the S\&P~500. We dynamically model the daily distributions of one-minute percentage log-returns. For each trading day, there are approximately 390 one-minute returns. Let $\mathbf{y}_{t} = (y_{t,1} \ddd y_{t, n_{t}})$ denote the one-minute returns for day $t$. For each $t \in \{1 \ddd T\}$, the return vector is computed by applying
\begin{equation*}
y_{t,i} = 100 \left[\log(\zeta_{t,i+1}) - \log(\zeta_{t,i})\right],
\end{equation*}
for each $i \in \{1 \ddd n_t - 1\}$. We then summarise each $\mathbf{y}_{t}$ by a QF-valued observation $X_{t} = S(\mathbf{y}_{t})$. The summarisation function $S$ here corresponds to the L-moment estimator of $g$-and-$h$ parameters. \change{The L-moment estimator matches the first four L-moments of the $g$-and-$h$ distribution to the sample L-moments, which are computed using all of the one-minute returns within a day. The implementation details of the L-moment method are in Appendix~\ref{app:lmoment}.} The four-dimensional mapped vector is obtained via $\boldsymbol{\xi}_{t} = \mathcal{M}(X_{t})$ for each $t \in \{1 \ddd T\}$. The conditional joint distribution of $\boldsymbol{\xi}_{t}$ is given by the model in Section~\ref{sec:xit_dist}. The parameters are estimated using the adaptive MCMC algorithm described in Section~\ref{sec:mcmc} whose configuration is identical to that for the simulation study.

\begin{figure}[h]
    \centering
    \begin{subfigure}[t]{0.33\textwidth}
        \centering
        \includegraphics[height=4.5cm]{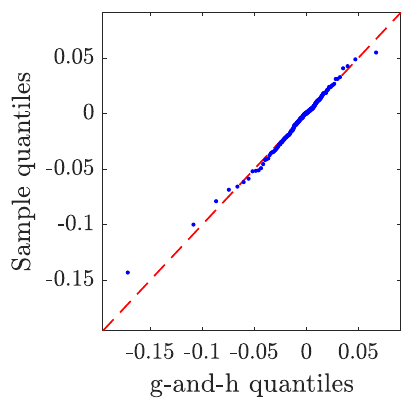}
        \caption{December 31, 2009}
    \end{subfigure}%
    ~
    \begin{subfigure}[t]{0.33\textwidth}
        \centering
        \includegraphics[height=4.5cm]{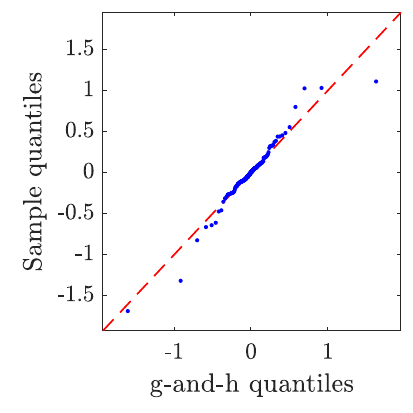}
        \caption{May 6, 2010}
    \end{subfigure}%
    ~
    \begin{subfigure}[t]{0.33\textwidth}
        \centering
        \includegraphics[height=4.5cm]{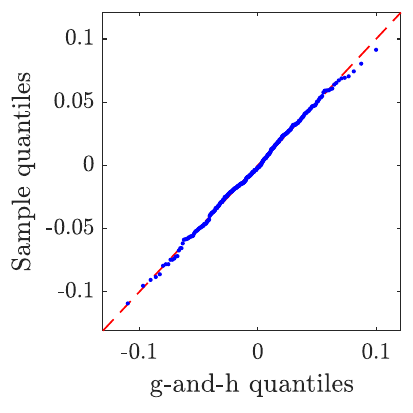}
        \caption{April 4, 2014}
    \end{subfigure}
    \caption{\small QQ-plots of one-minute returns of three specific trading days against fitted $g$-and-$h$ quantiles.}
    \label{fig:ghqq}
\end{figure}

To illustrate that the $g$-and-$h$ quantile functions are adequate summaries of the distributional features of one-minute returns, QQ-plots are shown in Figure~\ref{fig:ghqq} where sample quantiles of one-minute returns are plotted against the estimated $g$-and-$h$ quantiles for three stylised days. On the last trading day of a very volatile year, December 31, 2009, the returns are strongly negatively skewed. An infamous ``flash crash" occurred on May 6, 2010, which resulted in one-minute returns being extremely heavy-tailed. On April 4, 2014, the intra-daily returns are approximately normally distributed. From the QQ-plots, it can be seen that the $g$-and-$h$ quantile functions can provide reasonable approximations to these distributional shapes. 

\FloatBarrier

Summary statistics of the estimated posterior are reported in Table~\ref{tab:par_spx}. The parameters driving the conditional mean dynamics $\psi_{1} \ddd \psi_{4}$ and $\phi_{1} \ddd \phi_{4}$ are all estimated to be nonzero, as indicated by their credible intervals. The fact that the estimates of $\psi_{1}$ and $\psi_{3}$ are much closer to zero than those of $\psi_{2}$ and $\psi_{4}$ indicates that the observed values of $a_{t}$ and $g_{t}$ are much less informative than those of $b^{*}_{t}$ and $h_{t}$ about the respective conditional means at period $t + 1$. The estimates of the logistic link function parameters $\gamma^{*}$ and $c$ confirm that the Exponential component of the Apatosaurus distribution is only needed when $h_{t}$ is small. The fact that the copula parameters $\mathbf{R}_{2,1}$ and $\mathbf{R}_{3,2}$ are both estimated to be negative suggests that an increase in volatility is more likely to be accompanied by negative returns. This observation is in accordance with the well-documented ``leverage effect" observed in daily equity returns. However, it is interesting to observe a rather large negative estimate for $\mathbf{R}_{4,2}$, which indicates a negative relationship between volatility and kurtosis.

\begin{figure}[h]
\hspace{-2.8cm}
\includegraphics[width = 1.3\textwidth]{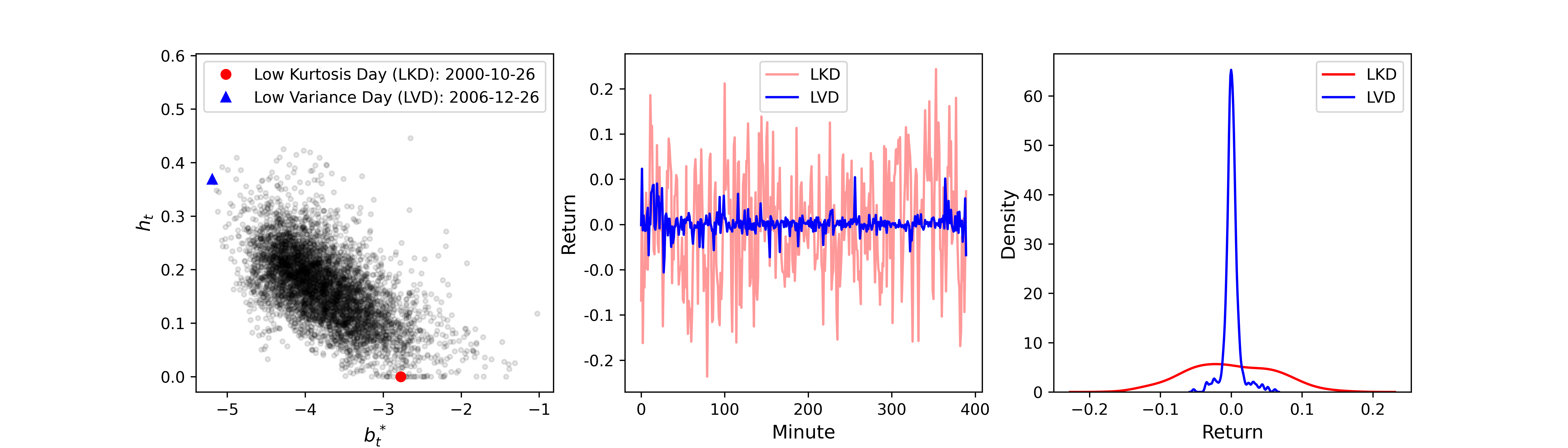}
\caption{\change{Left: scatter plot of realised values of $h_t$ against $b^*_t$ for the S\&P~500 index; red circle indicates the day with the lowest value of $h_t$ (LKD) while blue triangle indicates the day with lowest value of $b^*_t$ (LVD). Centre: one-minute returns of LKD and LVD. Right: kernel density estimates of the of one-minute returns of LKD and LVD.}
\label{fig:lkd_vs_lvd}}
\end{figure}

\begin{figure}[h]
    \centering
    \begin{subfigure}[t]{1.0\textwidth}
        \centering
        \includegraphics[height=4cm]{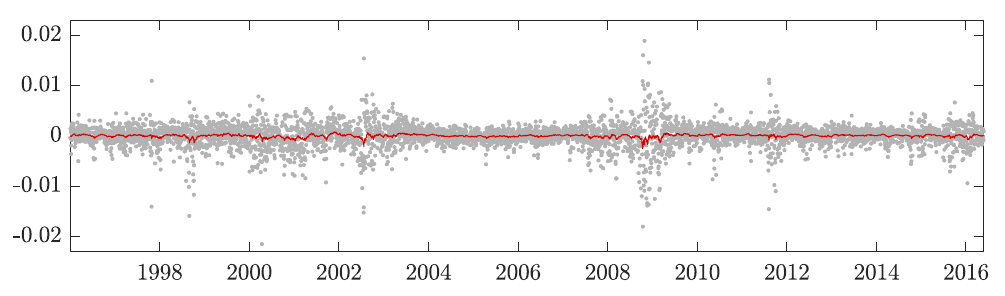}
        \caption{$a_{t}$}
    \end{subfigure}
    
    \begin{subfigure}[t]{1.0\textwidth}
        \centering
        \includegraphics[height=4cm]{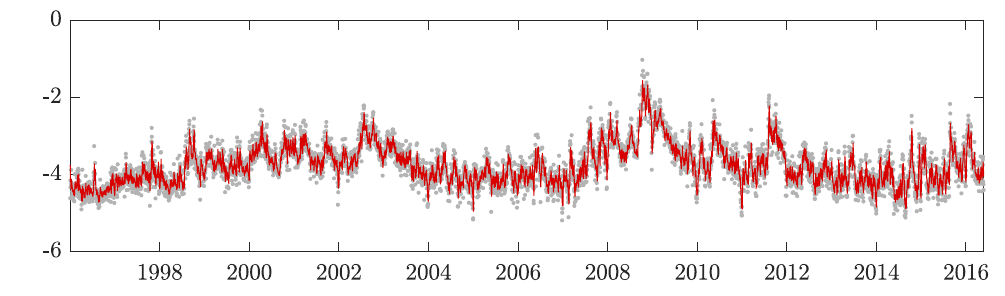}
        \caption{$b^{*}_{t}$}
    \end{subfigure}
    
    \begin{subfigure}[t]{1.0\textwidth}
        \centering
        \includegraphics[height=4cm]{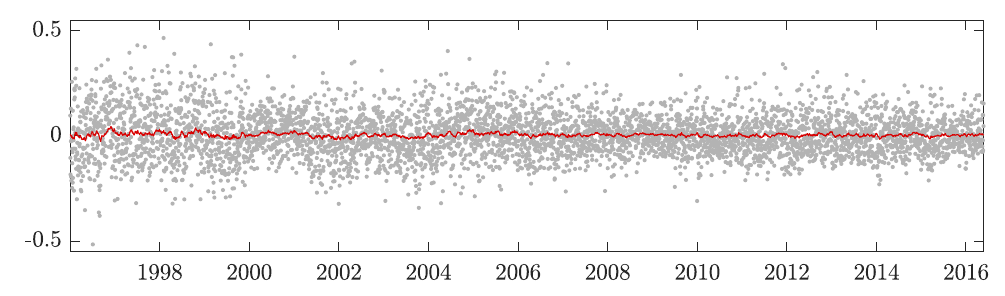}
        \caption{$g_{t}$}
    \end{subfigure}
    
    \begin{subfigure}[t]{1.0\textwidth}
        \centering
        \includegraphics[height=4cm]{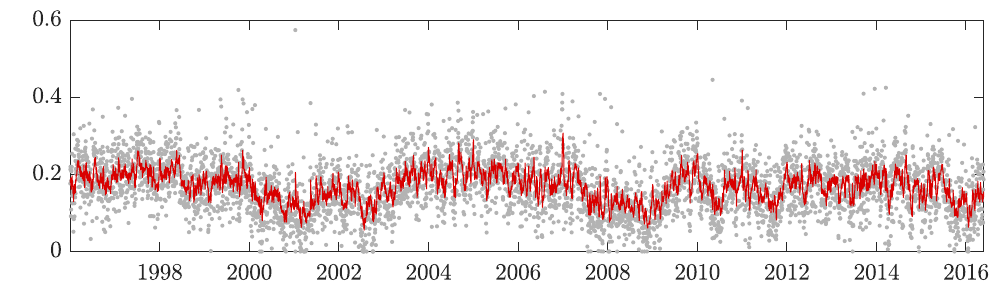}
        \caption{$h_{t}$}
    \end{subfigure}
    \caption{\small Posterior mean estimates of $\{\E(\boldsymbol{\xi}_{t} \vbar \mathcal{F}_{t-1})\}$ (red line) plotted over $\{\boldsymbol{\xi}_{t}\}$ (grey dots).}
    \label{fig:xi}
\end{figure}

\change{The negative correlation between scale $b_t$ and kurtosis $h_t$ suggests that the distributions of intra-daily returns have thinner tails on high volatility days than those on low volatility days. This phenomenon can be observed for the S\&P~500 index on sub-plots (b) and (d) of Figure~\ref{fig:xi}. To further investigate, we compare the one-minute return series of the day with the lowest kurtosis (measured by $h_t$) to the day with the lowest variance (measured by $b^*_t$). These two days are marked on the scatter plot in Figure~\ref{fig:lkd_vs_lvd} (left). From the time-series and the kernel density plots, it can be seen that there are many outlying observations on the low-variance day (LVD) resulting in a heavy-tailed distribution, whereas the low-kurtosis day (LKD) has a much higher variance but no extreme observations. In other words, compared to a volatile day, an investor is more likely experience occasional large price jumps on a quiet day. To the best of our knowledge, the negative dependency between volatility and heavy-tailedness for intra-daily returns is not documented in the literature. It is important that this negative dependency is captured by the DQF model, as it will contribute to the performance in forecasting tail-risk measures.}

\begin{table}[h]
\centering
\small
\begin{tabular}{cccccccccc}
\toprule
$\boldsymbol{\theta}_{1}$ & $\delta_{1}$ & $\psi_{1}$ & $\phi_{1}$ & $\omega_{1}$ & $\alpha_{1}$ & $\beta_{1}$ & $\eta_{1}$ & $\lambda_{1}$ &  \\ 
\midrule
Mean & 1.741E-06 & 0.034 & 0.924 & 6.492E-08 & 0.144 & 0.846 & 7.332 & -0.162 &  \\ 
Lower & -1.248E-06 & 0.024 & 0.892 & 4.456E-08 & 0.120 & 0.820 & 6.097 & -0.200 &  \\ 
Upper & 5.369E-06 & 0.046 & 0.951 & 8.959E-08 & 0.170 & 0.870 & 8.943 & -0.123 &  \\ 
\midrule
$\boldsymbol{\theta}_{2}$ & $\delta_{2}$ & $\psi_{2}$ & $\phi_{2}$ & $\omega_{2}$ & $\alpha_{2}$ & $\beta_{2}$ & $\eta_{2}$ & $\lambda_{2}$ &  \\ 
\midrule
Mean & -0.124 & 0.435 & 0.533 & 4.161E-03 & 0.047 & 0.890 & 21.702 & 0.059 &  \\ 
Lower & -0.149 & 0.411 & 0.505 & 2.209E-03 & 0.032 & 0.832 & 13.826 & 0.022 &  \\ 
Upper & -0.100 & 0.460 & 0.560 & 7.143E-03 & 0.066 & 0.931 & 34.779 & 0.096 &  \\ 
\midrule
$\boldsymbol{\theta}_{3}$ & $\delta_{3}$ & $\psi_{3}$ & $\phi_{3}$ & $\omega_{3}$ & $\alpha_{3}$ & $\beta_{3}$ & $\eta_{3}$ & $\lambda_{3}$ &  \\ 
\midrule
Mean & 2.002E-04 & 0.023 & 0.932 & 1.982E-05 & 0.020 & 0.978 & 19.546 & 0.145 &  \\ 
Lower & 9.734E-06 & 0.012 & 0.872 & 5.323E-10 & 0.013 & 0.967 & 13.028 & 0.089 &  \\ 
Upper & 5.249E-04 & 0.035 & 0.970 & 5.366E-05 & 0.030 & 0.986 & 31.264 & 0.185 &  \\ 
\midrule
$\boldsymbol{\theta}_{4}$ & $\delta_{4}$ & $\psi_{4}$ & $\phi_{4}$ & $\gamma^{*}$ & $c$ & $\sigma$ & $\eta_{4}$ & $\lambda_{4}$ & $\iota$ \\ 
\midrule
Mean & 2.737E-03 & 0.193 & 0.773 & 3.743 & 0.014 & 0.061 & 6.815 & 0.134 & 6.337E-05 \\ 
Lower & 1.152E-03 & 0.175 & 0.749 & 3.569 & 0.001 & 0.059 & 5.569 & 0.087 & 4.109E-05 \\ 
Upper & 4.433E-03 & 0.212 & 0.796 & 3.956 & 0.032 & 0.063 & 8.507 & 0.181 & 9.450E-05 \\ 
\midrule
$\boldsymbol{\theta}_{\mathrm{c}}$ & $\mathbf{R}_{2,1}$ & $\mathbf{R}_{3,1}$ & $\mathbf{R}_{4,1}$ & $\mathbf{R}_{3,2}$ & $\mathbf{R}_{4,2}$ & $\mathbf{R}_{4,3}$ & $\nu$ &  &  \\ 
\midrule
Mean & -0.288 & -0.065 & 0.176 & -0.229 & -0.524 & 0.086 & 20.120 &  &  \\ 
Lower & -0.315 & -0.094 & 0.147 & -0.256 & -0.545 & 0.058 & 15.843 &  &  \\ 
Upper & -0.262 & -0.037 & 0.203 & -0.202 & -0.503 & 0.114 & 26.117 &  &  \\ 
\bottomrule
\end{tabular}
\caption{\small DQF posterior summary for S\&P~500, showing the posterior mean (Mean) and the 95\% credible interval (Lower, Upper) for each parameter.}
\label{tab:par_spx}
\end{table}
\FloatBarrier

\FloatBarrier

The posterior mean estimates of the conditional means $\{\E(\boldsymbol{\xi}_{t} \vbar \mathcal{F}_{t-1})\}$ are plotted in Figure~\ref{fig:xi}, together with the realised values of $\{\boldsymbol{\xi}_{t}\}$. Firstly, compared to $\{a_{t}\}$ and $\{g_{t}\}$, the unconditional variances of $\{b^{*}_{t}\}$ and $\{h_{t}\}$ appear to be much better explained by the variations in the conditional means. As expected, the values of $h_{t}$ are clearly above zero for most of the days, indicating that one-minute returns are heavy-tailed. Finally, the negative relationship between $b^{*}_{t}$ and $h_{t}$ is apparent in sub-plots (b) and (d); during high volatility periods, $h_{t}$ can become vary close to zero.

\FloatBarrier

\begin{figure}[h]
\centering
\includegraphics[width=1.0\textwidth]{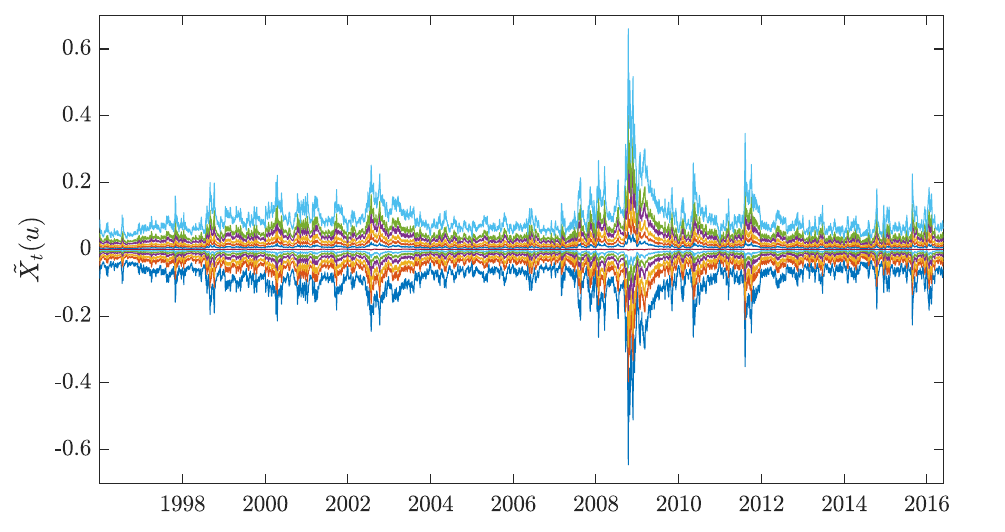}
\caption{\small Posterior mean estimates of $\{\tilde{X}_{t}(u)\}$ for $u \in \{0.01, 0.05, 0.25, 0.5, 0.75, 0.95, 0.99\}$.}
\label{fig:q1min}
\end{figure}

Recall that the filtered quantile function (i.e., one-step-ahead forecast) can be obtained from the conditional mean of $\boldsymbol{\xi}_{t}$ by applying the inverse mapping 
\begin{equation*}
\tilde{X}_{t} = \mathcal{M}^{-1}(\E[\boldsymbol{\xi}_{t} \vbar \mathcal{F}_{t-1}]).
\end{equation*}
To illustrate, in Figure~\ref{fig:q1min}, we plot the \change{in-sample estimates} $\tilde{X}_{1}(u) \ddd \tilde{X}_{T}(u)$ for various values of $u$. Notice that evaluating $\tilde{X}_{t}$ at multiple quantile levels does not require multiple estimations of the DQF model, and the quantile estimates do not cross over time.

\FloatBarrier

\begin{figure}[h]
\centering
\includegraphics[width=1.0\textwidth]{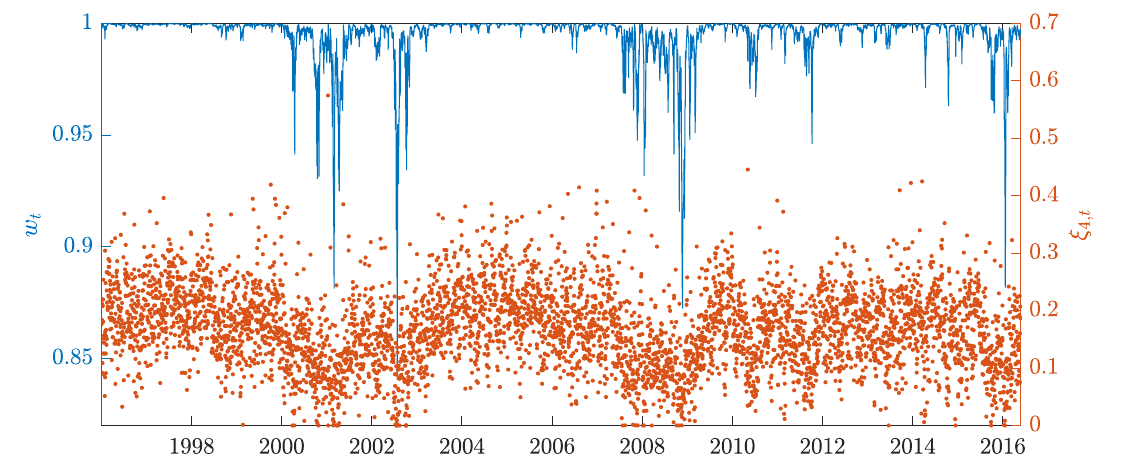}
\caption{\small Posterior mean estimates of $\{w_{t}\}$ (blue line, left axis) plotted together with $\{h_{t}\}$ (orange dots, right axis).}
\label{fig:w}
\end{figure}

\begin{figure}[h]
    \centering
    \begin{subfigure}[t]{0.35\textwidth}
        \centering
        \includegraphics[height=5cm]{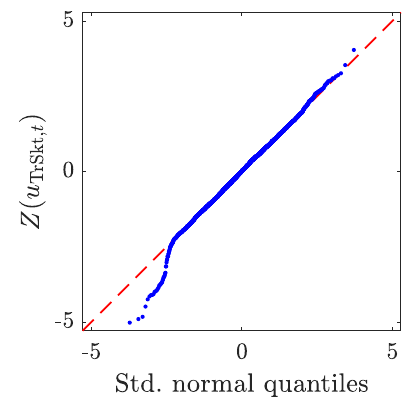}
        \caption{Truncated-skewed-$t$}
    \end{subfigure}%
    ~
    \begin{subfigure}[t]{0.35\textwidth}
        \centering
        \includegraphics[height=5cm]{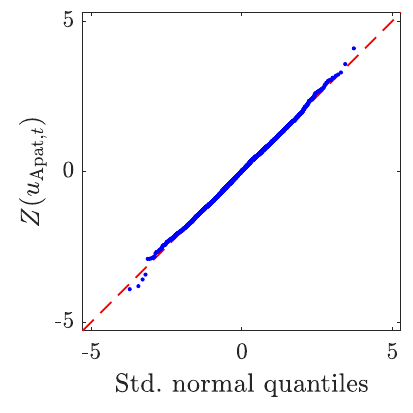}
        \caption{Apatosaurus}
    \end{subfigure}
    \caption{\small QQ-plots of transformed $\{h_{t}\}$ under truncated-skewed-$t$ and Apatosaurus models against standard normal quantiles.}
    \label{fig:qq_h}
\end{figure}

A considerable effort was spent on constructing the conditionally Apatosaurus marginal model with time-varying weights for $\{h_{t}\}$. The posterior mean estimates of the conditional weights are plotted in Figure~\ref{fig:w}, together with the realised $\{h_{t}\}$. As expected, the weights are close to one for most days; the Exponential component only plays a role for when $h_{t}$ is close to zero. As the weights are closed to one on average, it is worth knowing whether an advantage is gained over a simpler truncated-skewed-$t$ alternative. To see this, we estimate a truncated-skewed-$t$ model $h_{t} \sim F_{\mathrm{TrSkt}}(\cdot; \mu_{t}, \sigma, \eta, \lambda)$, where $\mu_{t} = \delta + \psi h_{t-1} + \phi \mu_{t-1}$. Let $u_{\mathrm{TrSkt},t} = F_{\mathrm{TrSkt}}(h_{t}; \hat{\mu}_{t}, \hat{\sigma}, \hat{\eta}, \hat{\lambda})$ be the probability integral transform (PIT) of $h_{t}$, where $\hat{\mu}_{t}, \hat{\sigma}, \hat{\eta}$, and $\hat{\lambda}$ are the posterior mean estimates. If the truncated-skewed-$t$ model is adequate, $Z(u_{\mathrm{TrSkt},t})$ will be a draw from the standard normal distribution, where $Z$ denotes the standard normal quantile function. Similarly, let $u_{\mathrm{Apat},t}$ denote the PIT of $h_{t}$ under the Apatosaurus model given by \eqref{eq:apatmar} where the posterior mean estimates are also used for the time-varying and constant parameters. We plot both $\{Z(u_{\mathrm{TrSkt},t})\}$ and $\{Z(u_{\mathrm{Apat},t})\}$ against the standard normal quantiles in Figure~\ref{fig:qq_h}. It is apparent that without the added Exponential component, the truncated-skewed-$t$ model is not flexible enough for the left tail of the $\{h_{t}\}$.

\FloatBarrier

\FloatBarrier

It is worth remarking that the proposed DQF model is highly flexible, and that such level of flexibility is necessary to accurately model the real data. In terms of model adequacy, it is shown in Appendix~\ref{app:dic} that the proposed model performs substantially better than a simpler model based on independent AR(1) margins.

\subsection{An Investigation into Time-Series Informativeness}
\label{sec:rsig}
\begin{figure}[h]
    \centering
    \begin{subfigure}[t]{0.35\textwidth}
        \centering
                \caption{$a_{t}$}
        \includegraphics[height=5cm]{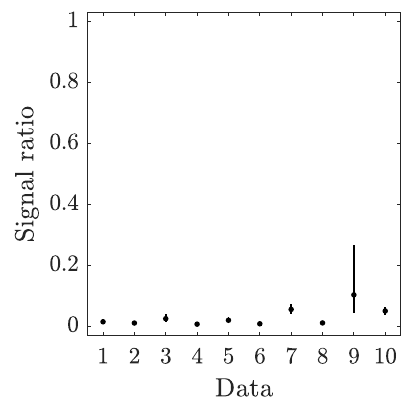}
    \end{subfigure}%
    ~
    \begin{subfigure}[t]{0.35\textwidth}
        \centering
                \caption{$b^{*}_{t}$}
        \includegraphics[height=5cm]{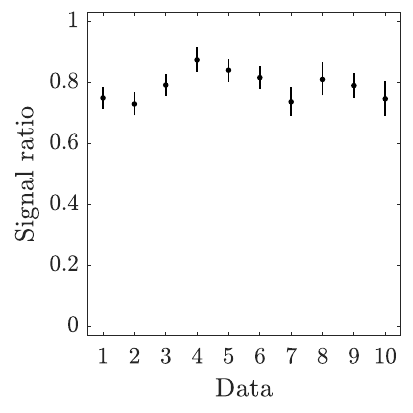}
    \end{subfigure}
    
    \begin{subfigure}[t]{0.35\textwidth}
        \centering
                \caption{$g_{t}$}
        \includegraphics[height=5cm]{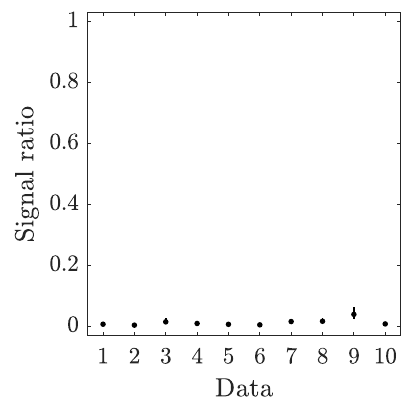}
    \end{subfigure}%
    ~
    \begin{subfigure}[t]{0.35\textwidth}
        \centering
                \caption{$h_{t}$}
        \includegraphics[height=5cm]{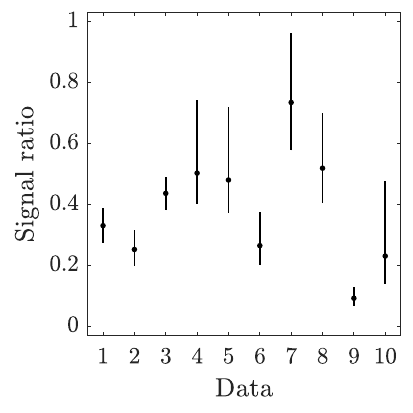}
    \end{subfigure}
    \caption{\small Posterior mean estimates (dots) and credible intervals (bars) of signal ratios for each marginal model. The numbers $1 \ddd 10$ are used to identify the ten indices: 1--SPX, 2--DJIA, 3--Nasdaq, 4--FTSE, 5--DAX, 6--CAC, 7--Nikkei, 8--HSI, 9--SSEC, 10--AORD.}
    \label{fig:rsig}
\end{figure}

One advantage of our approach is that it enables us to separately study the time-series predictability of various characteristics of intra-daily return distributions. By examining the plots in Figure~\ref{fig:xi} and the parameter estimates of $\psi_{1} \ddd \psi_{4}$ and $\phi_{1} \ddd \phi_{4}$, it seems apparent that some marginal processes are more ``informative" than others. For example, it seems reasonable to state that the time-series of $b^{*}_{t}$ and $h_{t}$ appear to be more predictable than those of $a_{t}$ and $g_{t}$. Here we formally quantify such informativeness in time-series, by proposing a model based measure, called \emph{signal ratio}. Let $\{\xi_{t}\colon\, t \in \mathbb{Z}\}$ be a real-valued covariance stationary process. The signal ratio, denoted by $R_{\mathrm{Sig}}$, is then defined as
\begin{equation*}
R_{\mathrm{Sig}} = \frac{\Var[\E(\xi_{t} \vbar \mathcal{F}_{t-1})]}{\Var(\xi_{t})},
\end{equation*}
where $\mathcal{F}_{t-1} = \sigma(\{\xi_{s}\colon\, s \le t-1\})$ is the natural filtration. Intuitively, $R_{\mathrm{Sig}}$ measures the proportion of unconditional variance explained by the variation in conditional means. The signal ratio nomenclature derives from the interpretation of conditional means as unobserved signals of a noisy process. It is easily checked that an i.i.d.~process has a signal ratio of zero, while a fully deterministic process has a signal ratio of one. The expression of $R_{\mathrm{Sig}}$ is available in closed form for the marginal model for $\{a_{t}\}$, $\{b^{*}_{t}\}$, and $\{g_{t}\}$. It can be computed numerically using simulation for the Apatosaurus model for $\{h_{t}\}$. Additional material on signal ratio is given in Appendix~\ref{app:rsig}. The posterior mean estimates of $R_{\mathrm{Sig}}$ and 95\% credible intervals are plotted in Figure~\ref{fig:rsig} for each margin. The $R_{\mathrm{Sig}}$ estimates are consistently high for $\{b^{*}_{t}\}$ across all indices, which justifies models that make use of realised measures of dispersion. For $\{h_{t}\}$, the estimates vary considerably across indices, with Nikkei being the highest and SSEC being the lowest; for some indices, the posteriors are much more diffused compared to those for $\{b^{*}_{t}\}$. All the $R_{\mathrm{Sig}}$ estimates for $\{a_{t}\}$ and $\{g_{t}\}$ are close to zero, except perhaps for SSEC, which suggests that it is in general difficult to predict the location and asymmetry of one-minute returns.

\FloatBarrier

\subsection{\change{Forecasting Value-at-Risk of Intra-daily Returns}}
\label{sec:var_1min}
\noindent \change{We focus on the empirical application of tail risk forecasting. To assess the out-of-sample performance, the proposed DQF model is applied to forecast the daily VaR measures of intra-daily returns, i.e., the quantiles of one-minute returns. The VaR forecasts for any selected probability level $u$ is computed by evaluating the one-step-ahead forecasts of the $g$-and-$h$ quantile function at $u$. The lower tail VaR forecasts for $u \in \{1\%, 5\%\}$ are compared to those produced by the state-of-the-art models for interval- and histogram-valued time-series, proposed by \cite{ArroyoGonzalezRiveraMate2010}, \cite{ArroyoEtAl2011}, and \cite{GonzalezRiveraArroyo2012}.}

\vspace{11pt}
\noindent \change{Given an interval-valued time-series (ITS) $\{[x]_t\}_{t=1}^T$, the exponential smoothing (ES) forecast \citep{ArroyoGonzalezRiveraMate2010, GonzalezRiveraArroyo2012} is written as:
\[
[\tilde{x}]_{t} = \alpha [x]_{t-1} + (1-\alpha) [\tilde{x}]_{t-1},
\]
where an interval $[x] := [x_L, x_U]$ is defined by an ordered pair of endpoints $(x_L, x_U)$ for $x_L < x_U$. The smoothing parameter $\alpha \in [0,1]$ is obtained using a simple one-dimensional grid-search, minimising the mean distance error $\frac{1}{T}\sum_{t=1}^T D_2([x]_t, [\tilde{x}]_t)$ where $D_2([x], [\tilde{x}]) := [(x_L - \tilde{x}_L)^2 + (x_U - \tilde{x}_U)^2]^{1/2}$.}

\vspace{11pt}
\noindent \change{Analogously, for a histogram-valued time-series (HTS) $\{h_{X_t}\}_{t=1}^{T}$, the ES forecast \citep{ArroyoEtAl2011, GonzalezRiveraArroyo2012} is given by:
\[
\tilde{h}_{X_t} = \alpha h_{X_{t-1}} + (1-\alpha) \tilde{h}_{X_{t-1}},
\]
where a histogram $h_{X} := \{([x]_i, \pi_i)\}_{i=1}^n$ is defined by a set of bins $\{[x]_i\}_{i=1}^n$ and the corresponding frequencies $\{\pi_i\}_{i=1}^n$. The weighted average of histograms $\tilde{h}_{X_t}$ is defined as the ``barycentric" histogram. The smoothing parameter $\alpha \in [0,1]$ is obtained using a grid-search, minimising the mean distance error $\frac{1}{T}\sum_{t=1}^T D_\mathrm{M}(h_{X_t}, \tilde{h}_{X_t})$ where $D_{\mathrm{M}}$ is the Mallows distance.}

\vspace{11pt}
\noindent \change{Following \cite{ArroyoGonzalezRiveraMate2010, ArroyoEtAl2011} and \cite{GonzalezRiveraArroyo2012}, the bin boundaries of each daily histogram are given by the sample quantiles of one-minute returns at levels $\{1\%, 5\%, 10\%, 20\%, \ddd 90\%, 95\%, 99\%\}$. Similarly, for ITS, the endpoints of each daily interval correspond to the 1\% and 5\% sample quantiles of intra-daily returns. As argued in \cite{ArroyoEtAl2011}, although we are mostly interested in the lower-tail of the return distribution, the idea of constructing the full histogram (with mid- and upper-quantiles) is to borrow information across many quantiles. On the other hand, the ITS model targets directly the quantiles of interest.}

\vspace{11pt}
\noindent \change{The one-minute returns are carefully cleaned according to the rules documented in Section~5.1. Each return series contains approximately 5,000 trading days of intra-daily observations, spanning from January 1996 to May 2016. We employ a sliding window of past 3,000 days for every one-day-ahead forecast. The last approximately 2,000 days of each series are used for out-of-sample assessment. To keep the computation cost manageable, each model is re-estimated after every 10 consecutive forecasts.}

\change{For the DQF model, the point forecasts are given by the the posterior mean of the predicted quantile at level $u \in \{5\%, 1\%\}$, obtained by integrating over the posterior distribution of model parameters.
\[
\int_{\mathbb{A}} \tilde{X}_t(u) \pi(\boldsymbol{\theta}) \mathrm{d}\boldsymbol{\theta}.
\]
Notice that conditional on observed data and a selected quantile level $u$, $\tilde{X}_t(u)$ is just a function of the model parameters $\boldsymbol{\theta}$. The posterior mean forecast given by the above integral accounts for parameter uncertainty by averaging over all possible parameter values weighted by the posterior distribution, and is computed using the output given by the MCMC algorithm. Specifically, the posterior means computed for $t \in \{3001 \ddd \allowbreak T\}$ using the most updated posterior sample. The adaptive MCMC algorithm is run every 10 days to update the posterior draws, at estimation point $t' \in \{3000, 3010 \ddd T-[(T-3000) \bmod 10]\}$ using the past 3000 QF-valued observations $\{X_{t'-3000+1} \ddd X_{t'}\}$. }

\vspace{11pt}
\noindent \change{To evaluate the out-of-sample performance, the mean absolute forecast errors (MAFE) over the forecasting period is computed for each model at a specified quantile level $u \in \{1\%, 5\%\}$:
\[
\mathrm{MAFE}_u := \frac{1}{T - 3000} \sum_{t=3001}^T |q_{u,t} - \tilde{q}_{u,t}|,
\]
where $q_{u,t}$ is the observed $u$-quantile of one-minute returns on day $t$, and $\tilde{q}_{u,t}$ is the one-day-ahead $u$-quantile forecast retrieved from the one-day-ahead symbol-valued forecast (either interval, histogram, or $g$-and-$h$ quantile function).}

\vspace{11pt}
\noindent \change{The proposed DQF model (DQF-Full) is compared against the exponential smoothing approach based on interval-valued (ITS-ES) and histogram-valued (HTS-ES) time-series. A much simpler specification (DQF-AR1) for the conditional joint distribution of $\boldsymbol{\xi}_t \vbar \mathcal{F}_{t-1}$ is also added for comparison, where each margin is assumed to independently follow an AR(1) process with Gaussian innovations. More details on this simplified model are provided in Appendix D. The $\mathrm{MAFE}_{5\%}$ and $\mathrm{MAFE}_{1\%}$ are reported in Tables~\ref{tab:mafe_5} and \ref{tab:mafe_1}, respectively.}

\vspace{11pt}
\noindent \change{The difference in MAFE between DQF-Full and each of the competing models is statistically assessed by conducting a heteroscedasticity-and-autocorrelation-consistent (HAC) Diebold-Mariano (DM) test \citep{DieboldMariano1995}. The $p$-values of the DM tests are reported in Tables~\ref{tab:mafe_dm_5} and \ref{tab:mafe_dm_1}. A small $p$-value indicates that, on average, the forecasting error of one model is smaller than that of the other over the out-of-sample period.}

\vspace{11pt}
\noindent \change{For 5\% VaR forecasts, the DQF-Full model delivers the smallest realised MAFE for nine of the ten market indices, with the exception of CAC. Compared to DQF-AR1 and HTS-ES, the DM null hypothesis is rejected at the customary 0.05 level in 5/10 markets, with DQF-Full being the preferred method. Compared to the ITS-ES model, DQF-Full is significantly preferred in 4/10 markets. In those cases where the DM null is not rejected at the 0.05 level, the proposed DQF-Full model is performing at least as well as the benchmark models.}

\vspace{11pt}
\noindent \change{For 1\% VaR forecasts, the DQF-Full model achieves the smallest realised MAFE for all ten market indices. Compared against the DQF-AR1 and HTS-ES models, the null hypothesis of the DM test is strongly rejected at the 0.025 level for 10/10 and 8/10 markets, respectively. Compared to the ITS-ES model, the DM null is rejected for 9/10 market indices. Out of a total of 30 pairwise DM tests, the null hypothesis is strongly rejected at levels much less than 0.01 in 22 cases, with DQF-Full being the favoured model.}

\vspace{11pt}
\noindent \change{In summary, via an extensive forecasting study with a long out-of-sample period of approximately 2,000 days, across ten international markets, the proposed DQF model (DQF-Full) is shown to be the overall best performing method in providing daily forecasts of the 5\% and 1\% VaR of intra-daily returns. The out-performance of the DQF model is especially pronounced when forecasting the more extreme 1\% VaR, which suggests that the DQF model is able to capture the conditional tail-shape of the high-frequency returns more accurately than the competing models. This is not surprising because (1) a wide variety of distributions can be approximated by the $g$-and-$h$ distribution to a very high degree, including the generalised Pareto distribution used in extreme value theory methods \citep{DuttaPerry2006}, and (2) our purposely designed conditional Apatosaurus marginal model is able to accurately capture the time-series dynamics of $\{h_t\}$, which controls the tail behavior of the $g$-and-$h$ distribution. Furthermore, the fact that DQF-Full delivers a significantly smaller forecast error than DQF-AR1 in most cases confirms that the additional flexibility of the full specification for the conditional distribution of $\boldsymbol{\xi}_t \vbar \mathcal{F}_{t-1}$ contributes to a more accurate model for the underlying data generating process.}

\begin{table}[h]
\hspace{-1.8cm}
\begin{tabular}{lcccccccccc}
\toprule
 & SPX & DJIA & NASDAQ & FTSE & DAX & CAC & NIKKEI & HSI & SSEC & AORD \\
\midrule
ITS-ES & 0.0140 & 0.0130 & 0.0135 & \textbf{0.0106} & 0.0137 & \textbf{0.0144} & 0.0176 & 0.0126 & 0.0174 & 0.0090 \\ 
HTS-ES & 0.0140 & 0.0130 & 0.0136 & \textbf{0.0106} & 0.0137 & 0.0145 & 0.0177 & 0.0127 & 0.0175 & 0.0089 \\ 
DQF-AR1 & 0.0140 & 0.0129 & 0.0135 & 0.0109 & 0.0138 & 0.0150 & 0.0177 & 0.0138 & 0.0177 & 0.0094 \\ 
DQF-Full & \textbf{0.0138} & \textbf{0.0127} & \textbf{0.0133} & \textbf{0.0106} & \textbf{0.0136} & 0.0145 & \textbf{0.0171} & \textbf{0.0121} & \textbf{0.0167} & \textbf{0.0085} \\
\bottomrule
\end{tabular}
\caption{\change{Mean absolute forecast errors for 5\% VaR forecasts of intra-daily returns. Bold texts indicate the most favoured models.}}
\label{tab:mafe_5}
\end{table}
\FloatBarrier

\begin{table}[h]
\hspace{-1.8cm}
\begin{tabular}{lcccccccccc}
\toprule
 & SPX & DJIA & NASDAQ & FTSE & DAX & CAC & NIKKEI & HSI & SSEC & AORD \\ 
\midrule
ITS-ES & 0.0249 & 0.0240 & 0.0231 & 0.0195 & 0.0288 & 0.0279 & 0.0383 & 0.0279 & 0.0321 & 0.0276 \\ 
HTS-ES & 0.0249 & 0.0240 & 0.0231 & 0.0195 & 0.0288 & 0.0279 & 0.0384 & 0.0280 & 0.0321 & 0.0276 \\ 
DQF-AR1 & 0.0249 & 0.0239 & 0.0236 & 0.0202 & 0.0294 & 0.0289 & 0.0381 & 0.0298 & 0.0312 & 0.0266 \\ 
DQF-Full & \textbf{0.0240} & \textbf{0.0232} & \textbf{0.0224} & \textbf{0.0191} & \textbf{0.0283} & \textbf{0.0274} & \textbf{0.0361} & \textbf{0.0268} & \textbf{0.0300} & \textbf{0.0252} \\ 
\bottomrule
\end{tabular}
\caption{\change{Mean absolute forecast errors for 1\% VaR forecasts of intra-daily returns. Bold texts indicate the most favoured models.}}
\label{tab:mafe_1}
\end{table}
\FloatBarrier

\begin{table}[h]
\hspace{-0.9cm}
\begin{tabular}{lcccccccccc}
\toprule
 & SPX & DJIA & NASDAQ & FTSE & DAX & CAC & NIKKEI & HSI & SSEC & AORD \\ 
\midrule
ITS-ES & 0.295 & \textbf{0.047} & 0.200 & 0.736 & 0.262 & 0.651 & 0.090 & \textbf{0.000} & \textbf{0.001} & \textbf{0.000} \\ 
HTS-ES & 0.277 & \textbf{0.038} & 0.082 & 0.534 & 0.187 & 0.822 & \textbf{0.041} & \textbf{0.000} & \textbf{0.000} & \textbf{0.001} \\ 
DQF-AR1 & 0.190 & 0.313 & 0.297 & 0.058 & 0.166 & \textbf{0.012} & \textbf{0.024} & \textbf{0.000} & \textbf{0.000} & \textbf{0.000} \\ 
\bottomrule
\end{tabular}
\caption{\change{$p$-values of Diebold-Mariano tests against ``DQF-Full" for 5\% VaR forecasts of intra-daily returns. Bold texts indicate $p$-values being less than the customary threshold of 0.05.}}
\label{tab:mafe_dm_5}
\end{table}
\FloatBarrier

\begin{table}[h]
\hspace{-0.9cm}
\begin{tabular}{lcccccccccc}
\toprule
 & SPX & DJIA & NASDAQ & FTSE & DAX & CAC & NIKKEI & HSI & SSEC & AORD \\ 
\midrule
ITS-ES & \textbf{0.002} & \textbf{0.001} & \textbf{0.001} & 0.141 & \textbf{0.048} & \textbf{0.025} & \textbf{0.000} & \textbf{0.000} & \textbf{0.000} & \textbf{0.000} \\ 
HTS-ES & \textbf{0.002} & \textbf{0.001} & \textbf{0.001} & 0.117 & 0.053 & \textbf{0.019} & \textbf{0.000} & \textbf{0.000} & \textbf{0.000} & \textbf{0.000} \\ 
DQF-AR1 & \textbf{0.013} & \textbf{0.024} & \textbf{0.000} & \textbf{0.000} & \textbf{0.001} & \textbf{0.000} & \textbf{0.000} & \textbf{0.000} & \textbf{0.000} & \textbf{0.000} \\ 
\bottomrule
\end{tabular}
\caption{\change{$p$-values of Diebold-Mariano tests against ``DQF-Full" for 1\% VaR forecasts of intra-daily returns. Bold texts indicate $p$-values being less than the customary threshold of 0.05.}}
\label{tab:mafe_dm_1}
\end{table}
\FloatBarrier

\subsection{\change{Lower Frequency Value-at-Risk via Quantile Regression}}
\label{sec:scaling}
\change{In addition to forecasting daily VaR of intra-daily returns, we demonstrate that the QF-valued forecasts provided by the DQF model can be used to forecast VaR measures at the daily timescale via a simple quantile regression model on daily returns. The simple quantile regression approach is used because it does not change the dynamics of the quantile forecasts of one-minute returns (except for the scale), which allows us to study whether the DQF forecasts can be useful in forecasting daily-scale VaR measures without too much additional effort (i.e., with a very simple model for daily returns). We term this quantile regression model the QR-DQF model.}

\change{Let $q^{\mathrm{M}}_{u,t}$ and $q^{\mathrm{D}}_{u,t}$ be the $u$-level quantiles of one-minute and daily returns, respectively. Let $\mathbf{y}^{\mathrm{D}} := \{y^{\mathrm{D}}_{1} \ddd y^{\mathrm{D}}_{T}\}$ be a sequence of daily close-to-close returns. We assume that a quantile of daily returns can be modelled by the linear relationship
\begin{equation}
\label{eq:qscale}
q^{\mathrm{D}}_{u,t} = s_{u} q^{\mathrm{M}}_{u,t}.
\end{equation}
As the DQF model provides a filtered value of the $u$-level quantile of one-minute returns for each day, $q^{\mathrm{M}}_{u,t}$ can be treated as observed and given by $\tilde{X}_{t}(u)$. If $s_u$ is constant across all $u \in (0,1)$, the coefficient $s_u$ has the interpretation of being the distributional scaling factor of a unifractal or multifractal process, where the distribution of the process at a given timescale is related to that at any other timescale through a scaling law. For examples, see \citet{HallamOlmo2014b} and \citet{HallamOlmo2014a} for recent approaches in estimating the density of daily returns from intra-daily data via distributional scaling laws. Here, we do not assume any scaling properties between distributions of returns at different timescales, such as those discussed in \citet{DiMatteo2007}, and allow the scaling factor to vary across quantile levels. The estimate of the coefficient $\hat{s}_{u}$ can be computed by solving the quantile regression minimisation problem \citep{KoenkerBassett1978}}
\begin{equation}
\label{eq:qreg_objfun}
\hat{s}_{u} = \argmin_{s_{u}} \sum_{t=1}^{T}\rho_{u}\left(y^{\mathrm{D}}_{t} - s_{u}q^{\mathrm{M}}_{u,t}\right),
\end{equation}
where the loss function $\rho_{u}$ is defined by
\begin{equation}
\label{eq:qreg_lossfun}
\rho_{u}(\varepsilon) := \varepsilon[u - I_{(-\infty,0)}(\varepsilon)].
\end{equation}
It is shown by \citet{YuMoyeed2001} that minimising the objective function in \eqref{eq:qreg_objfun} is numerically equivalent to maximising a likelihood function where the observations are assumed to follow the asymmetric Laplace (AL) distributions. The AL family has the following density function
\begin{equation}
f_{\mathrm{AL}}(\varepsilon; \mu, \sigma, u) = \frac{u(1 - u)}{\sigma}\exp\left[-\rho_{u}\left(\frac{\varepsilon - \mu}{\sigma}\right)\right],
\end{equation}
where $\mu \in \mathbb{R}$, $\sigma \in (0, \infty)$, and $u \in (0, 1)$ are the location, scale, and asymmetry parameters, respectively.

When $y^{\mathrm{D}}_{t}$ is assumed to follow an AL distribution with the location parameter being $s_{u}q^{\mathrm{M}}_{u,t}$, the likelihood function is then given by
\begin{equation}
\label{eq:qreg_like}
\begin{aligned}
f(\mathbf{y}^{\mathrm{D}}; s_{u}, \sigma) &= \prod_{t=1}^{T}f_{\mathrm{AL}}(y^{\mathrm{D}}_{t}; s_{u}q^{\mathrm{M}}_{u,t}, \sigma, u) \\
& \propto \sigma^{-T}\exp\left[-\sigma^{-1}\sum_{t=1}^{T}\rho_{u}(y^{\mathrm{D}}_{t} - s_{u}q^{\mathrm{M}}_{u,t})\right].
\end{aligned}
\end{equation}
The prior density $p$ is then defined by placing an improper flat prior on $s_{u}$ and an inverse prior on $\sigma$,
\begin{equation}
\label{eq:qreg_prior}
s_{u}, \sigma \sim p(s_{u}, \sigma) \propto \sigma^{-1}.
\end{equation}
The posterior density $\pi$ is then given by
\begin{equation}
\label{eq:qreg_post}
\begin{aligned}
s_{u}, \sigma \vbar \mathbf{y}^{\mathrm{D}} \sim \pi(s_{u}, \sigma) &\propto f(\mathbf{y}^{\mathrm{D}}; s_{u}, \sigma) p(s_{u}, \sigma) \\
&\propto \sigma^{-(T+1)}\exp\left[-\sigma^{-1}\sum_{t=1}^{T}\rho_{u}(y^{\mathrm{D}}_{t} - s_{u}q^{\mathrm{M}}_{u,t})\right].
\end{aligned}
\end{equation}
As we are only interested in the coefficient $s_{u}$, we integrate out the scale parameter $\sigma$ to obtain the marginal posterior density of $s_{u}$. Using the fact that \eqref{eq:qreg_post} has the form of the kernel of an Inverse Gamma density in $\sigma$, and that a density function must integrate to one, the marginal posterior of $s_{u}$ can be obtained in closed-form \citep{GerlachChenChan2011};
\begin{equation}
\label{eq:qreg_marpost}
\int_{0}^{\infty} \pi(s_{u},\sigma) d\sigma = \left[\sum_{i=1}^{T}\rho_{u}(y^{\mathrm{D}}_{t} - s_{u}q^{\mathrm{M}}_{u,t})\right]^{-T}.
\end{equation}

To compute the VaR estimates of daily returns at probability level $u$, we first sample from the univariate posterior in \eqref{eq:qreg_marpost} using the adaptive MCMC sampler described in Section~\ref{sec:mcmc}, where we plug-in the posterior mean estimates of $\{\tilde{X}_{t}(u)\}_{t=1}^T$ for $\{q^{\mathrm{M}}_{u,t}\}_{t=1}^T$. The posterior distribution of $\{q^{\mathrm{D}}_{u,t}\}_{t=1}^T$ conditional on $\{q^{\mathrm{M}}_{u,t}\}_{t=1}^T$ is then approximated via the MCMC output for $s_{u}$.

\subsection{Forecasting Value-at-Risk of Daily Returns}
\label{sec:fore_var}
\change{In this final part of this empirical study, the QR-DQF model described in Section~\ref{sec:scaling} is applied to forecast one-day-ahead the VaR measures of daily returns, at the 5\% and 1\%  probability levels. All the model parameters, including those for the DQF model and the quantile regression coefficient $s_u$, are estimated using the adaptive MCMC algorithm, based protocol employed in Section~\ref{sec:var_1min}, where each VaR forecast is computed with the parameters estimated using a sliding window of past 3,000 days.}

\begin{figure}[h]
\centering
\includegraphics[width=0.8\textwidth]{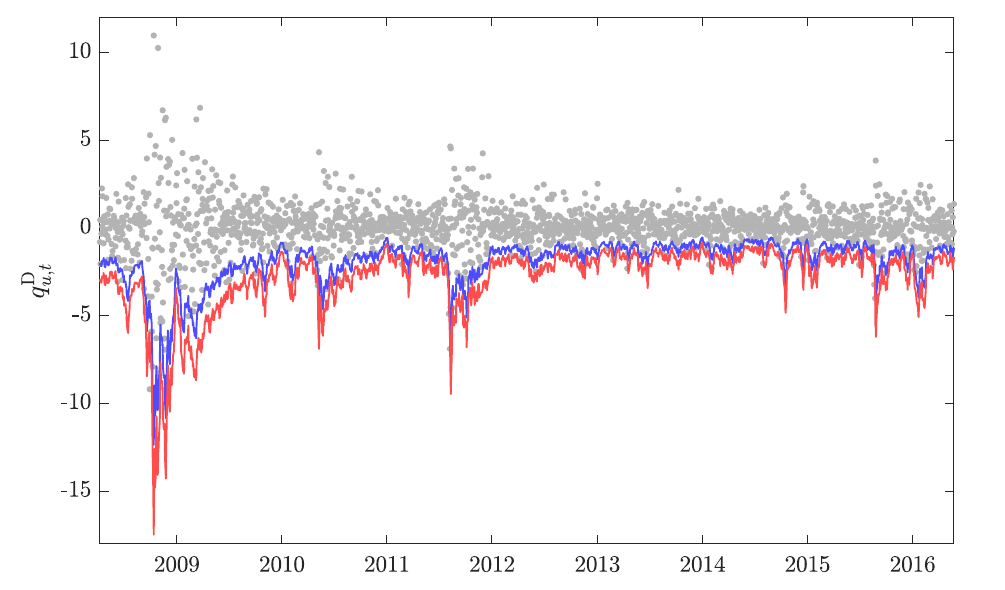}
\caption{\small One-step-ahead forecasts of $q^{\mathrm{D}}_{u,t}$ for $u \in \{5\%, 1\%\}$ given by the posterior mean for the S\&P~500 index.}
\label{fig:var}
\end{figure}

\begin{figure}[h]
    \centering
    \begin{subfigure}[t]{0.8\textwidth}
        \centering
        \includegraphics[height=3.5cm]{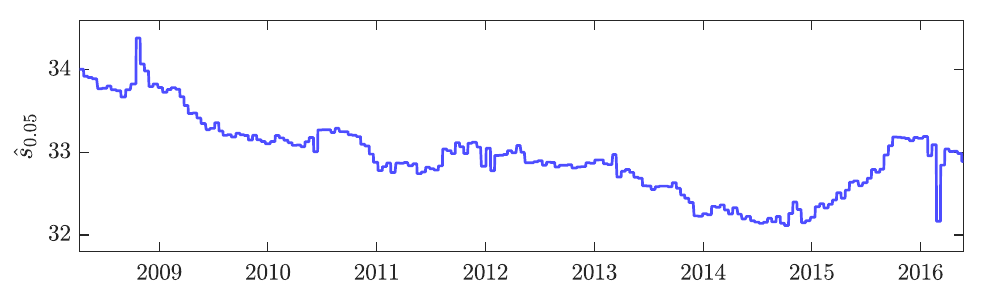}
        \caption{$u = 5\%$}
    \end{subfigure}
    \begin{subfigure}[t]{0.8\textwidth}
        \centering
        \includegraphics[height=3.5cm]{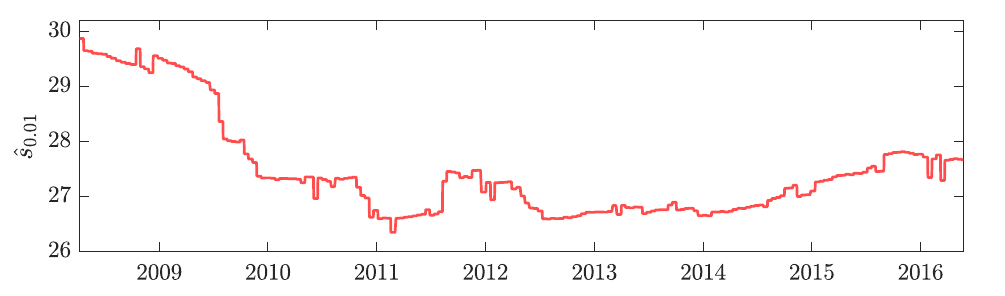}
        \caption{$u = 1\%$}
    \end{subfigure}
\caption{\small Scaling factors used for the forecasts of $q^{\mathrm{D}}_{u,t}$ for $u \in \{5\%, 1\%\}$ given by the posterior mean for the S\&P~500 index.}
\label{fig:s}
\end{figure}

As an illustration, the one-day-ahead out-of-sample VaR forecasts are plotted over daily returns for S\&P~500 in Figure~\ref{fig:var}; the corresponding estimates of the quantile regression coefficients, $\hat{s}_{5\%}$ and $\hat{s}_{1\%}$, are plotted in Figure~\ref{fig:s}. We make some observations: (i) The VaR forecasts follow closely the bottom shoulder of the daily return data, and react instantaneously to changes in volatility. This suggests that the tail dynamics of one-minute returns can be scaled to approximate that of daily returns. I.e., projecting the tail of intra-daily returns is a sensible method for obtaining daily VaR estimates. (ii) The fact that $\hat{s}_{5\%} > \hat{s}_{1\%}$ for the entire forecast period indicates that daily returns are potentially lighter-tailed than intra-daily returns, which is consistent with observations in the literature. (iii) The value of $\hat{s}_{u}$ is higher for high volatility periods, which suggests that $s_{u}$ may not be constant over time.

The theory of \emph{elicitability} provides a decision-theoretic framework of comparative backtesting of risk measure forecasts; see, e.g., \citet{Gneiting2011}, \citet{KouPeng2016}, \citet{Brehmer2017} and \citet{NoldeZiegel2017}. Consider a probability distribution $F \in \mathbb{F}$ on a state space $\mathcal{Y}$. A risk measure can be viewed as a functional $\mathscr{K}:\mathbb{F} \to \mathcal{Y}'$, where $\mathcal{Y}'$ is the action domain. A scoring function $\mathscr{S}: \mathcal{Y}' \times \mathcal{Y} \to [0,\infty)$ is \emph{strictly $\mathbb{F}$-consistent} for $\mathscr{K}$ if
\begin{equation}
\label{eq:consistent}
\mathscr{K}(F) = \argmin_{y'} \int_{\mathcal{Y}'} \mathscr{S}(y', \cdot) dF, \quad \forall F \in \mathbb{F}.
\end{equation}
A risk measure $\mathscr{K}$ is called \emph{elicitable} if there exists a strictly $\mathbb{F}$-consistent scoring function for it \citep{Gneiting2011}. When the risk measure is the $u$-quantile functional (i.e, VaR), a strictly $\mathbb{F}$-consistent scoring function is given by
\begin{equation}
\label{eq:scoring_func}
\mathscr{S}(y', y) := [I_{[y,\infty)}(y') - u](y' - y).
\end{equation}
This suggests that sequences of VaR forecasts can be ranked via an empirical approximation of the integral in Eq.~\eqref{eq:consistent}
\begin{equation}
\label{eq:ave_score}
\bar{\mathscr{S}} := \frac{1}{T - 3000}\sum_{t=3001}^{T}\mathscr{S}(q^{\mathrm{D}}_{u,t}, y^{\mathrm{D}}_{t}).
\end{equation}

\change{Based on the scoring function in Eq.~\eqref{eq:scoring_func}, the QR-DQF model is ranked against an array of popular models in the literature of forecasting VaR of daily returns: Symmetric Absolute Value CAViaR of \citet{EngleManganelli2004} (CAViaR), GJR-GARCH of \citet{GlostenJagannathanRunkle1993} with $t$ (GJR-t) and skewed-$t$ (GJR-skt) distributions, and Realized-GARCH of \citet{HansenHuangShek2012} with $t$ (Real-t) and skewed-$t$ (Real-skt) distributions. For the Real-t and Real-skt models, the ``log-linear" specification is used. For each sequence of VaR forecasts $\{q^{\mathrm{D}}_{u,t}\}_{t=3001}^T$, the value of $\bar{\mathscr{S}}$ is reported in Table~\ref{tab:s5} for $u = 5\%$ and Table~\ref{tab:s1} for $u = 1\%$, respectively. Across ten market indices that span different geographic regions, the QR-DQF model is most favoured in 6/10 markets, for both 5\% and 1\% VaR forecasts. Upon closer inspection, QR-DQF appears to be the dominant model for North-American and European markets (SPX, DJIA, Nasdaq, FTSE, DAX, CAC). For the Asia-Pacific indices (Nikkei, HSI, SSEC, and AORD) however, GJR-skt is the best suited model on average.}

\FloatBarrier

\FloatBarrier

\begin{table}[h]
\hspace{-1.2cm}
\small
\begin{tabular}{lcccccccccc}
\toprule
 & SPX & DJIA & Nasdaq & FTSE & DAX & CAC & Nikkei & HSI & SSEC & AORD \\ 
\midrule
CAViaR & 0.1434 & 0.1295 & 0.1543 & 0.1316 & 0.1498 & 0.1643 & 0.1839 & 0.1607 & 0.1896 & 0.1158 \\ 
GJR-t & 0.1368 & 0.1238 & 0.1503 & 0.1307 & 0.1493 & 0.1631 & 0.1785 & 0.1582 & 0.1896 & 0.1167 \\ 
GJR-skt & 0.1355 & 0.1227 & 0.1495 & 0.1298 & 0.1477 & 0.1619 & 0.1775 & \bf{0.1579} & \bf{0.1893} & 0.1151 \\ 
Real-t & 0.1325 & 0.1193 & 0.1443 & 0.1323 & 0.1477 & 0.1604 & 0.1807 & 0.1637 & 0.1914 & 0.1147 \\ 
Real-skt & 0.1313 & 0.1184 & 0.1445 & 0.1313 & \bf{0.1468} & 0.1603 & 0.1795 & 0.1629 & 0.1939 & \bf{0.1144} \\ 
QR-DQF & \bf{0.1306} & \bf{0.1182} & \bf{0.1442} & \bf{0.1291} & 0.1469 & \bf{0.1601} & \bf{0.1772} & 0.1638 & 0.1926 & 0.1163 \\ 
\bottomrule
\end{tabular}
\caption{\small Values of $\bar{\mathscr{S}}$ for 5\% VaR forecasts. Bold texts indicate the most favoured models.}
\label{tab:s5}
\end{table}
\FloatBarrier

\begin{table}[h]
\hspace{-1.2cm}
\small
\begin{tabular}{lcccccccccc}
\toprule
 & SPX & DJIA & Nasdaq & FTSE & DAX & CAC & Nikkei & HSI & SSEC & AORD \\ 
\midrule
CAViaR & 0.0411 & 0.0346 & 0.0417 & 0.0365 & 0.0388 & 0.0446 & 0.0523 & 0.0442 & \bf{0.0557} & 0.0313 \\ 
GJR-t & 0.0372 & 0.0316 & 0.0398 & 0.0352 & 0.0401 & 0.0441 & 0.0505 & 0.0429 & 0.0566 & 0.0304 \\ 
GJR-skt & 0.0367 & 0.0312 & 0.0388 & 0.0344 & 0.0399 & 0.0442 & \bf{0.0498} & \bf{0.0427} & 0.0558 & \bf{0.0297} \\ 
Real-t & 0.0360 & 0.0306 & 0.0390 & 0.0362 & 0.0388 & 0.0441 & 0.0508 & 0.0454 & 0.0582 & 0.0301 \\ 
Real-skt & 0.0358 & 0.0306 & 0.0392 & 0.0354 & 0.0383 & 0.0439 & \bf{0.0498} & 0.0450 & 0.0595 & 0.0305 \\ 
QR-DQF & \bf{0.0349} & \bf{0.0296} & \bf{0.0378} & \bf{0.0342} & \bf{0.0379} & \bf{0.0435} & 0.0514 & 0.0473 & 0.0581 & 0.0314 \\ 
\bottomrule
\end{tabular}
\caption{\small Values of $\bar{\mathscr{S}}$ for 1\% VaR forecasts. Bold texts indicate the most favoured models.}
\label{tab:s1}
\end{table}
\FloatBarrier

\change{It is possible to conduct pairwise DM tests with the quantile loss function to statistically assess the difference between the reported values of $\bar{\mathscr{S}}$ in Tables~\ref{tab:s5} and \ref{tab:s1}. However, we find that the tail-index estimates (from the Hill estimator) of the realised loss differentials are well below 2 for many pairs, suggesting that the variances of the loss differentials are highly likely to be unbounded in many cases. Since finite variance is a necessary condition for validity of the DM test \citep{Diebold2015}, we will turn to the violation-based tests that have become standard in the VaR backtesting literature.}

\change{The violation rate (VRate) is a frequently employed measure for assessing the VaR forecast accuracy. It is defined as the proportion of returns in the forecasting period that exceed the VaR forecasts. I.e.,
\begin{equation}
\mathrm{VRate}_u := \frac{1}{T-3000} \sum_{t=3001}^T I_{(-\infty, y^{\mathrm{D}}_t)}(q^{\mathrm{D}}_{u,t}).
\end{equation}
Models with $\mathrm{VRate}_u$ being closer to the nominal probability level $u$ are preferred. The unconditional coverage (UC) test of \citet{Kupiec1995} aims to test the null hypothesis that $\mathrm{VRate}_u = u$. The $p$-values of the UC tests are reported in Tables~\ref{tab:uc5} and \ref{tab:uc1} for $u = 5\%$ and $u = 1\%$, respectively. A small $p$-value indicates that the empirical VRate is significantly different from the nominal VaR threshold. On the other hand, a large $p$-value implies that $\mathrm{VRate}_u$ is close to $u$.} 

\change{For 5\% VaR forecasts (Table~\ref{tab:uc5}), the QR-DQF model is best-ranked in SPX, DJIA, FTSE, and DAX in terms of VRate, while being rejected by the UC test in the four Asia-Pacific markets. Across all ten indices, the CAViaR model is performing the best on average, without any rejections of the UC null hypothesis.  The GJR-t and Real-t models are worst-ranked on average, with 7/10 rejections for both models.}

\change{For 1\% VaR forecasts (Table~\ref{tab:uc1}), the QR-DQF model has the most favoured VRates (i.e., closest to 1\%) in DJIA, NASDAQ, and SSEC, while being rejected by the UC test in 3/10 markets. On average, GJR-skt is the best performing model across the ten markets, with best-ranked VRates in 6/10 markets, followed by the Real-skt model, with best-ranked VRates in 3/10 markets. Both skewed-t models are not rejected by the UC test in any of the markets. Across the series, the GJR-t and Real-t models are the worst-ranked on average, with the UC null hypothesis rejected in 8/10 and 7/10 markets, respectively.}

\begin{table}[h]
\hspace{-1.2cm}
\small
\begin{tabular}{lcccccccccc}
\toprule
 & SPX & DJIA & NASDAQ & FTSE & DAX & CAC & NIKKEI & HSI & SSEC & AORD \\ 
\midrule
CAViaR & 0.128 & 0.101 & 0.085 & 0.123 & 0.463 & 0.184 & \bf 0.394 & \bf 0.496 & 0.898 & 0.082 \\ 
GJR-t & \rd{0.001} & \rd{0.001} & \rd{0.013} & \rd{0.001} & \rd{0.002} & \rd{0.001} & \rd{0.018} & 0.060 & 0.386 & \rd{0.000} \\ 
GJR-skt & 0.128 & \bf 0.123 & 0.401 & 0.210 & \rd{0.023} & 0.087 & 0.290 & 0.236 & \bf 0.932 & \rd{0.034} \\ 
Real-t & \rd{0.000} & \rd{0.000} & 0.104 & \rd{0.002} & \rd{0.001} & 0.218 & \rd{0.001} & \rd{0.000} & \rd{0.000} & 0.148 \\ 
Real-skt & 0.128 & \rd{0.027} & \bf 0.634 & 0.101 & 0.295 & 0.724 & \rd{0.011} & \rd{0.001} & \rd{0.000} & \bf 0.891 \\ 
QR-DQF & \bf 0.154 & 0.067 & 0.127 & \bf 0.393 & \bf 0.530 & \bf 0.877 & \rd{0.024} & \rd{0.000} & \rd{0.000} & \rd{0.026} \\ 
\bottomrule
\end{tabular}
\caption{\small \change{$p$-values of the \citet{Kupiec1995} UC test of 5\% VaR forecasts. Larger $p$-values are preferred. Bold texts indicate the most favoured models based on violation rates, while red texts indicate that the violation rates are significantly different from 5\% according to the UC test.}}
\label{tab:uc5}
\end{table}
\FloatBarrier

\begin{table}[h]
\hspace{-1.2cm}
\small
\begin{tabular}{lcccccccccc}
\toprule
 & SPX & DJIA & NASDAQ & FTSE & DAX & CAC & NIKKEI & HSI & SSEC & AORD \\ 
\midrule
CAViaR & \rd{0.005} & 0.068 & \rd{0.010} & 0.124 & 0.131 & \rd{0.027} & \rd{0.019} & 0.410 & \rd{0.021} & 0.473 \\ 
GJR-t & \rd{0.003} & \rd{0.000} & \rd{0.000} & \rd{0.000} & \rd{0.011} & 0.067 & \rd{0.002} & 0.211 & \rd{0.036} & \rd{0.000} \\ 
GJR-skt & 0.428 & \bf 0.422 & 0.426 & \bf 0.473 & 0.280 & \bf 0.810 & \bf 0.494 & \bf 0.872 & 0.092 & \bf 0.942 \\ 
Real-t & \rd{0.002} & \rd{0.016} & \rd{0.016} & \rd{0.000} & \rd{0.003} & \rd{0.009} & \rd{0.003} & 0.096 & 0.185 & 0.771 \\ 
Real-skt & \bf 0.939 & 0.319 & \bf 0.941 & 0.124 & \bf 0.680 & 0.406 & 0.124 & 0.096 & 0.062 & 0.117 \\ 
QR-DQF & \rd{0.027} & \bf 0.422 & \bf 0.941 & 0.282 & \rd{0.003} & 0.067 & 0.051 & \rd{0.014} & \bf 0.571 & 0.069 \\ 
\bottomrule
\end{tabular}
\caption{\small \change{$p$-values of the \citet{Kupiec1995} UC test of 1\% VaR forecasts. Larger $p$-values are preferred. Bold texts indicate the most favoured models based on violation rates, while red texts indicate that the violation rates are significantly different from 1\% according to the UC test.}}
\label{tab:uc1}
\end{table}
\FloatBarrier

\change{In addition to VRate, independence of violations is another important factor to consider for backtesting VaR forecast accuracy. The dynamic quantile (DQ) test of \citet{EngleManganelli2004} aims to test jointly the VRate and independence of violations. Based on the series of hits $\{\mathrm{Hit}_{u,t}\}_{t=3001}^T$ during the forecasting period, where $\mathrm{Hit}_{u,t} := I_{(-\infty, y^{\mathrm{D}}_t)}(q^{\mathrm{D}}_{u,t}) - u$, the DQ null hypothesis is $\E(\mathrm{Hit}_{u,t}) = 0$ \emph{and} that $\mathrm{Hit}_{u,t}$ is uncorrelated with variables in an information set. For the information set, it is customary in the VaR forecasting literature to include five lagged hits and the contemporaneous VaR forecast. The $p$-values of the DQ tests are reported in Tables~\ref{tab:dq5} and \ref{tab:dq1} for $u = 5\%$ and $u = 1\%$, respectively. A small $p$-value indicates that the VaR forecasts are either incorrectly proportioned or serially correlated or both.}

\change{For 5\% VaR forecasts (Table~\ref{tab:dq5}),  the QR-DQF is the best performing model in SPX and FTSE, while being rejected by the DQ test in NIKKEI, HSI, and SSEC. On average, across the ten series, GJR-skt is the best-ranked model, with 2/10 rejections of the DQ null hypothesis. The GJR-t and Real-t models are bottom-ranked overall, with 8/10 and 7/10 rejections, respectively.}

\change{For 1\% VaR forecasts (Table~\ref{tab:dq1}),  QR-DQF is the most favoured model in SSEC with respect to the DQ statistic, while being rejected in 6/10 markets. GJR-skt is the top-ranked model in 8/10 markets, with a single DQ null rejection in NIKKEI. The Real-skt model has the second least number of rejections (3/10). Across the ten markets, the GJR-t and Real-t models are least favoured on average, with the DQ null hypothesis rejected in 9/10 markets for both models.}

\begin{table}[h]
\hspace{-1.2cm}
\small
\begin{tabular}{lcccccccccc}
\toprule
 & SPX & DJIA & NASDAQ & FTSE & DAX & CAC & NIKKEI & HSI & SSEC & AORD \\ 
\midrule
CAViaR & \rd{0.000} & \rd{0.025} & \rd{0.003} & 0.621 & 0.263 & \bf 0.673 & \rd{0.000} & 0.578 & 0.370 & 0.206 \\ 
GJR-t & \rd{0.001} & \rd{0.003} & \rd{0.016} & \rd{0.004} & \rd{0.032} & \rd{0.033} & \rd{0.002} & 0.394 & 0.889 & \rd{0.000} \\ 
GJR-skt & 0.231 & \bf 0.121 & 0.319 & 0.451 & \bf 0.279 & 0.501 & \bf \rd{0.004} & \bf 0.659 & \bf 0.917 & \rd{0.035} \\ 
Real-t & \rd{0.000} & \rd{0.001} & \rd{0.083} & \rd{0.001} & \rd{0.000} & 0.472 & \rd{0.000} & \rd{0.000} & 0.079 & 0.445 \\ 
Real-skt & \rd{0.041} & 0.053 & \bf 0.573 & 0.222 & 0.080 & 0.823 & \rd{0.000} & \rd{0.007} & \rd{0.027} & \bf 0.815 \\ 
QR-DQF & \bf 0.240 & 0.057 & 0.459 & \bf 0.812 & 0.183 & 0.584 & \rd{0.001} & \rd{0.000} & \rd{0.039} & 0.422 \\ 
\bottomrule
\end{tabular}
\caption{\small \change{$p$-values of the \citet{EngleManganelli2004} DQ test of 5\% VaR forecasts. Larger $p$-values are preferred. Bold texts indicate the most favoured models based on the DQ statistic, while red texts indicate rejections of the null hypothesis at the 0.05 threshold.}}
\label{tab:dq5}
\end{table}
\FloatBarrier

\begin{table}[h]
\hspace{-1.2cm}
\small
\begin{tabular}{lcccccccccc}
\toprule
 & SPX & DJIA & NASDAQ & FTSE & DAX & CAC & NIKKEI & HSI & SSEC & AORD \\ 
\midrule
CAViaR & \rd{0.000} & \rd{0.000} & \rd{0.000} & 0.145 & \bf 0.367 & \rd{0.001} & \rd{0.009} & 0.433 & 0.159 & 0.744 \\ 
GJR-t & \rd{0.001} & \rd{0.000} & \rd{0.002} & \rd{0.001} & \rd{0.005} & \rd{0.034} & \rd{0.002} & \rd{0.005} & 0.177 & \rd{0.000} \\ 
GJR-skt & \bf 0.497 & \bf 0.362 & \bf 0.428 & \bf 0.928 & 0.084 & \bf 0.487 & \bf \rd{0.013} & \bf 0.524 & 0.314 & \bf 0.850 \\ 
Real-t & \rd{0.000} & \rd{0.000} & \rd{0.015} & \rd{0.000} & \rd{0.001} & \rd{0.001} & \rd{0.000} & \rd{0.002} & 0.965 & \rd{0.025} \\ 
Real-skt & 0.110 & 0.121 & 0.292 & \rd{0.015} & 0.113 & 0.142 & \rd{0.000} & \rd{0.002} & 0.868 & 0.211 \\ 
QR-DQF & \rd{0.000} & \rd{0.000} & 0.305 & 0.078 & \rd{0.008} & \rd{0.015} & \rd{0.000} & \rd{0.005} & \bf 0.992 & 0.849 \\ 
\bottomrule
\end{tabular}
\caption{\small \change{$p$-values of the \citet{EngleManganelli2004} DQ test of 1\% VaR forecasts. Larger $p$-values are preferred. Bold texts indicate the most favoured models based on the DQ statistic, while red texts indicate rejections of the null hypothesis at the 0.05 threshold.}}
\label{tab:dq1}
\end{table}
\FloatBarrier

\change{In summary, the proposed QR-DQF model is ranked against some of the most popular models in VaR forecasting literature, including those employing realised variance. The forecasting study consists of return series across ten geographically diverse market indices and forecasting periods of approximately 2,000 days. The models are compared in terms of VaR forecasting accuracy, using a strictly consistent scoring function (i.e., the quantile loss function) and standard tests based on violations. No one model is consistently outperforming the others across all ten markets at both 5\% and 1\% probability levels. According to the scoring function, the QR-DQF model is best-ranked in 6/10 markets for both 5\% and 1\% VaR forecasts. Based on both UC and DQ tests, the QR-DQF model is most favoured in SPX and FTSE for 5\% VaR forecasts, and best-ranked in SSEC at the 1\% level. The discrepancies in model ranking between accuracy measures are likely due to the following reasons. (1) The UC and DQ tests consider only the proportion of violations, while the scoring function takes into account both proportion and magnitude of violations. (2) The quantile loss function is a strictly consistent scoring function for VaR, whereas the distance from $\mathrm{VRate}_u$ to $u$ and $\E(\mathrm{Hit}_{u,t})$ are not. (3) As found by authors such as \citet{GiotLaurent2004} and \citet{ChenGerlach2013}, the violation based tests are more sensitive to the specification of the conditional distribution than that of the volatility dynamics. This is also evident in the UC and DQ test results (Tables~\ref{tab:uc5}, \ref{tab:uc1}, \ref{tab:dq5}, \ref{tab:dq1}), where the GJR-skt model consistently outperforms the GJR-t model, while the two skewed-t models have very similar results despite the fact that the Real-skt model uses intra-daily data (realised variance) and GJR-skt relies only on daily returns.  We can conclude that, in certain markets, the QR-DQF model is able to provide competitive VaR forecasts for daily returns.}

\section{Conclusion}
\label{sec:conclusion}
\change{Motivated by the recent development in time-series models for histogram-valued data in the SDA literature, we propose to consider a flexible parametric quantile function as a new symbol-type for summarising intra-daily returns. A new time-series model for QF-valued observations is developed based on the four-parameter $g$-and-$h$ quantile function. We call this model the DQF model. To account for parameter uncertainty and take advantage of the inherent numerical stability associated with sampling based procedures, a Bayesian formulation is proposed for the DQF model, together with a carefully designed adaptive MCMC algorithm. Via an extensive forecasting study, the DQF model is shown to significantly outperform the previously proposed ITS-ES and HTS-ES models in terms of forecasting 5\% and 1\% VaR of intra-daily returns. The out-performance of the DQF model is more prominent at the more extreme 1\% probability level, indicating that the DQF model is able to more accurately capture the dynamic tail-behaviour of the intra-daily returns. Through an additional forecasting experiment, it is demonstrated that the output (i.e., QF-valued forecasts) from the DQF model can be used by a simple quantile regression model (QR-DQF) to forecast VaR of daily returns. Compared to an array of popular models from the VaR forecasting literature, using both a strictly consistent scoring function and standard violation-based tests, the QR-DQF model is consistently best-ranked in SPX and FTSE for 5\% VaR forecasts, and SSEC for 1\% VaR forecasts.}

\section{Code}
MATLAB code to reproduce the experiments is available from:
\begin{center}
\href{https://github.com/wilson-ye-chen/aqua}{\texttt{github.com/wilson-ye-chen/aqua}}
\end{center}

\ifnotblinded
\section{Acknowledgement}
\label{sec:acknowledgement}
We thank Chris J. Oates for his careful reading of the manuscript and helpful comments. WYC and SAS were supported by the Australian Research Council through the Australian Centre of Excellence for Mathematical and Statistical Frontiers (ACEMS, CE140100049), and SAS through the Discovery Project Scheme (FT170100079).
\fi

\bibliographystyle{chicago}
\bibliography{symbolic}

\begin{thebibliography}{}

\bibitem[\protect\citeauthoryear{Akaike}{Akaike}{1998}]{Akaike1998}
Akaike, H. (1998).
\newblock {Information Theory and an Extension of the Maximum Likelihood
  Principle}.
\newblock In {\em Selected Papers of Hirotugu Akaike}, pp.\  199--213.
  Springer.

\bibitem[\protect\citeauthoryear{Allingham, King, and Mengersen}{Allingham
  et~al.}{2009}]{AllinghamKingMengersen2009}
Allingham, D., R.~King, and K.~L. Mengersen (2009).
\newblock {Bayesian Estimation of Quantile Distributions}.
\newblock {\em Statistics and Computing\/}~{\em 19\/}(2), 189--201.

\bibitem[\protect\citeauthoryear{Andersen and Bollerslev}{Andersen and
  Bollerslev}{1998}]{AndersenBollerslev1998}
Andersen, T.~G. and T.~Bollerslev (1998).
\newblock Answering the skeptics: Yes, standard volatility models do provide
  accurate forecasts.
\newblock {\em International economic review\/}, 885--905.

\bibitem[\protect\citeauthoryear{Andersen, Bollerslev, Diebold, and
  Labys}{Andersen et~al.}{2001}]{AndersenEtAl2001}
Andersen, T.~G., T.~Bollerslev, F.~X. Diebold, and P.~Labys (2001).
\newblock The distribution of realized exchange rate volatility.
\newblock {\em Journal of the American statistical association\/}~{\em
  96\/}(453), 42--55.

\bibitem[\protect\citeauthoryear{Andersen, Bollerslev, Diebold, and
  Labys}{Andersen et~al.}{2003}]{AndersenEtAl2003}
Andersen, T.~G., T.~Bollerslev, F.~X. Diebold, and P.~Labys (2003).
\newblock {Modeling and Forecasting Realized Volatility}.
\newblock {\em Econometrica\/}~{\em 71\/}(2), 579--625.

\bibitem[\protect\citeauthoryear{Arroyo, Gonz{\'a}lez-Rivera, and
  Mat{\'e}}{Arroyo et~al.}{2010}]{ArroyoGonzalezRiveraMate2010}
Arroyo, J., G.~Gonz{\'a}lez-Rivera, and C.~Mat{\'e} (2010).
\newblock Forecasting with interval and histogram data. some financial
  applications.
\newblock {\em Handbook of empirical economics and finance\/}, 247--280.

\bibitem[\protect\citeauthoryear{Arroyo, Gonz{\'a}lez-Rivera, Mat{\'e}, and
  San~Roque}{Arroyo et~al.}{2011}]{ArroyoEtAl2011}
Arroyo, J., G.~Gonz{\'a}lez-Rivera, C.~Mat{\'e}, and A.~M. San~Roque (2011).
\newblock {Smoothing Methods for Histogram-Valued Time Series: An Application
  to Value-at-Risk}.
\newblock {\em Statistical Analysis and Data Mining\/}~{\em 4\/}(2), 216--228.

\bibitem[\protect\citeauthoryear{Arroyo and Mat{\'e}}{Arroyo and
  Mat{\'e}}{2009}]{ArroyoMate2009}
Arroyo, J. and C.~Mat{\'e} (2009).
\newblock Forecasting histogram time series with k-nearest neighbours methods.
\newblock {\em International Journal of Forecasting\/}~{\em 25\/}(1), 192--207.

\bibitem[\protect\citeauthoryear{Barndorff-Nielsen and
  Shephard}{Barndorff-Nielsen and
  Shephard}{2002}]{BarndorffNielsenShephard2002}
Barndorff-Nielsen, O.~E. and N.~Shephard (2002).
\newblock Econometric analysis of realized volatility and its use in estimating
  stochastic volatility models.
\newblock {\em Journal of the Royal Statistical Society: Series B (Statistical
  Methodology)\/}~{\em 64\/}(2), 253--280.

\bibitem[\protect\citeauthoryear{Bauwens and Lubrano}{Bauwens and
  Lubrano}{1998}]{BauwensLubrano1998}
Bauwens, L. and M.~Lubrano (1998).
\newblock {Bayesian Inference on GARCH Models Using the Gibbs Sampler}.
\newblock {\em The Econometrics Journal\/}~{\em 1\/}(1), 23--46.

\bibitem[\protect\citeauthoryear{Beranger, Lin, and Sisson}{Beranger
  et~al.}{2020}]{BerangerLinSisson2020}
Beranger, B., H.~Lin, and S.~A. Sisson (2020).
\newblock New models for symbolic data analysis.
\newblock {\em arXiv preprint arXiv:1809.03659\/}.

\bibitem[\protect\citeauthoryear{Billard}{Billard}{2011}]{Billard2011}
Billard, L. (2011).
\newblock Brief overview of symbolic data and analytic issues.
\newblock {\em Statistical Analysis and Data Mining: The ASA Data Science
  Journal\/}~{\em 4\/}(2), 149--156.

\bibitem[\protect\citeauthoryear{Billard and Diday}{Billard and
  Diday}{2003}]{BillardDiday2003}
Billard, L. and E.~Diday (2003).
\newblock From the statistics of data to the statistics of knowledge: symbolic
  data analysis.
\newblock {\em Journal of the American Statistical Association\/}~{\em
  98\/}(462), 470--487.

\bibitem[\protect\citeauthoryear{Bosq}{Bosq}{2015}]{Bosq2015}
Bosq, D. (2015).
\newblock {Models Associated with Extended Exponential Smoothing}.
\newblock {\em Communications in Statistics-Theory and Methods\/}~{\em
  44\/}(3), 468--475.

\bibitem[\protect\citeauthoryear{Brehmer}{Brehmer}{2017}]{Brehmer2017}
Brehmer, J. (2017).
\newblock {Elicitability and its Application in Risk Management}.

\bibitem[\protect\citeauthoryear{Briol, Oates, Cockayne, Chen, and
  Girolami}{Briol et~al.}{2017}]{BriolEtAl2017}
Briol, F.-X., C.~J. Oates, J.~Cockayne, W.~Y. Chen, and M.~Girolami (2017).
\newblock On the sampling problem for kernel quadrature.
\newblock In {\em International Conference on Machine Learning}, pp.\
  586--595. PMLR.

\bibitem[\protect\citeauthoryear{Brito and Duarte~Silva}{Brito and
  Duarte~Silva}{2012}]{BritoDuarteSilva2012}
Brito, P. and A.~P. Duarte~Silva (2012).
\newblock Modelling interval data with normal and skew-normal distributions.
\newblock {\em Journal of Applied Statistics\/}~{\em 39\/}(1), 3--20.

\bibitem[\protect\citeauthoryear{Brownlees and Gallo}{Brownlees and
  Gallo}{2006}]{BrownleesGallo2006}
Brownlees, C.~T. and G.~M. Gallo (2006).
\newblock {Financial Econometric Analysis at Ultra-High Frequency: Data
  Handling Concerns}.
\newblock {\em Computational Statistics \& Data Analysis\/}~{\em 51\/}(4),
  2232--2245.

\bibitem[\protect\citeauthoryear{Chen and Gerlach}{Chen and
  Gerlach}{2013}]{ChenGerlach2013}
Chen, Q. and R.~H. Gerlach (2013).
\newblock The two-sided weibull distribution and forecasting financial tail
  risk.
\newblock {\em International Journal of Forecasting\/}~{\em 29\/}(4), 527--540.

\bibitem[\protect\citeauthoryear{Clements, Galv{\~a}o, and Kim}{Clements
  et~al.}{2008}]{ClementsGalvaoKim2008}
Clements, M.~P., A.~B. Galv{\~a}o, and J.~H. Kim (2008).
\newblock Quantile forecasts of daily exchange rate returns from forecasts of
  realized volatility.
\newblock {\em Journal of Empirical Finance\/}~{\em 15\/}(4), 729--750.

\bibitem[\protect\citeauthoryear{Corsi}{Corsi}{2009}]{Corsi2009}
Corsi, F. (2009).
\newblock A simple approximate long-memory model of realized volatility.
\newblock {\em Journal of Financial Econometrics\/}~{\em 7\/}(2), 174--196.

\bibitem[\protect\citeauthoryear{Cuevas}{Cuevas}{2014}]{Cuevas2014}
Cuevas, A. (2014).
\newblock {A Partial Overview of the Theory of Statistics with Functional
  Data}.
\newblock {\em Journal of Statistical Planning and Inference\/}~{\em 147\/}(0),
  1--23.

\bibitem[\protect\citeauthoryear{Delaigle and Hall}{Delaigle and
  Hall}{2010}]{DelaigleHall2010}
Delaigle, A. and P.~Hall (2010).
\newblock {Defining Probability Density for a Distribution of Random
  Functions}.
\newblock {\em The Annals of Statistics\/}, 1171--1193.

\bibitem[\protect\citeauthoryear{Demarta and McNeil}{Demarta and
  McNeil}{2005}]{DemartaMcNeil2005}
Demarta, S. and A.~J. McNeil (2005).
\newblock {The t Copula and Related Copulas}.
\newblock {\em International Statistical Review\/}~{\em 73\/}(1), 111--129.

\bibitem[\protect\citeauthoryear{Di~Matteo}{Di~Matteo}{2007}]{DiMatteo2007}
Di~Matteo, T. (2007).
\newblock Multi-scaling in finance.
\newblock {\em Quantitative finance\/}~{\em 7\/}(1), 21--36.

\bibitem[\protect\citeauthoryear{Dias and Brito}{Dias and
  Brito}{2015}]{DiasBrito2015}
Dias, S. and P.~Brito (2015).
\newblock Linear regression model with histogram-valued variables.
\newblock {\em Statistical Analysis and Data Mining: The ASA Data Science
  Journal\/}~{\em 8\/}(2), 75--113.

\bibitem[\protect\citeauthoryear{Diebold}{Diebold}{2015}]{Diebold2015}
Diebold, F.~X. (2015).
\newblock Comparing predictive accuracy, twenty years later: A personal
  perspective on the use and abuse of diebold--mariano tests.
\newblock {\em Journal of Business \& Economic Statistics\/}~{\em 33\/}(1),
  1--1.

\bibitem[\protect\citeauthoryear{Diebold and Mariano}{Diebold and
  Mariano}{1995}]{DieboldMariano1995}
Diebold, F.~X. and R.~S. Mariano (1995).
\newblock Comparing predictive accuracy.
\newblock {\em Journal of Business \& Economic Statistics\/}~{\em 13\/}(3).

\bibitem[\protect\citeauthoryear{Dutta and Perry}{Dutta and
  Perry}{2006}]{DuttaPerry2006}
Dutta, K. and J.~Perry (2006).
\newblock A tale of tails: an empirical analysis of loss distribution models
  for estimating operational risk capital.
\newblock Technical Report 06-13, Federal Reserve Bank of Boston.

\bibitem[\protect\citeauthoryear{Engle and Manganelli}{Engle and
  Manganelli}{2004}]{EngleManganelli2004}
Engle, R.~F. and S.~Manganelli (2004).
\newblock {CAViaR: Conditional Autoregressive Value at Risk by Regression
  Quantiles}.
\newblock {\em Journal of Business \& Economic Statistics\/}~{\em 22\/}(4),
  367--381.

\bibitem[\protect\citeauthoryear{Gelman, Roberts, and Gilks}{Gelman
  et~al.}{1996}]{GelmanRobertsGilks1996}
Gelman, A., G.~Roberts, and W.~Gilks (1996).
\newblock {Efficient Metropolis Jumping Rules}.
\newblock {\em Bayesian statistics\/}~{\em 5\/}(599-608), 42.

\bibitem[\protect\citeauthoryear{Gerlach and Wang}{Gerlach and
  Wang}{2020}]{GerlachWang2020}
Gerlach, R. and C.~Wang (2020).
\newblock Bayesian semi-parametric realized conditional autoregressive
  expectile models for tail risk forecasting.
\newblock {\em Journal of Financial Econometrics\/}, 1--34.

\bibitem[\protect\citeauthoryear{Gerlach, Chen, and Chan}{Gerlach
  et~al.}{2011}]{GerlachChenChan2011}
Gerlach, R.~H., C.~W. Chen, and N.~Y. Chan (2011).
\newblock {Bayesian Time-Varying Quantile Forecasting for Value-at-Risk in
  Financial Markets}.
\newblock {\em Journal of Business \& Economic Statistics\/}~{\em 29\/}(4),
  481--492.

\bibitem[\protect\citeauthoryear{Ghysels, Santa-Clara, and Valkanov}{Ghysels
  et~al.}{2006}]{GhyselsSantaClaraValkanov2006}
Ghysels, E., P.~Santa-Clara, and R.~Valkanov (2006).
\newblock Predicting volatility: getting the most out of return data sampled at
  different frequencies.
\newblock {\em Journal of Econometrics\/}~{\em 131\/}(1-2), 59--95.

\bibitem[\protect\citeauthoryear{Giot and Laurent}{Giot and
  Laurent}{2004}]{GiotLaurent2004}
Giot, P. and S.~Laurent (2004).
\newblock Modelling daily value-at-risk using realized volatility and arch type
  models.
\newblock {\em Journal of empirical finance\/}~{\em 11\/}(3), 379--398.

\bibitem[\protect\citeauthoryear{Glosten, Jagannathan, and Runkle}{Glosten
  et~al.}{1993}]{GlostenJagannathanRunkle1993}
Glosten, L.~R., R.~Jagannathan, and D.~E. Runkle (1993).
\newblock {On the Relation Between the Expected Value and the Volatility of the
  Nominal Excess Return on Stocks}.
\newblock {\em The Journal of Finance\/}~{\em 48\/}(5), 1779--1801.

\bibitem[\protect\citeauthoryear{Gneiting}{Gneiting}{2011}]{Gneiting2011}
Gneiting, T. (2011).
\newblock {Making and Evaluating Point Forecasts}.
\newblock {\em Journal of the American Statistical Association\/}~{\em
  106\/}(494), 746--762.

\bibitem[\protect\citeauthoryear{Gonz{\'a}lez-Rivera and
  Arroyo}{Gonz{\'a}lez-Rivera and Arroyo}{2012}]{GonzalezRiveraArroyo2012}
Gonz{\'a}lez-Rivera, G. and J.~Arroyo (2012).
\newblock {Time Series Modeling of Histogram-Valued Data: The Daily Histogram
  Time Series of S\&P500 Intradaily Returns}.
\newblock {\em International Journal of Forecasting\/}~{\em 28\/}(1), 20--33.

\bibitem[\protect\citeauthoryear{Greenwood, Landwehr, Matalas, and
  Wallis}{Greenwood et~al.}{1979}]{GreenwoodEtAl1979}
Greenwood, J.~A., J.~M. Landwehr, N.~C. Matalas, and J.~R. Wallis (1979).
\newblock {Probability Weighted Moments: Definition and Relation to Parameters
  of Several Distributions Expressable in Inverse Form}.
\newblock {\em Water Resources Research\/}~{\em 15\/}(5), 1049--1054.

\bibitem[\protect\citeauthoryear{Hallam and Olmo}{Hallam and
  Olmo}{2014a}]{HallamOlmo2014b}
Hallam, M. and J.~Olmo (2014a).
\newblock Forecasting daily return densities from intraday data: A multifractal
  approach.
\newblock {\em International Journal of Forecasting\/}~{\em 30\/}(4), 863--881.

\bibitem[\protect\citeauthoryear{Hallam and Olmo}{Hallam and
  Olmo}{2014b}]{HallamOlmo2014a}
Hallam, M. and J.~Olmo (2014b).
\newblock Semiparametric density forecasts of daily financial returns from
  intraday data.
\newblock {\em Journal of Financial Econometrics\/}~{\em 12\/}(2), 408--432.

\bibitem[\protect\citeauthoryear{Hansen}{Hansen}{1994}]{Hansen1994}
Hansen, B.~E. (1994, Aug).
\newblock {Autoregressive Conditional Density Estimation}.
\newblock {\em International Economic Review\/}~{\em 35\/}(3), 705--730.

\bibitem[\protect\citeauthoryear{Hansen, Huang, and Shek}{Hansen
  et~al.}{2012}]{HansenHuangShek2012}
Hansen, P.~R., Z.~Huang, and H.~H. Shek (2012).
\newblock {Realized GARCH: A Joint Model for Returns and Realized Measures of
  Volatility}.
\newblock {\em Journal of Applied Econometrics\/}~{\em 27\/}(6), 877--906.

\bibitem[\protect\citeauthoryear{Harvey and Trimbur}{Harvey and
  Trimbur}{2003}]{HarveyTrimbur2003}
Harvey, A. and T.~Trimbur (2003).
\newblock {Trend Estimation, Signal-Noise Ratios and the Frequency of
  Observations}.
\newblock In {\em Proceedings of the 4th Colloquium on Modern Tools for
  Business Cycle Analysis, EUROSTAT}.

\bibitem[\protect\citeauthoryear{Harvey}{Harvey}{1993}]{Harvey1993}
Harvey, A.~C. (1993).
\newblock {\em {Time Series Models}}.
\newblock MIT Press.

\bibitem[\protect\citeauthoryear{Haynes and Mengersen}{Haynes and
  Mengersen}{2005}]{HaynesMengersen2005}
Haynes, M. and K.~Mengersen (2005).
\newblock {Bayesian Estimation of g-and-k Distributions using MCMC}.
\newblock {\em Computational Statistics\/}~{\em 20\/}(1), 7--30.

\bibitem[\protect\citeauthoryear{Headrick, Kowalchuk, and Sheng}{Headrick
  et~al.}{2008}]{HeadrickKowalchukSheng2008}
Headrick, T.~C., R.~K. Kowalchuk, and Y.~Sheng (2008).
\newblock {Parametric Probability Densities and Distribution Functions for
  Tukey g-and-h Transformations and Their Use for Fitting Data}.
\newblock {\em Applied Mathematical Sciences\/}~{\em 2\/}(9), 449--462.

\bibitem[\protect\citeauthoryear{Herrholz}{Herrholz}{2010}]{Herrholz2010}
Herrholz, E. (2010).
\newblock {\em Parsimonious Histograms}.
\newblock Ph.\ D. thesis, {Mathematisch-Naturwissenschaftlichen Fakult\"{a}t,
  Ernst-Moritz-Arndt-Universit\"{a}t Greifswald}.

\bibitem[\protect\citeauthoryear{Hoaglin}{Hoaglin}{1985}]{Hoaglin1985}
Hoaglin, D.~C. (1985).
\newblock {\em {Summarizing Shape Numerically: The g-and-h Distributions}}.
\newblock Wiley Online Library.

\bibitem[\protect\citeauthoryear{Hosking}{Hosking}{1990}]{Hosking1990}
Hosking, J.~R. (1990).
\newblock {L-moments: Analysis and Estimation of Distributions using Linear
  Combinations of Order Statistics}.
\newblock {\em Journal of the Royal Statistical Society. Series B
  (Methodological)\/}~{\em 52\/}(1), 105--124.

\bibitem[\protect\citeauthoryear{Hossain and Hossain}{Hossain and
  Hossain}{2009}]{HossainHossain2009}
Hossain, M.~A. and S.~S. Hossain (2009).
\newblock {Numerical Maximum Likelihood Estimation for the g-and-k Distribution
  Using Ranked Set Sample}.
\newblock {\em Journal of Statistics\/}~{\em 16\/}(1).

\bibitem[\protect\citeauthoryear{Hron, Brito, and Filzmoser}{Hron
  et~al.}{2017}]{HronBritoFilzmoser2017}
Hron, K., P.~Brito, and P.~Filzmoser (2017).
\newblock Exploratory data analysis for interval compositional data.
\newblock {\em Advances in Data Analysis and Classification\/}~{\em 11\/}(2),
  223--241.

\bibitem[\protect\citeauthoryear{Jondeau and Rockinger}{Jondeau and
  Rockinger}{2003}]{JondeauRockinger2003}
Jondeau, E. and M.~Rockinger (2003).
\newblock {Conditional Volatility, Skewness, and Kurtosis: Existence,
  Persistence, and Comovements}.
\newblock {\em Journal of Economic Dynamics and Control\/}~{\em 27\/}(10),
  1699--1737.

\bibitem[\protect\citeauthoryear{Koenker and Bassett}{Koenker and
  Bassett}{1978}]{KoenkerBassett1978}
Koenker, R. and G.~Bassett (1978).
\newblock {Regression Quantiles}.
\newblock {\em Econometrica\/}~{\em 46\/}(1), 33--50.

\bibitem[\protect\citeauthoryear{Kou and Peng}{Kou and
  Peng}{2016}]{KouPeng2016}
Kou, S. and X.~Peng (2016, September--October).
\newblock {On the Measurement of Economic Tail Risk}.
\newblock {\em Operations Research\/}~{\em 64\/}(5), 1056--1072.

\bibitem[\protect\citeauthoryear{Kupiec}{Kupiec}{1995}]{Kupiec1995}
Kupiec, P. (1995).
\newblock Techniques for verifying the accuracy of risk measurement models.
\newblock {\em The J. of Derivatives\/}~{\em 3\/}(2).

\bibitem[\protect\citeauthoryear{Le-Rademacher and Billard}{Le-Rademacher and
  Billard}{2011}]{LeRademacherBillard2011}
Le-Rademacher, J. and L.~Billard (2011).
\newblock {Likelihood Functions and Some Maximum Likelihood Estimators for
  Symbolic Data}.
\newblock {\em Journal of Statistical Planning and Inference\/}~{\em 141\/}(4),
  1593--1602.

\bibitem[\protect\citeauthoryear{Li, Munk, Sieling, and Walther}{Li
  et~al.}{2020}]{LiEtAl2020}
Li, H., A.~Munk, H.~Sieling, and G.~Walther (2020).
\newblock The essential histogram.
\newblock {\em Biometrika\/}~{\em 107\/}(2), 347--364.

\bibitem[\protect\citeauthoryear{Maheu and McCurdy}{Maheu and
  McCurdy}{2011}]{MaheuMcCurdy2011}
Maheu, J.~M. and T.~H. McCurdy (2011).
\newblock Do high-frequency measures of volatility improve forecasts of return
  distributions?
\newblock {\em Journal of Econometrics\/}~{\em 160\/}(1), 69--76.

\bibitem[\protect\citeauthoryear{Martens, Van~Dijk, and De~Pooter}{Martens
  et~al.}{2004}]{MartensVanDijkDePooter2004}
Martens, M., D.~Van~Dijk, and M.~De~Pooter (2004).
\newblock Modeling and forecasting s\&p 500 volatility: Long memory, structural
  breaks and nonlinearity.
\newblock Technical report, Tinbergen Institute discussion paper.

\bibitem[\protect\citeauthoryear{Martinez and Iglewicz}{Martinez and
  Iglewicz}{1984}]{MartinezIglewicz1984}
Martinez, J. and B.~Iglewicz (1984).
\newblock {Some Properties of the Tukey g and h Family of Distributions}.
\newblock {\em Communications in Statistics-Theory and Methods\/}~{\em
  13\/}(3), 353--369.

\bibitem[\protect\citeauthoryear{McDonald and Michelfelder}{McDonald and
  Michelfelder}{2016}]{McDonaldMichelfelder2016}
McDonald, J. and R.~Michelfelder (2016).
\newblock Partially adaptive and robust estimation of asset models:
  accommodating skewness and kurtosis in returns.
\newblock {\em Journal of Mathematical Finance\/}~{\em 7\/}(1), 219--237.

\bibitem[\protect\citeauthoryear{Meddahi}{Meddahi}{2002}]{Meddahi2002}
Meddahi, N. (2002).
\newblock A theoretical comparison between integrated and realized volatility.
\newblock {\em Journal of Applied Econometrics\/}~{\em 17\/}(5), 479--508.

\bibitem[\protect\citeauthoryear{Nolde and Ziegel}{Nolde and
  Ziegel}{2017}]{NoldeZiegel2017}
Nolde, N. and J.~Ziegel (2017).
\newblock {Elicitability and Backtesting: Perspectives for Banking Regulation}.
\newblock {\em Annals of Applied Statistics\/}.
\newblock Forthcoming.

\bibitem[\protect\citeauthoryear{Perreault, Bob{\'e}e, and Rasmussen}{Perreault
  et~al.}{1999a}]{PerreaultBobeeRasmussen1999}
Perreault, L., B.~Bob{\'e}e, and P.~Rasmussen (1999a).
\newblock {Halphen Distribution System. I: Mathematical and Statistical
  Properties}.
\newblock {\em Journal of Hydrologic Engineering\/}~{\em 4\/}(3), 189--199.

\bibitem[\protect\citeauthoryear{Perreault, Bob{\'e}e, and Rasmussen}{Perreault
  et~al.}{1999b}]{PerreaultBobeeRasmussen1999b}
Perreault, L., B.~Bob{\'e}e, and P.~Rasmussen (1999b).
\newblock {Halphen Distribution System. II: Parameter and Quantile Estimation}.
\newblock {\em Journal of Hydrologic Engineering\/}~{\em 4\/}(3), 200--208.

\bibitem[\protect\citeauthoryear{Peters and Sisson}{Peters and
  Sisson}{2006}]{PetersSisson2006}
Peters, G. and S.~Sisson (2006).
\newblock {Bayesian Inference, Monte Carlo Sampling and Operational Risk}.
\newblock {\em Journal of Operational Risk\/}~{\em 1\/}(3), 27--50.

\bibitem[\protect\citeauthoryear{Peters, Chen, and Gerlach}{Peters
  et~al.}{2016}]{PetersChenGerlach2016}
Peters, G.~W., W.~Y. Chen, and R.~H. Gerlach (2016).
\newblock {Estimating Quantile Families of Loss Distributions for Non-Life
  Insurance Modelling via L-Moments}.
\newblock {\em Risks\/}~{\em 4\/}(2), 14.

\bibitem[\protect\citeauthoryear{Rayner and MacGillivray}{Rayner and
  MacGillivray}{2002}]{RaynerMacGillivray2002}
Rayner, G. and H.~MacGillivray (2002).
\newblock {Numerical Maximum Likelihood Estimation for the g-and-k and
  Generalized g-and-h Distributions}.
\newblock {\em Statistics and Computing\/}~{\em 12\/}(1), 57--75.

\bibitem[\protect\citeauthoryear{Roberts and Rosenthal}{Roberts and
  Rosenthal}{2001}]{RobertsRosenthal2001}
Roberts, G.~O. and J.~S. Rosenthal (2001).
\newblock {Optimal Scaling for Various Metropolis-Hastings Algorithms}.
\newblock {\em Statistical Science\/}~{\em 16\/}(4), 351--367.

\bibitem[\protect\citeauthoryear{Spiegelhalter, Best, Carlin, and Van
  Der~Linde}{Spiegelhalter et~al.}{2002}]{SpiegelhalterEtAl2002}
Spiegelhalter, D.~J., N.~G. Best, B.~P. Carlin, and A.~Van Der~Linde (2002).
\newblock {Bayesian Measures of Model Complexity and Fit}.
\newblock {\em Journal of the Royal Statistical Society: Series B (Statistical
  Methodology)\/}~{\em 64\/}(4), 583--639.

\bibitem[\protect\citeauthoryear{Taylor}{Taylor}{2011}]{Taylor2011}
Taylor, S.~J. (2011).
\newblock {\em {Asset Price Dynamics, Volatility, and Prediction}}.
\newblock Princeton University Press.

\bibitem[\protect\citeauthoryear{Tepper and Sapiro}{Tepper and
  Sapiro}{2012}]{TepperSapiro2012}
Tepper, M. and G.~Sapiro (2012).
\newblock {L1 Splines for Robust, Simple, and Fast Smoothing of Grid Data}.

\bibitem[\protect\citeauthoryear{Tepper and Sapiro}{Tepper and
  Sapiro}{2013}]{TepperSapiro2013}
Tepper, M. and G.~Sapiro (2013).
\newblock {Fast L1 Smoothing Splines with an Application to Kinect Depth Data}.
\newblock In {\em Image Processing (ICIP), 2013 20th IEEE International
  Conference on}, pp.\  504--508. IEEE.

\bibitem[\protect\citeauthoryear{Tsay}{Tsay}{2010}]{Tsay2010}
Tsay, R.~S. (2010).
\newblock {\em {Analysis of Financial Time Series}}.
\newblock John Wiley \& Sons.

\bibitem[\protect\citeauthoryear{Tukey}{Tukey}{1977}]{Tukey1977}
Tukey, J.~W. (1977).
\newblock {Modern Techniques in Data Analysis}.
\newblock In {\em Proceedings of the NSF-Sponsored Regional Research
  Conference}.

\bibitem[\protect\citeauthoryear{Xu, Iglewicz, and Chervoneva}{Xu
  et~al.}{2014}]{XuIglewiczChervoneva2014}
Xu, Y., B.~Iglewicz, and I.~Chervoneva (2014).
\newblock {Robust Estimation of the Parameters of g-and-h Distributions, with
  Applications to Outlier Detection}.
\newblock {\em Computational Statistics \& Data Analysis\/}~{\em 75\/}(0), 66
  -- 80.

\bibitem[\protect\citeauthoryear{Yu and Moyeed}{Yu and
  Moyeed}{2001}]{YuMoyeed2001}
Yu, K. and R.~A. Moyeed (2001).
\newblock {Bayesian Quantile Regression}.
\newblock {\em Statistics \& Probability Letters\/}~{\em 54\/}(4), 437--447.

\bibitem[\protect\citeauthoryear{Zhang and Sisson}{Zhang and
  Sisson}{2017}]{ZhangSisson2017}
Zhang, X. and S.~A. Sisson (2017).
\newblock {Constructing Likelihood Functions for Interval-Valued Random
  Variables}.

\end{thebibliography}

\appendix
\section{L-moment Method for Estimating $g$-and-$h$ Parameters}
\label{app:lmoment}
In this section, we briefly summarise the L-moment method of \citet{PetersChenGerlach2016}. Since we are concerned with the estimation of a single quantile function for a given time-period, to simplify notation, the subscript $t$ is dropped for the rest of the section whenever clarity is not lost.

L-moments are defined by \citet{Hosking1990} to be certain linear combinations of expectations of order statistics. Specifically, let $y_{(1)} \le y_{(2)} \le \cdots \le y_{(n)}$ denote a sample of ordered observations. For $k \in \{1, 2, \,\ldots\}$, the $k$-th L-moment is defined as
\begin{equation}
\label{eq:lmom_sum}
l_{k} = \frac{1}{k}\sum_{i=0}^{k-1}(-1)^{i}\binom{k-1}{i}\E\left[y_{(k-i)}\right].
\end{equation}
The connection between L-moments and a quantile function becomes apparent when L-moments are expressed as projections of a quantile function onto a sequence of orthogonal polynomials that forms a basis of $L^{2}$;
\begin{equation}
\label{eq:lmom_int}
l_{k} = \int_{0}^{1} X(u)L_{k-1}(u) \, du,
\end{equation}
where $L_{k}$ is the $k$-th shifted Legendre polynomial in the sequence. Compared to classical moments, L-moments are able to characterise a wider range of distributions as all L-moments of a distribution exist if and only if the mean exits. Furthermore, a distribution with finite mean is uniquely characterised by its sequence of L-moments. Using the representation of \eqref{eq:lmom_int}, the first four L-moments are given by
\begin{equation}
\label{eq:lmom}
\begin{aligned}
l_{1} &= \int_{0}^{1} X(u) \, du, \\
l_{2} &= \int_{0}^{1} X(u)(2u - 1) \, du, \\
l_{3} &= \int_{0}^{1} X(u)(6u^{2} - 6u + 1) \, du, \\
l_{4} &= \int_{0}^{1} X(u)(20u^{3} - 30u^{2} + 12u - 1) \, du.
\end{aligned}
\end{equation}
The location and scale invariant L-moment ratios, $\tau_{3}$ and $\tau_{4}$, analogous to the classical skewness and kurtosis, respectively termed L-skewness and L-kurtosis in \citet{Hosking1990}, are defined as
\begin{equation}
\begin{aligned}
\tau_{3} &= l_{3} / l_{2}, \\
\tau_{4} &= l_{4} / l_{2}.
\end{aligned}
\end{equation}
Unlike the classical skewness and kurtosis, L-skewness and L-kurtosis are bounded, with $\tau_{3} \in (-1, 1)$ and $\tau_{4} \in [\frac{1}{4}(5 \tau_{3}^{2} - 1), 1)$. The boundedness of L-moment ratios makes them easy to interpret.

The sample L-moments, also known as L-statistics, are unbiased estimates of L-moments based on the order statistics of an observed sample. In particular, the first four sample L-moments are given by
\begin{equation}
\label{eq:lstat}
\begin{aligned}
\hat{l}_{1} &= \hat{M}_{0}, \\
\hat{l}_{2} &= 2\hat{M}_{1} - \hat{M}_{0}, \\
\hat{l}_{3} &= 6\hat{M}_{2} - 6\hat{M}_{1} + \hat{M}_{0}, \\
\hat{l}_{4} &= 20\hat{M}_{3} - 30\hat{M}_{2} + 12\hat{M}_{1} - \hat{M}_{0},
\end{aligned}
\end{equation}
where $\hat{M}_{k}$ is the $k$-th sample probability weighted moment \citep{GreenwoodEtAl1979}, given by
\begin{equation}
\label{eq:pwm}
\hat{M}_{k} =
\begin{dcases}
\frac{1}{n}\sum_{i=1}^{n}y_{(i)} & \text{if $k = 0$} \\
\frac{1}{n}\sum_{i=1}^{n}\frac{(i - 1)(i - 2) \cdots (i - k)}{(n - 1)(n - 2) \cdots (n - k)} y_{(i)} & \text{if $k > 0$}.
\end{dcases}
\end{equation}

The estimates of $g$ and $h$ are simultaneously found by iteratively minimising the objective
\begin{equation}
\label{eq:mlmom_gnh}
(\tau_{3} - \hat{\tau}_{3})^{2} + (\tau_{4} - \hat{\tau}_{4})^{2},
\end{equation}
subject to $0 \le h < 1$, where $\hat{\tau}_{3} = \hat{l}_{3} / \hat{l}_{2}$ is the sample L-skewness and $\hat{\tau}_{4} = \hat{l}_{4} / \hat{l}_{2}$ is the sample L-kurtosis. The integrals in \eqref{eq:lmom} are available in closed-form for the $g$-and-$h$ quantile function \citep{PetersChenGerlach2016}, or they can be obtained numerically using one-dimensional adaptive quadrature. Given the estimates of $g$ and $h$, the estimates of $b$ and $a$ are given by
\begin{equation}
\label{eq:mlmom_bna}
\begin{aligned}
b &= \hat{l}_{2} / l_{2}, \\
a &= \hat{l}_{1} - bl_{1}.
\end{aligned}
\end{equation}

\section{Additional Material on Apatosaurus Distributions}
\label{app:apatosaurus}
\paragraph{Density:}
\label{app:apat_pdf}
The random variable $h \sim F_{\mathrm{Apat}}(h; \mu, \sigma, \eta, \lambda, \iota, w)$ has a density function given by
\begin{equation}
\label{eq:apat_pdf}
f_{\mathrm{Apat}}(h; \mu, \sigma, \eta, \lambda, \iota, w) = wf_{\mathrm{TrSkt}}(h; \mu,\sigma,\eta,\lambda) + (1 - w)f_{\mathrm{Exp}}(h; \iota)
\end{equation}
for $h \in [0, \infty)$, where $f_{\mathrm{TrSkt}}$ and $f_{\mathrm{Exp}}$ are the density functions of a truncated-skewed-$t$ distribution and an Exponential distribution, and $w \in [0,1]$ is the mixing weight.

The truncated-skewed-$t$ distribution has the following density function.
\begin{equation}
\label{eq:trskt_pdf}
f_{\mathrm{TrSkt}}(h; \mu,\sigma,\eta,\lambda) = \frac{f_{\mathrm{Skt}}(h; \mu,\sigma,\eta,\lambda)}{1-F_{\mathrm{Skt}}(0; \mu,\sigma,\eta,\lambda)},
\end{equation}
where $f_{\mathrm{Skt}}$ and $F_{\mathrm{Skt}}$ are the density and distribution functions of the skewed-$t$ distribution of \citet{Hansen1994}, parameterised by its mode $\mu$ and scale $\sigma$ in addition to the asymmetry and degrees-of-freedom parameters, $\eta$ and $\lambda$. The density function of skewed-$t$ distribution $f_{\mathrm{Skt}}$ is given by
\begin{equation}
\label{eq:skt_pdf}
f_{\mathrm{Skt}}(h; \mu, \sigma, \eta, \lambda) =
\begin{dcases}
\frac{\varphi}{\sigma}\left[1 + \frac{1}{\eta - 2}\left(\frac{h - \mu}{\sigma(1 - \lambda)}\right)^{2}\right]^{-(\eta + 1) / 2} & \text{if $h < \mu$}, \\
\frac{\varphi}{\sigma}\left[1 + \frac{1}{\eta - 2}\left(\frac{h - \mu}{\sigma(1 + \lambda)}\right)^{2}\right]^{-(\eta + 1) / 2} & \text{if $h \ge \mu$},
\end{dcases}
\end{equation}
where $\sigma \in (0, \infty)$, $\eta \in (2, \infty)$, $\lambda \in (-1, 1)$, and
\begin{equation}
\label{eq:skt_varphi}
\varphi = \frac{\Gamma\left(\frac{\eta + 1}{2}\right)}{\sqrt{\pi(\eta - 2)}\Gamma\left(\frac{\eta}{2}\right)}.
\end{equation}
The distribution function $F_{\mathrm{Skt}}$ can be derived in a similar manner to Proposition 1 of \citet{JondeauRockinger2003}, and is given by
\begin{equation}
\label{eq:skt_cdf}
F_{\mathrm{Skt}}(h; \mu, \sigma, \eta, \lambda) =
\begin{dcases}
(1 - \lambda)F_{\mathrm{t},\eta}\left(\frac{h - \mu}{\sigma(1 - \lambda)}\sqrt{\frac{\eta}{\eta - 2}}\right) & \text{if $h < \mu$}, \\
(1 + \lambda)F_{\mathrm{t},\eta}\left(\frac{h - \mu}{\sigma(1 + \lambda)}\sqrt{\frac{\eta}{\eta - 2}}\right) - \lambda & \text{if $h \ge \mu$},
\end{dcases}
\end{equation}
where $F_{\mathrm{t},\eta}$ is the distribution function of the $t$-distribution with $\eta$ degrees-of-freedom.

The density function of the Exponential distribution in equation~\eqref{eq:apat_pdf} is given by
\begin{equation}
\label{eq:exp_pdf}
f_{\mathrm{Exp}}(h; \iota) = \frac{1}{\iota}\exp\left(\frac{-h}{\iota}\right),
\end{equation}
where $\iota \in (0, \infty)$ is the mean parameter.

\paragraph{Mean:}
\label{app:apat_mean}
The mean of the truncated-skewed-$t$ distribution can be derived by 
%noticing the following equivalence.
noting that
\begin{equation}
\label{eq:trskt_mean_id}
\begin{aligned}
m_{\mathrm{TrSkt}} &= \frac{1}{1 - F_{\mathrm{Skt}}(0; \mu, \sigma, \eta, \lambda)}\int_{0}^{\infty}hf_{\mathrm{Skt}}(h; \mu, \sigma, \eta, \lambda)\,dh \\
&= \frac{1}{1 - F_{\mathrm{Skt}}(-\mu; 0, \sigma, \eta, \lambda)}\int_{-\mu}^{\infty}hf_{\mathrm{Skt}}(h; 0, \sigma, \eta, \lambda)\,dh + \mu.
\end{aligned}
\end{equation}
The integral can then be written as
\begin{equation}
\label{eq:int_sum}
\int_{-\mu}^{\infty}hf_{\mathrm{Skt}}(h; 0, \sigma, \eta, \lambda)\,dh = \int_{-\mu}^{0}hf_{\mathrm{Skt}}(h; 0, \sigma, \eta, \lambda)\,dh + \int_{0}^{\infty}hf_{\mathrm{Skt}}(h; 0, \sigma, \eta, \lambda)\,dh.
\end{equation}
Assuming that $\mu \in [0, \infty)$ and using the substitution
\begin{equation}
\label{eq:sub_left}
u(h) = 1 + \frac{1}{\eta - 2}\left(\frac{h}{\sigma(1 - \lambda)}\right)^{2},
\end{equation}
the first integral is given by
\begin{equation}
\label{eq:int_left}
\begin{aligned}
\int_{-\mu}^{0}hf_{\mathrm{Skt}}(h; 0, \sigma, \eta, \lambda)\,dh &= \frac{1}{2}\varphi\sigma(1 - \lambda)^{2}(\eta - 2)\int_{u(-\mu)}^{1}u^{-(\eta + 1)/2}\,du \\
&= -\frac{1}{2}\varphi\sigma(1 - \lambda)^{2}(\eta - 2)\int_{1}^{u(-\mu)}u^{-(\eta + 1)/2}\,du \\
&= -\varphi\sigma(1 - \lambda)^{2}\left(\frac{\eta - 2}{\eta - 1}\right)\left[1 - u(-\mu)^{(1 - \eta)/2}\right].
\end{aligned}
\end{equation}
Using the substitution
\begin{equation}
\label{eq:sub_right}
u(h) = 1 + \frac{1}{\eta - 2}\left(\frac{h}{\sigma(1 + \lambda)}\right)^{2},
\end{equation}
the second integral in \eqref{eq:int_sum} is given by
\begin{equation}
\label{eq:int_right}
\begin{aligned}
\int_{0}^{\infty}hf_{\mathrm{Skt}}(h; 0, \sigma, \eta, \lambda)\,dh &= \frac{1}{2}\varphi\sigma(1 + \lambda)^{2}(\eta - 2)\int_{1}^{\infty}u^{-(\eta + 1)/2}\,du \\
&= \varphi\sigma(1 + \lambda)^{2}\left(\frac{\eta - 2}{\eta - 1}\right).
\end{aligned}
\end{equation}
Thus,
\begin{equation}
\label{eq:trskt_mean}
m_{\mathrm{TrSkt}} = \frac{\varphi\sigma\left(\frac{\eta - 2}{\eta - 1}\right)\left\{(1 + \lambda)^{2} - (1 - \lambda)^{2}\left[1 - u(-\mu)^{(1 - \eta)/2}\right]\right\}}{1 - F_{\mathrm{Skt}}(-\mu; 0, \sigma, \eta, \lambda)} + \mu,
\end{equation}
where
\begin{equation}
\label{eq:u_left_mu}
u(-\mu) = 1 + \frac{1}{\eta - 2}\left(\frac{-\mu}{\sigma(1 - \lambda)}\right)^{2}.
\end{equation}
The mean for $\mu \in (-\infty, 0)$ can also be derived using the substitution given by \eqref{eq:sub_right}, however it is natural for most applications to restrict the mode the the \emph{pre-truncated} skewed-$t$ distribution to be non-negative, i.e. $\mu \in [0, \infty)$.

The mean of the Exponential distribution is simply
\begin{equation}
\label{eq:exp_mean}
m_{\mathrm{Exp}} = \iota.
\end{equation}
It follows from the density function in \eqref{eq:apat_pdf} that the mean of the Apatosaurus distribution is given by
\begin{equation}
\label{eq:apat_mean}
m_{\mathrm{Apat}} = wm_{\mathrm{TrSkt}} + (1 - w)m_{\mathrm{Exp}}.
\end{equation}

\paragraph{Distribution function:}
\label{app:apat_cdf}
The distribution function of the Apatosaurus distribution is given by
\begin{equation}
\label{eq:apat_cdf}
F_{\mathrm{Apat}}(h; \mu, \sigma, \eta, \lambda, \iota, w) = wF_{\mathrm{TrSkt}}(h; \mu, \sigma, \eta, \lambda) + (1 - w)F_{\mathrm{Exp}}(h; \iota),
\end{equation}
where $F_{\mathrm{TrSkt}}$ and $F_{\mathrm{Exp}}$ are the distribution functions of a truncated-skewed-$t$ distribution and an Exponential distribution. We can straightforwardly derive the distribution function of the truncated-skewed-$t$ distribution as follows.
\begin{equation}
\label{eq:trskt_cdf}
\begin{aligned}
F_{\mathrm{TrSkt}}(h; \mu, \sigma, \eta, \lambda) &= \int_{0}^{h}f_{\mathrm{TrSkt}}(y; \mu, \sigma, \eta, \lambda)\,dy \\
&= \frac{\int_{0}^{h}f_{\mathrm{Skt}}(y; \mu, \sigma, \eta, \lambda)\,dy}{1 - F_{\mathrm{Skt}}(0; \mu, \sigma, \eta, \lambda)} \\
&= \frac{F_{\mathrm{Skt}}(h; \mu, \sigma, \eta, \lambda) - F_{\mathrm{Skt}}(0; \mu, \sigma, \eta, \lambda)}{1 - F_{\mathrm{Skt}}(0; \mu, \sigma, \eta, \lambda)}.
\end{aligned}
\end{equation}
The distribution function of the Exponential distribution is given by
\begin{equation}
\label{eq:exp_cdf}
F_{\mathrm{Exp}}(h; \iota) = 1 - \exp\left(\frac{-h}{\iota}\right).
\end{equation}

\paragraph{Random number:}
\label{app:apat_rnd}
We can generate a random number $h$ from an Apatosaurus distribution by first generating a component label $l$ from a Bernoulli distribution with parameter $w$, and then generating $(h \vbar l = 1) \sim F_{\mathrm{TrSkt}}(h; \mu, \sigma, \eta, \lambda)$ or $(h \vbar l = 0) \sim F_{\mathrm{Exp}}(h; \iota)$. A truncated-skewed-$t$ random number can be generated using the quantile function technique, by first generating $u$ from a uniform distribution on the interval $(0, 1)$, and then applying the transformation $h = F_{\mathrm{TrSkt}}^{-1}(u; \mu, \sigma, \eta, \lambda)$. We obtain the quantile function of the truncated-skewed-$t$ distribution $F_{\mathrm{TrSkt}}^{-1}$ by inverting the distribution function in \eqref{eq:trskt_cdf}. Thus,
\begin{equation}
\label{eq:trskt_inv}
F_{\mathrm{TrSkt}}^{-1}(u; \mu, \sigma, \eta, \lambda) = F_{\mathrm{Skt}}^{-1}\left(u[1 - F_{\mathrm{Skt}}(0; \mu, \sigma, \eta, \lambda)] + F_{\mathrm{Skt}}(0; \mu, \sigma, \eta, \lambda); \mu, \sigma, \eta, \lambda\right),
\end{equation}
where $F_{\mathrm{Skt}}^{-1}$ is the quantile function of the skewed-$t$ distribution found by inverting the skewed-$t$ distribution function in \eqref{eq:skt_cdf}. Thus,
\begin{equation}
\label{eq:skt_inv}
F_{\mathrm{Skt}}^{-1}(u; \mu, \sigma, \eta, \lambda) =
\begin{dcases}
\sigma(1 - \lambda)\sqrt{\frac{\eta - 2}{\eta}}F_{\mathrm{St}, \eta}^{-1}\left(\frac{u}{1 - \lambda}\right) + \mu & \text{if $u < \frac{1 - \lambda}{2}$}, \\
\sigma(1 + \lambda)\sqrt{\frac{\eta - 2}{\eta}}F_{\mathrm{St}, \eta}^{-1}\left(\frac{u}{1 + \lambda}\right) + \mu & \text{if $u \ge \frac{1 - \lambda}{2}$},
\end{dcases}
\end{equation}
where $F_{\mathrm{St}, \eta}^{-1}$ is the quantile function of the Student $t$ distribution with $\eta$ degrees-of-freedom. An Exponential random number can be generated in a similar fashion by transforming an uniform random number on $(0, 1)$ using the Exponential quantile function $F_{\mathrm{Exp}}^{-1}$ given by
\begin{equation}
\label{eq:exp_inv}
F_{\mathrm{Exp}}^{-1}(u; \iota) = -\iota\log(1 - u).
\end{equation}

\section{Parameter Blocking Scheme for Sampling}
\label{app:blocking}
The entire parameter vector $\boldsymbol{\theta} = (\boldsymbol{\theta}_{[1]} \ddd \boldsymbol{\theta}_{[10]})$ is partitioned into ten blocks as follows:
\begin{equation*}
\begin{aligned}
\boldsymbol{\theta}_{[1]} &= (\delta_{1}, \psi_{1}, \phi_{1}), \\
\boldsymbol{\theta}_{[2]} &= (\omega_{1}, \alpha_{1}, \beta_{1}, \eta_{1}, \lambda_{1}), \\
\boldsymbol{\theta}_{[3]} &= (\delta_{2}, \psi_{2}, \phi_{2}), \\
\boldsymbol{\theta}_{[4]} &= (\omega_{2}, \alpha_{2}, \beta_{2}, \eta_{2}, \lambda_{2}), \\
\boldsymbol{\theta}_{[5]} &= (\delta_{3}, \psi_{3}, \phi_{3}), \\
\boldsymbol{\theta}_{[6]} &= (\omega_{3}, \alpha_{3}, \beta_{3}, \eta_{3}, \lambda_{3}), \\
\boldsymbol{\theta}_{[7]} &= (\delta_{4}, \psi_{4}, \phi_{4}, \gamma^{*}, c), \\
\boldsymbol{\theta}_{[8]} &= (\sigma, \eta_{4}, \lambda_{4}, \iota), \\
\boldsymbol{\theta}_{[9]} &= (\mathbf{R}_{2,1}, \mathbf{R}_{3,1}, \mathbf{R}_{4,1}, \mathbf{R}_{3,2}, \mathbf{R}_{4,2}, \mathbf{R}_{4,3}), \\
\boldsymbol{\theta}_{[10]} &= \nu.
\end{aligned}
\end{equation*}

\section{Model Adequacy Against Independent AR(1) Margins}
\label{app:dic}
In Section \ref{sec:xit_dist} of the main text, the conditional distribution of $\boldsymbol{\xi}_t$ is modelled by a Student-$t$ copula and carefully designed marginal models. A simpler alternative is to assume that each $\{\xi_{i,t}\}$ series independently follows a univariate AR(1) process with Gaussian innovations. For $i \in \{1 \ddd 4\}$,
\begin{equation}
\label{eq:xi_ar1}
\begin{aligned}
\xi_{i,t} &= \delta_i + \psi_i \xi_{i, t-1} + \epsilon_{i,t}, \\
\epsilon_{i,t} &\sim F_{\mathrm{Gauss}}(\cdot; 0, \sigma_i^2),
\end{aligned}
\end{equation}
where $ F_{\mathrm{Gauss}}(\cdot; 0, \sigma_i^2)$ denotes the normal distribution with mean zero and variance $\sigma_i^2$. To show that the proposed full model offers a large improvement in model adequacy over the simpler independent AR(1) alternative given by Eq.~\eqref{eq:xi_ar1}, the two models are compared based on the deviance information criterion (DIC) of \citet{SpiegelhalterEtAl2002}.

The DIC can be seen as a Bayesian generalisation of the Akaike information criterion (AIC) \citep{Akaike1998}, and is given by
\begin{equation}
\label{eq:dic}
\mathrm{DIC} = \bar{D} + p_{D},
\end{equation}
where $\bar{D}$ is the posterior mean of deviance -- a Bayesian measure of model fit, and $p_{D}$ is the effective number of parameters -- a measure of model complexity. The model with the smallest DIC achieves the best fit-complexity trade-off. Both terms on the right hand side can be computed straightforwardly using the MCMC output.

In Table~\ref{tab:dic}, the estimates of the DIC, $\bar{D}$, and $p_{D}$ are reported for the two models. For each stock index, the DIC of the proposed full model is substantially lower than the simpler model, with the smallest gap being more than 6,000 (6117 for CAC).

\begin{table}[h]
\small
\hspace{-2.2cm}
\begin{tabular}{clcccccccccc}
\toprule
 &  & SPX & DJIA & Nasdaq & FTSE & DAX & CAC & Nikkei & HSI & SSEC & AORD \\ 
\midrule
\multirow{2}{*}{$\mathrm{DIC}$} & AR1 & -67929.5 & -71713.2 & -66498.2 & -71497.6 & -68253.0 & -75798.2 & -53264.5 & -60590.2 & -54752.3 & -64276.5 \\ 
 & Full & -76063.0 & -78347.9 & -75465.7 & -81530.9 & -75925.8 & -81915.7 & -61672.1 & -68352.4 & -63343.0 & -71600.9 \\ 
\midrule
\multirow{2}{*}{$\bar{D}$} & AR1 & -67941.5 & -71725.2 & -66510.2 & -71509.7 & -68264.9 & -75810.3 & -53276.5 & -60602.2 & -54764.4 & -64288.5 \\ 
 & Full & -76102.0 & -78386.0 & -75503.9 & -81566.1 & -75962.0 & -81949.8 & -61708.5 & -68389.5 & -63380.8 & -71639.1 \\ 
\midrule
\multirow{2}{*}{$p_{D}$} & AR1 & 12.0 & 12.0 & 12.1 & 12.0 & 11.9 & 12.1 & 12.0 & 12.1 & 12.1 & 12.0 \\ 
 & Full & 39.0 & 38.0 & 38.2 & 35.3 & 36.2 & 34.1 & 36.3 & 37.1 & 37.8 & 38.2 \\ 
\bottomrule
\end{tabular}
\caption{Estimates of $\mathrm{DIC}$, $\bar{D}$, and $p_{D}$. The row label ``AR1" corresponds to the simple model with independent AR(1)-Gaussian margins, and ``Full" corresponds to the proposed full model based on the Student-$t$ copula.}
\label{tab:dic}
\end{table}

\section{Additional Material on Signal Ratio}
\label{app:rsig}
As mentioned in Section~\ref{sec:rsig} of the main text, we propose a criteria for measuring the amount of predictable information present in the data conditional on a model. The criteria is similar in spirit to the signal-to-noise ratio \citep{Harvey1993, HarveyTrimbur2003} and is termed the signal ratio. Let $\{\xi_{t}: t \in \mathbb{Z}\}$ be a real-valued covariance stationary process. The signal ratio, denoted by $R_{\mathrm{Sig}}$, is then defined as
\begin{equation}
\label{eq:rsig}
R_{\mathrm{Sig}} = \frac{\Var[\E(\xi_{t} \vbar \mathcal{F}_{t-1})]}{\Var(\xi_{t})},
\end{equation}
where $\mathcal{F}_{t-1} = \sigma(\{\xi_{s}\colon\, s \le t-1\})$ is the natural filtration. As the conditional mean of an observed noisy process can be interpreted as the underlying single of the process, the numerator of \eqref{eq:rsig} represents the long-run (unconditional) variance of the signal; the denominator is the unconditional variance of the data. It can be shown in general by the \emph{law of total variance} that $R_{\mathrm{Sig}} \in [0, 1)$ if $\Var(\xi_{t}) < \infty$. Notice that $R_{\mathrm{Sig}}$ can be computed for a large class of covariance stationary time-series models.

For the rest of the section, we show that the signal ratio is explicitly available for the family of stationary exponential smoothing models, which correspond to the ARMA($1,1$) model. Consider the following exponential smoothing model
\begin{equation}
\label{eq:md}
\begin{aligned}
\xi_{t} &= \mu_{t} + \epsilon_{t}, \\
\mu_{t} &= \delta + \psi \xi_{t-1} + \phi \mu_{t-1},
\end{aligned}
\end{equation}
where $\epsilon_{t}$ is a \emph{martingale difference}, with $\E(\epsilon_{t}) = \E(\epsilon_{t} \vbar \mathcal{F}_{t-1}) = 0$, $\Var(\epsilon_{t} \vbar \mathcal{F}_{t-1}) = \sigma^{2}_{t}$, and $\Var(\epsilon_{t}) < \infty$. Using the fact that $\mu_{t}$ and $\epsilon_{t}$ are uncorrelated, the signal ratio can be written as
\begin{equation}
\label{eq:rsig_md}
R_{\mathrm{Sig}} = \frac{\Var(\mu_{t})}{\Var(\xi_{t})} = \frac{\Var(\xi_{t}) - \Var(\epsilon_{t})}{\Var(\xi_{t})}.
\end{equation}
The unconditional variance of $\xi_{t}$ can be obtained by first rewriting \eqref{eq:md} as an ARMA$(1,1)$ model
\begin{equation}
\label{eq:arma}
\xi_{t} = \delta + (\psi + \phi)\xi_{t-1} - \phi \epsilon_{t-1} + \epsilon_{t}.
\end{equation}
Assuming without loss of generality that $\delta = 0$, and using the assumption that $\{\xi_{t}\}$ is covariance stationary, we can then take the variance of \eqref{eq:arma} to obtain the unconditional variance \citep{Tsay2010},
\begin{equation}
\label{eq:var_xit}
\Var(\xi_{t}) = \frac{1 - 2(\psi + \phi)\phi + \phi^{2}}{1 - (\psi + \phi)^{2}}\Var(\epsilon_{t}).
\end{equation}
By substituting \eqref{eq:var_xit} into \eqref{eq:rsig_md}, the explicit expression of the signal ratio is obtained for the model in \eqref{eq:md}.
\begin{equation}
\label{eq:rsig_md_exp}
R_{\mathrm{Sig}} = \frac{\psi^{2}}{1 - 2\psi\phi - \phi^{2}}.
\end{equation}
Notice that the unconditional variance of $\epsilon_{t}$ is cancelled out in the expression, leaving it as a function of only the parameters $\psi$ and $\phi$.

Let $\gamma = \psi + \phi$. Using the ARMA$(1,1)$ representation in \eqref{eq:arma}, it can be shown that the autocorrelation of $\{\xi_{t}\}$ at lag $k$, denoted by $\rho_{k}$, follows the recursive relationship $\rho_{k} = \gamma\rho_{k-1}$, for $k \ge 2$ \citep{Taylor2011}. It is clear that $\gamma$ controls the rate of decay of autocorrelations. Thus, the parameter $\gamma$ is referred to as the persistence level, or memory, of the time-series. If $|\gamma| < 1$, the process $\{\xi_{t}\}$ is mean stationary. This condition is implied to be true by the covariance stationarity assumption.

\begin{figure}[h]
\begin{center}
\includegraphics[width=0.7\textwidth]{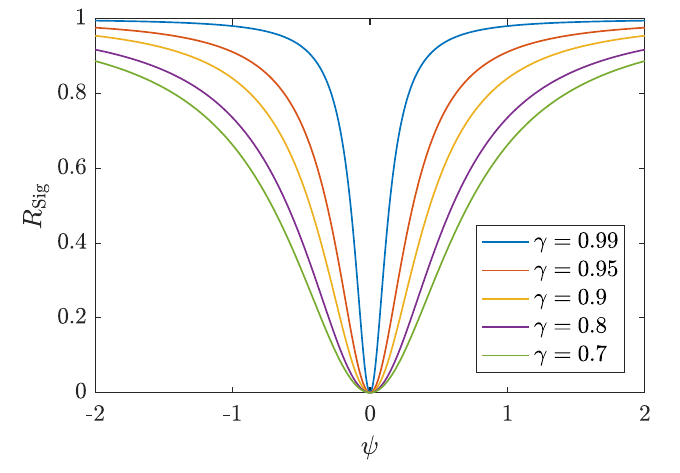}
\end{center}
\caption{\small Plot of $R_{\mathrm{Sig}}$ as a function of $\psi$ for $\psi \in [-2, 2]$ and $\gamma \in \{0.99, 0.95, 0.9, 0.8, 0.7\}$.}
\label{fig:rsig_func}
\end{figure}

By letting $\phi = \gamma - \psi$, we can express the signal ratio in \eqref{eq:rsig_md_exp} in terms of the persistence level $\gamma$,
\begin{equation}
\label{eq:rsig_md_gamma}
R_{\mathrm{Sig}} = \frac{\psi^{2}}{1 - \gamma^{2} + \psi^{2}}.
\end{equation}
It can be seen that
\begin{equation}
\label{eq:rsig_lim}
\lim_{\psi \to \pm\infty} \frac{\psi^{2}}{1 - \gamma^{2} + \psi^{2}} = 1.
\end{equation}
That is, conditional on a persistence level $\gamma$, with $|\gamma| < 1$, the signal ratio converges to 1 as $\psi$ moves away from 0. From \eqref{eq:rsig_md_gamma}, it is also apparent that $\gamma$ controls the rate at which the signal ratio converges to 1. Specifically, when viewed as a function of $\psi$, the larger the value of $\gamma$, the more quickly the signal ratio converges to 1 as $|\psi| \to \infty$. Figure~\ref{fig:rsig_func} shows a plot of the signal ratio in \eqref{eq:rsig_md_gamma} as a function of $\psi$ for various levels of persistence.

\FloatBarrier

The process $\{\xi_{t}\}$ is said to be \emph{invertible} if it admits the AR$(\infty)$ representation,
\begin{equation}
\label{eq:ar_inf}
\xi_{t} = \frac{\delta}{1 - \phi} + \psi\sum_{i=0}^{\infty}\phi^{i}\xi_{t-1-i} + \epsilon_{t}.
\end{equation}
By expanding the recursive definition in $\eqref{eq:md}$, it can be shown that $\{\xi_{t}\}$ is invertible if $|\phi| < 1$. By letting $\psi = \gamma - \phi$ in \eqref{eq:rsig_md_gamma}, and considering the cases $\phi = -1$ and $\phi = 1$, the \emph{upper bound} on $R_{\mathrm{Sig}}$ for which $\{\xi_{t}\}$ is invertible can be expressed as a piece-wise linear function of $\gamma$. Specifically, if $\gamma$ is fixed and $|\phi| < 1$, then
\begin{equation}
\label{eq:rsig_md_bound}
\sup \{R_{\mathrm{Sig}}\} = \frac{|\gamma| + 1}{2}.
\end{equation}

\section{Session Times}
\label{app:session_times}
\begin{table}[h]
\small
\centering
\begin{tabular}{lcccc}
\toprule
 & From & To & Start & End \\ 
\midrule
SPX & 1996-01-03 & 2016-05-24 & 09:30 & 16:00 \\ 
DJIA & 1996-01-03 & 2016-05-24 & 09:30 & 16:00 \\ 
Nasdaq & 1996-01-03 & 2016-05-24 & 09:30 & 16:00 \\ 
FTSE & 1996-01-03 & 1998-07-19 & 08:30 & 16:30 \\ 
 & 1998-07-20 & 1999-09-17 & 09:00 & 16:30 \\ 
 & 1999-09-18 & 2016-05-24 & 08:00 & 16:30 \\ 
DAX & 1996-01-03 & 1999-09-17 & 08:30 & 17:00 \\ 
 & 1999-09-18 & 2016-05-24 & 09:00 & 17:30 \\ 
CAC & 1996-01-03 & 1999-09-19 & 10:00 & 17:00 \\ 
 & 1999-09-20 & 2000-04-02 & 09:00 & 17:00 \\ 
 & 2000-04-03 & 2016-05-24 & 09:00 & 17:30 \\ 
Nikkei & 1996-01-03 & 2006-01-18 & 09:00 & 11:00 \\ 
 &  &  & 12:30 & 15:00 \\ 
 & 2006-01-19 & 2006-04-23 & 09:00 & 11:00 \\ 
 &  &  & 13:00 & 15:00 \\ 
 & 2006-04-24 & 2011-11-20 & 09:00 & 11:00 \\ 
 &  &  & 12:30 & 15:00 \\ 
 & 2011-11-21 & 2016-05-24 & 09:00 & 11:30 \\ 
 &  &  & 12:30 & 15:00 \\ 
HSI & 1996-01-03 & 2011-03-06 & 10:00 & 12:30 \\ 
 &  &  & 14:30 & 16:00 \\ 
 & 2011-03-07 & 2012-03-04 & 09:30 & 12:00 \\ 
 &  &  & 13:30 & 16:00 \\ 
 & 2012-03-05 & 2016-05-24 & 09:30 & 12:00 \\ 
 &  &  & 13:00 & 16:00 \\ 
SSEC & 1996-01-03 & 2016-05-24 & 09:30 & 11:30 \\ 
 &  &  & 13:00 & 15:00 \\ 
AORD & 1996-01-03 & 2016-05-24 & 10:00 & 16:00 \\ 
\bottomrule
\end{tabular}
\caption{\small History of session times for each of the ten indices. The sample period is from January 3, 1996 to May 24, 2016. For exchanges that have lunch breaks, morning and afternoon session times are recorded on separate lines.}
\label{tab:session_times}
\end{table}
\FloatBarrier

\section{Robust Outlier Score using Fast L1 Splines}
\label{app:outscr}
Let $\boldsymbol{\zeta}_{t} = (\zeta_{t,1} \ddd \zeta_{t,n_{t}})$ denote the $n_{t}$-dimensional vector of one-minute prices for day $t$. The corresponding vector of outlier scores $\boldsymbol{\delta}_{t} \in \mathbb{R}^{n_{t}}$ is computed by the following steps:
\begin{algorithmic}[1]
\State{$\hat{\boldsymbol{\zeta}}_{t} \gets \argmin_{\mathbf{z} \in \mathbb{R}^{n_{t}}} \lVert \mathbf{z} - \boldsymbol{\zeta}_{t} \rVert_{1} + \Lambda_{1} \lVert \mathbf{D}\mathbf{z} \rVert_{2}^{2}$}
\State{$\boldsymbol{\sigma}_{t} \gets \log \left|\boldsymbol{\zeta}_{t} - \hat{\boldsymbol{\zeta}}_{t}\right|$}
\State{$\hat{\boldsymbol{\sigma}}_{t} \gets \argmin_{\mathbf{z} \in \mathbb{R}^{n_{t}}} \lVert \mathbf{z} - \boldsymbol{\sigma}_{t} \rVert_{1} + \Lambda_{2} \lVert \mathbf{D}\mathbf{z} \rVert_{2}^{2}$}
\State{$\boldsymbol{\varepsilon}_{t} \gets (\boldsymbol{\zeta}_{t} - \hat{\boldsymbol{\zeta}}_{t}) \oslash \hat{\boldsymbol{\sigma}}_{t}$}
\State{$\boldsymbol{\delta}_{t} \gets \left| (\boldsymbol{\varepsilon}_{t} - \mathrm{Median}(\boldsymbol{\varepsilon}_{t})) \oslash \mathrm{IQR}(\boldsymbol{\varepsilon}_{t}) \right|$}
\end{algorithmic}
In the above steps, $\lVert \cdot \rVert_{p}$ denotes the L$p$ norm of a vector, $\oslash$ denotes the Hadamard (element-wise) division, $\mathrm{Median}(\cdot)$ is the sample median operator, and $\mathrm{IQR}(\cdot)$ is the sample inter-quartile range operator. In steps 1 and 3, we use the fast L1 smoothing splines developed in \citet{TepperSapiro2012} and \citet{TepperSapiro2013} to compute robust nonparametric approximations to the the intra-daily trend and scale functions. The key feature of the L1 spline is its robustness against outliers. The $\mathbf{D}$ matrix is the standard discrete second-order differential operator, defined as
\begin{equation*}
\mathbf{D} =
\begin{bmatrix}
-1 &  1 & & & \\
1 & -2 & 1 & & \\
& \ddots & \ddots & \ddots & \\
& & 1 & -2 & 1 \\
& & & 1 & -2
\end{bmatrix}.
\end{equation*}
As pointed out by the authors, the L1 optimisation problems in steps 1 and 3 can be solved very efficiently using the split-Bregman method. The values of the smoothing parameters are $\Lambda_{1} = 50$ and $\Lambda_{2} = 50000$. Using experiments, we find these values to work well for one-minute equity prices. The important rule-of-thumb here is that the scale function needs to vary much more smoothly than the trend function, i.e., $\Lambda_{2} \gg \Lambda_{1}$, in order to avoid ``over-cleaning".

\end{onehalfspace}
\end{document}